\documentclass[12pt]{article}
\pdfoutput=1
\usepackage{epsf, amssymb,amsmath}
\usepackage{cite}
\usepackage{epsfig}

\usepackage{amsmath}
\usepackage{latexsym}
\usepackage{epsf}
\usepackage[english]{babel}
\usepackage{graphicx,color}
\usepackage{url}

\usepackage[colorlinks=true,linktocpage=true,linkcolor=blue,]{hyperref}
\newcommand{\arXiv}[1]{\href{http://www.arXiv.org/abs/#1}{#1}}

\makeatletter
\renewcommand\section{\@startsection {section}{1}{\z@}%
                               {-3.5ex \@plus -1ex \@minus -.2ex}
                               {2.3ex \@plus.2ex}%
                               {\normalfont\large\bfseries}}
\renewcommand\subsection{\@startsection{subsection}{2}{\z@}%
                                 {-3.25ex\@plus -1ex \@minus -.2ex}%
                                 {1.5ex \@plus .2ex}%
                                 {\normalfont\bfseries}}
\makeatother

\numberwithin{equation}{section}

\def\Im{{\rm Im ~}}

\def\d{\delta}

\def\eps{\epsilon}

\def\r{\rho}
\def\w{\omega}

\def\t{\tilde}

\def\F{{\cal F}}

\def\({\left (}
\def\){\right )}

\newcommand{\del}{\partial}

\newcommand{\be}{\begin{equation}}
\newcommand{\ee}{\end{equation}}
\newcommand{\beq}{\begin{equation}}
\newcommand{\eeq}{\end{equation}}
\newcommand{\bea}{\begin{eqnarray}}
\newcommand{\eea}{\end{eqnarray}}
 \newcommand{\ba}{\begin{eqnarray}}
\newcommand{\ea}{\end{eqnarray}}

\newcommand{\sores}{SO(2)_{\textrm{res}} }

\begin{document}

\begin{titlepage}
\begin{flushright}
\end{flushright}

\begin{center}
\vskip 1.5cm  
{\Large {\bf Holographic pump probe spectroscopy}}
\vskip 1cm

\renewcommand{\thefootnote}{\fnsymbol{footnote}}

 {\large A.~Bagrov,$^a$ B.~Craps,$^{b}$ F.~Galli,$^{c}$ V.~Ker\"anen,$^d$ \\ \vskip3mm
 E.~Keski-Vakkuri,$^d$ J.~Zaanen$^e$}
 \vskip5mm

{$^a$Institute for Molecules and Materials, Radboud University,\\ Nijmegen, The Netherlands \\
$^b$Theoretische Natuurkunde, Vrije Universiteit Brussel (VUB) and \\ The International Solvay Institutes, Brussels, Belgium \\
$^c$Perimeter Institute for Theoretical Physics, Waterloo, Ontario, Canada\\
$^d$Department of Physics, University of Helsinki, Helsinki, Finland\\
$^e$Instituut-Lorentz for Theoretical Physics, Universiteit Leiden, \\Leiden, The Netherlands}

\vskip 4mm
{\small\noindent  {\tt  abagrov@science.ru.nl, Ben.Craps@vub.be,  fgalli@perimeterinstitute.ca, vkeranen1@gmail.com, esko.keski-vakkuri@helsinki.fi, jan@lorentz.leidenuniv.nl }}

\end{center}
 \vfill

\begin{center}
{\bf ABSTRACT}
\vspace{3mm}
\end{center}
We study the non-linear response of a 2+1 dimensional holographic model with weak momentum relaxation and finite charge density to an oscillatory electric field pump pulse.  Following   the time evolution of one point functions after the pumping has ended, we find that deviations from thermality are   well captured within the linear response theory.  For electric pulses with  a negligible zero frequency component the response approaches the instantaneously thermalizing form typical of holographic Vaidya models.  We link this to the suppression of the amplitude of the quasinormal mode that governs the approach to equilibrium.  In the large frequency limit, we are also able to show analytically that the holographic geometry takes the Vaidya form. A simple toy model captures these features of our holographic setup. 
Computing the out-of-equilibrium probe optical conductivity after the pump pulse, we similarly find that for high-frequency pulses the optical conductivity reaches its final equilibrium value effectively instantaneously. Pulses with significant DC components show exponential relaxation governed by twice the frequency of the vector quasinormal mode that governs the approach to equilibrium for the background solution. We explain this numerical factor in terms of a simple  symmetry argument. 
\end{titlepage}

\tableofcontents

\section{Introduction}

The gauge/gravity duality applied within the context of strongly correlated many-body quantum systems started out as an interesting, yet limited, source of intuition on some properties of quantum critical matter. In the past decade, it has evolved into a powerful framework capable of taking into account a number of phenomenological aspects that should not be neglected when dealing with realistic models, such as crystal lattices, disorder, non-relativistic dispersion relations,  etc.  \cite{Ammon:2015wua,Zaanenetalbook}.

An important advantage of the holographic approach is its capacity of describing within a unique framework both equilibrium and out-of-equilibrium quantum systems by mapping them to tractable problems in general relativity, which can be systematically analyzed in real time without any need for conceptually new approaches.  In the past, most of the attention towards far-from-equilibrium situations in this framework has gone to the formation of quark gluon plasmas in heavy ion collisions. Holographic models relevant for this process incorporating a number of realistic features have been suggested and explored, with interesting results in relation with experiments \cite{DeWolfe:2013cua}. 
Studies of far from equilibrium situations directly relevant to condensed matter systems have been, on the other hand, relatively scarce and mostly limited to toy models (see e.g. \cite{Murata:2010dx,Bhaseen:2012gg,Chesler:2014gya,Sonner:2014tca,Callebaut:2014tva,Das:2014lda,Zeng:2016api,Zeng:2016gqj,Camilo:2015wea,Withers:2016lft}). 
At the same time, recent advances in ultrafast experimental techniques in condensed matter physics have put a demand for a theoretical framework capable of explaining and predicting observed phenomena \cite{Orenstein,DalConte,Giannettietal,Freericks}. 
One is therefore led to ask whether, given the current state-of-the-art,  time-dependent holographic models can make contact with experiments. 
In this paper we make a step in this direction by proposing a model for  pump-probe experiments in which one follows the optical response of a holographic strange metal after it has been  taken into a highly excited state by an electromagnetic pulse.  

Our starting point is the minimal model considered in \cite{Andrade:2013gsa}, which describes a 2+1 dimensional strange metal at finite temperature  and density, in  presence of a weak momentum relaxation mechanism obtained  through axion fields linear in the boundary spatial coordinates. This efficiently reproduces the effects of explicit translational symmetry breaking \cite{Horowitz:2012ky} while preserving a homogeneous and isotropic bulk geometry (see also \cite{Vegh:2013sk,Blake:2013owa,Donos:2013eha} for related holographic models).
To mimic a pump pulse, we   quench the holographic system by applying for a finite amount of time an oscillatory electric field. For simplicity we take it to be in the form of a modulated Gaussian wave packet of mean frequency $\omega_P$. In this way the system is driven  into a highly excited out-of-equilibrium state, which then relaxes towards a new equilibrium state at a higher temperature,  but equal charge density.

In contrast with the zero density case where, both with  \cite{Horowitz:2013mia} and without \cite{Bardoux:2012aw} momentum relaxation the bulk dynamics results into a simple Vaidya geometry, at finite density the response of the system to the external electric field becomes more complicated. In fact, although the electric field always sets the charges into motion, explicitly breaking spatial isotropy and inducing on the boundary non-trivial currents, at finite density this also causes non-zero momentum densities. Holographically this  corresponds to having additional metric components and field excitations, which generically make the problem not treatable analytically.  
 
We study the resulting non-linear bulk dynamics with numerical methods and follow the evolution of the boundary one point functions as $\omega_P$ is varied.

Although we work in the non-linear regime,  we find that deviations from thermality after the pump pulse ends are surprisingly well captured within the linear response theory  and their decay  controlled by quasinormal modes (QNMs). 
In particular, at zero frequency the purely imaginary longest lived QNM  of the vector sector  governs the decay toward the new equilibrium configuration.\footnote{For the specific case of zero mean frequency, a related analysis has previously been performed in \cite{Withers:2016lft}. There the one point functions of electric and heat currents, as well as the QNMs that control their decay were studied in detail, also  away from the weak momentum relaxation regime considered here.}  As the pump frequency is dialed up,  we find that the response of the bulk geometry is increasingly well approximated by a bulk solution of the Vaidya form. That is, we observe that as soon as the pump electric field is turned off  the boundary one point functions almost instantaneously approach their final equilibrium configurations.  In fact, in the limit  $\w_P  \to \infty$ we are also able to show analytically that the bulk solution takes precisely the Vaidya form, and  one point functions thermalize instantaneously.  

From the bulk point of view the origin of this dynamics can be understood from an analysis of QNM amplitudes.  As we show explicitly, the amplitude of each QNM contribution is determined by the Fourier transform of the electric pump pulse evaluated at the frequency of the mode in question.  The almost instantaneous approach to thermality at large frequencies is then explained by the absence of overlap between the pump spectrum and the frequency of long lived modes. A very simple toy model realized in terms of a  driven harmonic oscillator effectively captures the main features of the bulk solution. 

With the numerical background in hand we then proceed to compute the main observable of interest for a pump-probe experiment,  the probe optical conductivity after the quench.  In the same way as in a pump-probe experiment, we consider the optical response of the holographic strange metal to a probe pulse that is applied only after the pump has ended. This is incorporated in the definition of out-of-equilibrium conductivity we adopt.  

Similarly to what happens for the background solution, the conductivity thermalizes almost instantaneously whenever the pump pulse has a negligible DC component. This behavior, although surprising from the boundary point of view,  is completely natural  with the insight  provided by the  analysis of the bulk background  solution: If the geometry is described by a Vaidya solution, by causality in the bulk, the response to any perturbation applied after the light-like Vaidya shell will be insensitive to any detail of the quench other than the final equilibrium configuration.     
On the other hand, we find that for pump pulses with a DC component the optical conductivity relaxes with a rate set by twice the lowest vector QNM frequency. The appearance of this QNM, which governs momentum relaxation, can be understood from the fact that the zero frequency component in the pulse corresponds to a static electric field, which accelerates the finite density system. When the pulse is over, the resulting finite momentum has to relax in order to reach equilibrium. To explain the factor of two, which is less intuitive from a boundary point of view, we provide a careful but general analysis of linearized bulk fluctuations relying on the symmetries of the final equilibrium configuration.

A brief summary of our main results appeared before in \cite{Bagrov:2017tqn}. There we proposed this model as an idealized setup for realistic pump-probe experiments, and the almost instantaneous thermalization as an extreme limit of fast thermalization that might manifest itself experimentally in certain regimes, similarly to what has been observed in the creation of quark gluon plasma in heavy ion collisions.
In this paper, we present the computations behind them, as well as a number of new results, including the surprisingly good estimate of the size of non-thermality from a linear response analysis, and the explanation based on symmetry of the appearance of twice the lowest vector QNM frequency. 

The rest of the paper is organized as follows. In the next section, we define the bulk model of a strange metal with momentum dissipation and briefly review its equilibrium properties. In Sec.~\ref{sec:numerics}, details of the used numerical techniques are outlined. In Sec.~\ref{sec:noneqbst}, we provide the non-equilibrium background solution computed  numerically and discuss the behavior of the corresponding  boundary one point   functions. In Sec.~\ref{sec:toy}, we introduce a toy model of a rapidly driven oscillator, which captures some of the important features of our holographic model and makes the phenomenological picture more transparent. Sec.~\ref{sec:conduct} contains the main physical result of the paper, the time-dependent AC conductivity. Finally, in Sec.~\ref{sec:conjecture} we conclude with a general discussion of our results, including prospects and challenges for comparison with experiment.


\section{The model} \label{sec:model}

The model we want to consider is specified by the action
 \beq \label{eq:action}
S = \frac{1}{2\kappa_4^2}\int d^4x\,\sqrt{-g}\Big[R - 2\Lambda - \frac{1}{2}\sum_{I = 1}^2(\partial\phi_I)^2 
- \frac{1}{4} F^2\Big] \, , 
\eeq
with $\Lambda =-3 $ and equations of motion 
\begin{align} \label{eq:eqs}
&E_{\mu\nu} \equiv G_{\mu\nu} + g_{\mu\nu}  \Lambda -   \frac{1}{2}\( g^{\rho\sigma} F_{\mu\rho}F_{\nu\sigma} -\frac{1}{4}g_{\mu\nu}F^2 \)   -\frac{1}{2} \sum^{d-1}_{I}\( \del_{\mu} \phi_I \del_{\nu} \phi_I  - \frac{1}{2}g_{\mu\nu} (\del\phi_I)^2 \) =0   \nonumber \\ 
&M_{\nu} = \nabla_{\mu}F^{\mu}_{\phantom{\mu}\nu}  = \frac{1}{\sqrt{-g}}\del_{\mu}\(\sqrt{-g} F^{\mu\sigma} \)g_{\sigma\nu}   = 0\, ,\\
&\Box \phi_{I} \equiv  \frac{1}{\sqrt{-g}}\del_{\mu}\(\sqrt{-g} \del^{\mu} \phi_I \)    = 0 \, . \nonumber 
\end{align}

This admits a homogeneous and isotropic charged black brane configuration  with non-trivial scalar fields profiles  \cite{Bardoux:2012aw} that  was explored in \cite{Andrade:2013gsa}  as a simple holographic model for spatial translational symmetry breaking. Such a configuration has scalar fields with a linear dependence on the spatial coordinates   $x^i=x, y$  common to the dual field theory
\beq
\phi_1 = k x,\quad \phi_2 = k y, \label{eq:scalarsk}
\eeq
and  translationally invariant geometry and gauge field 
\begin{align} \label{eq:eqsol}
&ds^2 = \frac{1}{z^2}\Big( - f dt^2 + \frac{dz^2}{f} + dx^2 + dy^2\Big) \, ,\nonumber 
\\
&f(z) = 1 - \frac{1}{2}k^2 z^2 - m z^3 +\frac{1}{4}\rho^2 z^4 \, , 
\\
&A=  (- \mu + \rho z)dt \, . \nonumber 
\end{align}

The dual field theory state is a thermal state with finite charge density and with translational symmetry breaking, whose properties can be fully specified in terms of $T, \rho$  and $k$. 
The chemical potential $\mu$ is determined in terms of the charge density  $\rho$ by requiring the regularity condition that  $A$ should vanish at the horizon of the black  brane, leading to  $\mu = \rho z_0$, with  $z_0$ being the horizon location where $f(z_0)=0$. The  location of the horizon $z_0$ is associated with the  temperature  $T$ of the field theory state through 
\be
 T =   \frac{1}{4\pi z_0} \left( 3-\frac{k^2z^2_0}{2}-\frac{\mu^2z^2_0}{4}\right)  \,   ,
\ee
which gives the Hawking temperature of the black brane geometry. 
Notice that for fixed $k, \mu$ and $z_0$, the mass parameter $m$ appearing in the gravitational solution is not an independent quantity. It is fixed by the condition $f(z_0)=0$ and is directly related to the energy density of the dual equilibrium state 
\be
\epsilon = 2m = \frac{2}{z^3_0}\left(1-\frac{k^2z^2_0}{2}+\frac{\mu^2z^2_0}{4}\right) \,  \label{eq:epsilon} 
\ee
and the isotropic pressure
\be
p = \frac{\eps}{2} =m \, .
\ee
Finally,  the entropy density of this configuration is 
\be
s = \frac{4 \pi }{z_0^2} \label{eq:entropy} \, . 
\ee

The reason why such a holographic solution  with a completely homogeneous and isotropic geometry can be used to effectively describe momentum dissipation can be grasped  from the Ward identities 
\begin{align}
&\nabla_{\mu} \langle T^{\mu\nu} \rangle=  \nabla^{\nu}\varphi_{I} \langle \mathcal{ O}_I  \rangle  + \F^{\nu\mu}  \langle J_{\mu} \rangle\ , \label{eq:wardT} \\
&\nabla_{\mu}  \langle J^{\mu} \rangle  =0 \, . \label{eq:wardJ}
\end{align}
Following the standard AdS/CFT dictionary,  the operators  $\mathcal{ O}_{I}$  are dual to the bulk scalars and the couplings $\varphi_{I}$  are directly related to the asymptotic values of the bulk scalar profiles, that is  $\varphi_{I} = k x^i \delta_{i,I}$ in our case. Similarly the $U(1)$ current $J^{\mu}$ is dual to  the AdS gauge field $A_{\mu}$ and the boundary field strength  $\F^{\nu\mu}$ is determined in terms of the asymptotic value of  $A_{\mu}$.
 Let us first  notice that \eqref{eq:wardJ}  implies that the charge density $\rho  = \langle J^{t} \rangle$ is conserved. From \eqref{eq:wardT} instead it follows that spatial momenta $\langle T^{t i} \rangle$ will generically not be conserved whenever on the r.h.s. one has non-vanishing vevs.    The coupling between the scalar and the gauge field  in the bulk is such that  the  boundary electric field $E_i = \F_{it}$  induces a non-zero expectation value for $\mathcal{ O}_{I}$, and thus  %
\be
\del_t  \langle T^{t i} \rangle  =   k  \langle \mathcal{ O}_I \rangle \delta_{i,I}  + \rho E_{i} \, .
\ee

Before concluding this section, let  us quickly review the holographic result for the equilibrium optical conductivity computed in this model  \cite{Andrade:2013gsa}, which will be of use in the rest of the paper.  
 The probe optical conductivity $\sigma$ measures the linear response of  the boundary current $J_x$ to a  boundary probe electric field   $E_x  =  - \del_t  A^{0}_{x}$. To compute it holographically one can consider the minimal consistent set of ``vector" bulk fluctuations %
\bea \label{eq:linearized}
\delta A_{x} &=& e^{-i \w t } \delta a_x(z)   \, ,\nonumber \\
\delta g_{tx} &=& e^{-i \w t } z^{-2} \delta h_{tx}(z) \, , \\
\delta \phi_{1} &=& e^{-i \w t} k^{-1} \delta \varphi(z) \, , \nonumber 
\eea
around the equilibrium background (\ref{eq:scalarsk}\,--\,\ref{eq:eqsol}), and use the relation between the AdS asymptotic modes of   $\delta A_{x}$ and the boundary quantities
\be
\delta A_{x} \approx   A^{0}_{x} +   z \langle  J_{x} \rangle + \dots \, 
\ee
to write 
\be
\sigma(\w)  = \frac{\langle  J_{x} \rangle}{i \omega   A^{0}_{x} }   \, .
\ee
The computation of the optical conductivity therefore amounts to solving the following system of linearized equations for the fluctuations \eqref{eq:linearized}   
\begin{align}
& \del^2_{t}\delta \phi_{1}  - z^2 f^2  \del_z \( \frac{f}{z^2}\del_z \delta \phi_{1} \)   -k z^2  \del_{t}\delta g_{tx} = 0  \, , \nonumber  \\
&\del^2_{t}\delta A_x  -  f \del_z\( f \del_z  \delta A_x\)  - \rho  f   \del_{z}\(z^2 \delta g_{tx} \)  = 0 \,  , \label{eq:linfluc}   \\
&   \del_{z}\(z^2  \del_{t}\delta g_{tx} \) + \rho  z^2 \del_{t}\delta A_x - k f \del_z \delta \phi_{1}  =0   \nonumber\, ,
\end{align}
subject to  appropriate asymptotically AdS boundary condition for the metric, non-vanishing source for the gauge field,  and vanishing source for the scalar fluctuation.   

\begin{figure}[t]
  \begin{center}
      \includegraphics[width=0.48 \textwidth]{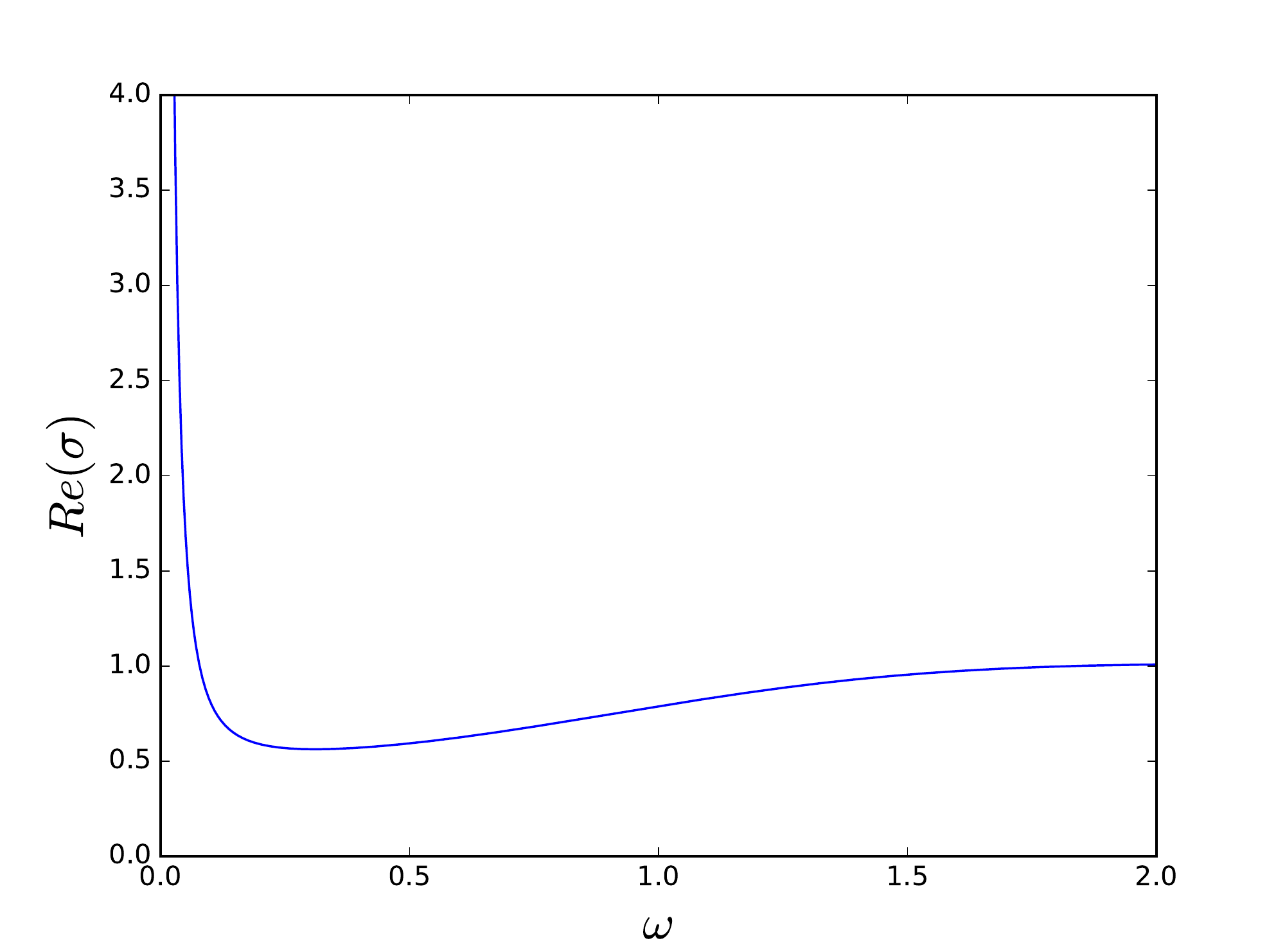}    \hfill \includegraphics[width=0.48 \textwidth]{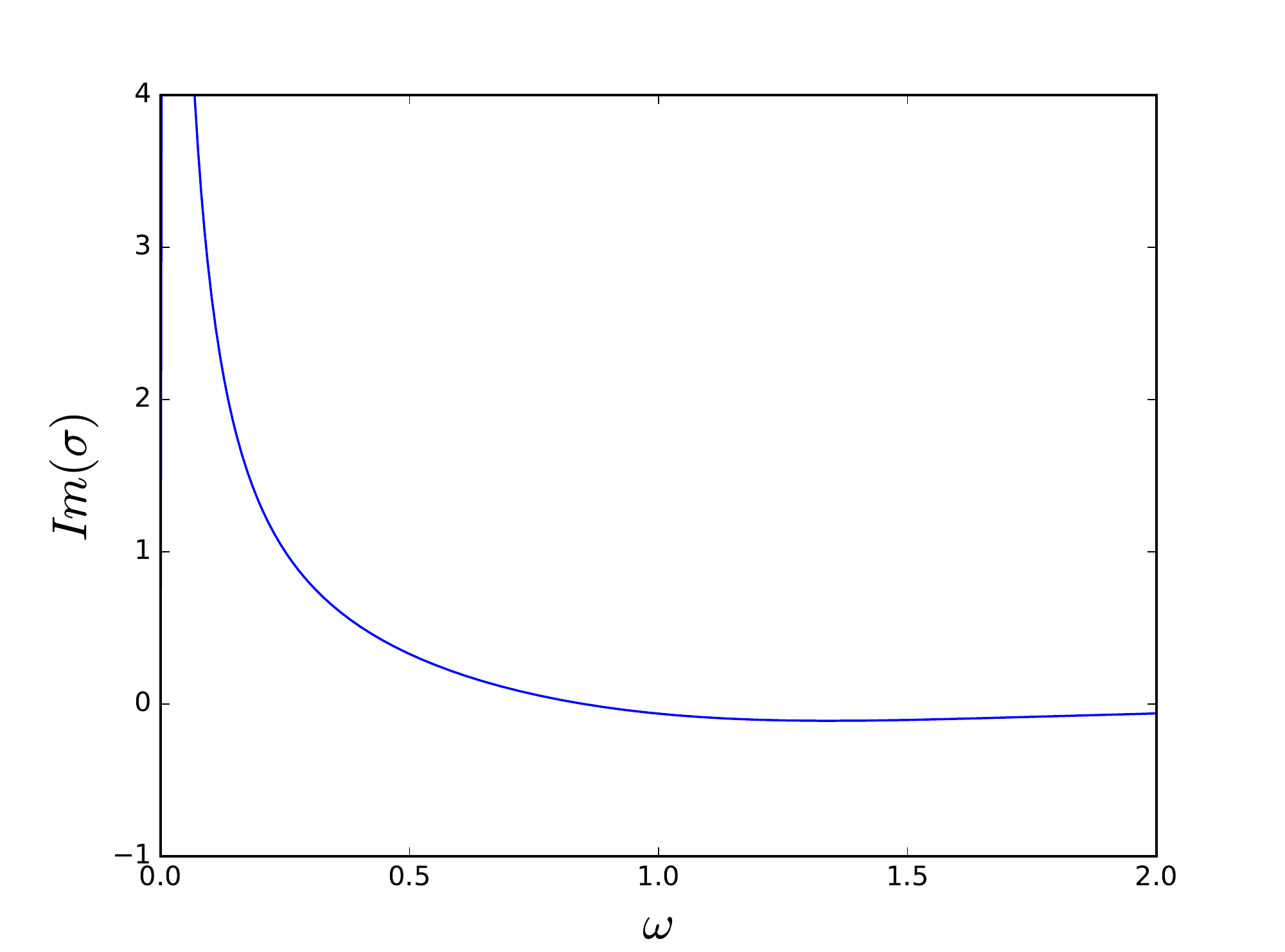}
 \end{center}
  \caption{The real (left) and imaginary part (right) of the optical conductivity for $k = 0.2$, $T = 0.2$ , and  $\mu = 1.0$.   }
  \label{fig:eqconductivity}
\end{figure}

Away from the zero-frequency limit these equations can be solved numerically, and one finds that for small enough values of $k$ as compared to the other parameters of the equilibrium solution  the resulting optical conductivity  has the low frequency Drude form  \cite{Davison:2013jba,Vegh:2013sk,Andrade:2013gsa}. In Fig.~\ref{fig:eqconductivity}, we reproduce a sample plot of the optical conductivity for  $k = 0.2$, $T = 0.2$, and  $\mu = 1.0$.  
The finite value of the zero-frequency DC conductivity can be obtained analytically  \cite{Andrade:2013gsa} 
 \be
 \sigma_{DC} = 1+ \frac{\mu^2}{k^2} \, .
 \ee
The relaxation rate $\tau_{Q}$ associated to the Drude peak corresponds to the purely imaginary frequency of the lowest lying quasinormal mode of the bulk vector preturbations. In the small $k$ regime we will be interested in, this has been obtained analytically in \cite{Davison:2013jba} and reads
\be  \label{eq:relax}
\frac{1}{\tau_{Q} }\approx  \frac{s k^2}{6 \pi \epsilon} \, .
\ee


\section{Setup and details on numerical calculation \label{sec:numerics}} 
 
To study the response of the model of the previous section to the boundary electric field,  we go to ingoing Eddington-Finkelstein coordinates and, following \cite{Withers:2016lft}, consider the ansatz 
\begin{align} \label{eq:ansatz}
ds^2 &= - F_z( z,v)dv^2 - \frac{2 dv dz}{z^2} + 2F_x(z, v) dx dv 
+ \Sigma(z, v)^2(e^{-B(z, v)}dx^2 + e^{B(z, v)}dy^2)\, .\nonumber
\\
A &= (E_x(v) x + a_v(z, v))dv + a_x(z, v)dx \, ,\nonumber
\\
\phi_1 &= k x + \Phi(z, v)\, ,
\\
\phi_2 &= k y \, .\nonumber
\end{align}
We solve the resulting system imposing appropriate asymptotically AdS boundary conditions under the assumption that for early enough times, when the  pulse $E_x(v)$ has not been turned on yet, the  solution coincides with the equilibrium configuration of Sec.~\ref{sec:model}. 
By now, there are standard methods for solving such numerical relativity systems (see e.g. \cite{Chesler:2008hg,Heller:2013oxa, Chesler:2013lia, Ecker:2015kna}). We will review the main ingredients of this procedure below.

Inspecting Einstein's equations, it is convenient to define derivative operators along ingoing and outgoing radial 
null geodesics, which act on a field $X(z, v)$ as follows:
\begin{align} 
X' &= \partial_z X \, ,  
\\
\dot{X} &= \partial_v X - \frac{z^2}{2}F_z\partial_z X \, . 
\end{align}
With this notation, the equations of motion become
{ \allowdisplaybreaks
\begin{align}
 0 &= \Sigma'' + \frac{2}{z}\Sigma' + \frac{1}{4}((B')^2 + (\Phi')^2)\Sigma + \frac{e^{B}(a_x')^2}{4\Sigma},
\label{eq:eom1}
\\
0 &= F_x'' + \Big(\frac{2}{z} + B'\Big) F_x' + \Big(B'' - \frac{2(\Sigma')^2}{\Sigma^2} + \frac{2B'\Sigma'}{\Sigma} + \frac{(\Phi')^2}{2} + \frac{(B')^2}{2} + \frac{2B'}{z} \nonumber
\\
&\hspace{5cm}  - \frac{e^B (a_x')^2}{2\Sigma^2}\Big) F_x + \frac{k}{z^2}\Phi' 
+ a_v'a_x'  \, ,
\label{eq:eom2}
\\
0 &= a_v'' + \Big(\frac{2}{z} + \frac{2\Sigma'}{\Sigma}\Big)a_v' - \frac{e^B F_x a_x''}{\Sigma^2} - \frac{e^B}{\Sigma^2}
\Big(F_x B' + F_x' + \frac{2 F_x}{z}\Big)a_x' \, ,
\label{eq:eom3}
\\
0 &= \dot{\Sigma}' + \frac{\Sigma'}{\Sigma}\dot{\Sigma} + \frac{3\Sigma}{2z^2} - \frac{z^2}{8}\Sigma (a_v')^2 - \frac{k^2 e^{-B}}{8z^2\Sigma} - \frac{e^{B}}{\Sigma}\Big(\frac{k^2}{8z^2} - \frac{z^2 e^B F_x^2(a_x')^2}{8\Sigma^2} \nonumber
\\
&+ \frac{z}{2} F_x^2B' + \frac{z^2}{8}F_x^2(B')^2 + z F_x F_x' + \frac{z^2}{2}B' F_x F_x' + \frac{z^2}{8}(F_x')^2 + \frac{1}{4}k F_x \Phi'  \nonumber
\\
&+ \frac{z^2}{2\Sigma}F_x^2 B'\Sigma' + \frac{z^2}{2\Sigma}F_x F_x'\Sigma' - \frac{z^2 F_x^2 (\Sigma')^2}{2\Sigma^2} + \frac{z^2}{4}F_x^2(B'' + F_x'')\Big)  \, ,
\label{eq:eom4}
\\
0&=\dot{B}' +\frac{\dot{B} \Sigma'}{\Sigma} -\frac{e^{-B} k^2}{4 z^2 \Sigma^2}+\frac{e^{B} k^2}{4 z^2 \Sigma^2}+\frac{e^{2 B} z^2 F_x^2 a_x'^2}{4 \Sigma^4}-\frac{e^{B} z^2 F_x^2 B'^2}{4 \Sigma^2}\nonumber
\\
&+\frac{e^{B} z^2 F_x'^2}{4 \Sigma^2}+\frac{e^{B} z^2 F_x^2 \Sigma'^2}{\Sigma^4}-\frac{e^{B} \dot{a}_x a_x'}{2 \Sigma^2}+\frac{e^{B} E_x(v) a_x'}{2 \Sigma^2}+\frac{\dot{\Sigma} B'}{\Sigma}-\frac{e^{B} z F_x^2 B'}{\Sigma^2}
\nonumber
\\
&-\frac{e^{B} z^2 F_x B' F_x'}{2 \Sigma^2}-\frac{e^{B} z^2 F_x^2 B' \Sigma'}{\Sigma^3}-\frac{e^{B} z^2 F_x F_x' \Sigma'}{\Sigma^3}-\frac{e^{B} z^2 F_x^2 B''}{2 \Sigma^2} \, ,
\label{eq:eom5}
\\
0&=\dot{a}_x'+\frac{1}{2} \dot{a}_x B' +\frac{1}{2} \dot{B} a_x'-\frac{1}{2} F_x a_v' B' z^2-\frac{1}{2} a_v' F_x' z^2
-\frac{1}{2} F_x a_v'' z^2
\nonumber
\\
&-F_x a_v' z -\frac{1}{2} E_x(v) B' \, ,
\label{eq:eom6}
\\
0&= \dot{\Phi}'+\frac{\dot{\Phi} \Sigma'}{\Sigma} -\frac{e^{B} F_x^2 B' \Phi' z^2}{2 \Sigma^2}-\frac{e^{B} F_x F_x' \Phi' z^2}{\Sigma^2}-\frac{e^{B} F_x^2 \Phi'' z^2}{2 \Sigma^2}-\frac{e^{B} F_x^2 \Phi' z}{\Sigma^2}\nonumber
\\
&-\frac{e^{B} k F_x B'}{2 \Sigma^2}-\frac{e^{B} k F_x'}{2 \Sigma^2}+\frac{\dot{\Sigma} \Phi'}{\Sigma} \, ,
\label{eq:eom7}
\\
0&=F_z'' +\frac{2 F_z'}{z} + \frac{e^{-B} k^2}{2 z^4 \Sigma^2}-\frac{e^{B} k^2}{2 z^4 \Sigma^2}+\frac{e^{B} F_x \Phi' k}{z^2 \Sigma^2}-\frac{1}{2} a_v'^2-\frac{e^{2 B} F_x^2 a_x'^2}{\Sigma^4}\nonumber
\\
&+\frac{e^{B} F_x^2 B'^2}{\Sigma^2}+\frac{e^{B} F_x'^2}{2 \Sigma^2}+\frac{e^{B} F_x^2 \Phi'^2}{2 \Sigma^2}-\frac{4 e^{B} F_x^2 \Sigma'^2}{\Sigma^4}+\frac{e^{B} F_x a_v' a_x'}{\Sigma^2}+\frac{e^{B} \dot{a}_x a_x'}{z^2 \Sigma^2}\nonumber
\\
&-\frac{e^{B} E_x(v) a_x'}{z^2 \Sigma^2}-\frac{\dot{B} B'}{z^2}-\frac{2 \dot{\Sigma} B'}{z^2 \Sigma}+\frac{4 e^{B} F_x^2 B'}{z \Sigma^2}-\frac{2 \dot{B}'}{z^2}+\frac{3 e^{B} F_x B' F_x'}{\Sigma^2} 
\nonumber
\\
&+\frac{4 e^{B} F_x F_x'}{z \Sigma^2}-\frac{\dot{\Phi} \Phi'}{z^2}+\frac{4 e^{B} F_x^2 B' \Sigma'}{\Sigma^3}+\frac{2 e^{B} F_x F_x' \Sigma'}{\Sigma^3}-\frac{2 \dot{B} \Sigma'}{z^2 \Sigma}-\frac{4 \dot{\Sigma}'}{z^2 \Sigma} 
\nonumber
\\
&+\frac{2 e^{B} F_x^2 B''}{\Sigma^2}+\frac{2 e^{B} F_x F_x''}{\Sigma^2}-\frac{6}{z^4}\, .
\label{eq:eom8}
\end{align}}
Notice that we are denoting, for example, $\dot{\Phi}' = \partial_z(\dot{\Phi})$, i.e.,  the dot derivatives are taken
before the $z$ derivatives.  The above set of equations represents  a convenient set of non-redundant equations that can be  obtained from all the non identically vanishing equations of motion  following from our ansatz. More precisely, the first three equations correspond to the $E_{zz}$ and  $E_{zx}$ components of Einstein's equations and to the  $M_{v}$ component of Maxwell's equations respectively. The fourth equation  is given by $E_{zv}$ once  $E_{zz}$ is used to eliminate $\Sigma''$.  The fifth equation is  given by the linear combination $e^{2B}E_{xx} + E_{yy} $. The remaining three equations are  respectively the $M_x$ component of Maxwell's equation, the $E_{xx}$  component of Einstein's equations and the equation for the scalar field $\phi_1$.
 
The strategy for numerically solving the equations from the specified initial and boundary conditions  proceeds iteratively as follows. The fields  $(B, \Phi, a_x)$ represents the free initial data. All the other fields can then be solved from the equations of motion on a constant time slice. In more detail, after specifying the initial data at a given lightcone time, we solve (\ref{eq:eom1}) for the field $\Sigma$. This is a non-linear ordinary differential equation as no time derivatives appear. Next, equations (\ref{eq:eom2}) and (\ref{eq:eom3}) provide two coupled linear ordinary differential equations for $F_x$ and $a_v$. Given $\Sigma, F_x$ and $a_v$ on the fixed time slice, one can proceed similarly   to solve for the dotted fields. First we can solve (\ref{eq:eom4}) for $\dot{\Sigma}$. Then, we can solve the two coupled linear differential equations (\ref{eq:eom5}) and (\ref{eq:eom6}) for $\dot{B}$ and $\dot{a}_x$ and, subsequently, the linear ordinary differential equation (\ref{eq:eom7}) for $\dot{\Phi}$. Finally, via the linear ordinary differential equation (\ref{eq:eom8})  we determine  $F_z$. This way we obtain all the fields and dotted fields on the initial time slice. 
The time evolution is obtained by undoing the definition of the dotted derivative fields, to obtain a set of dynamical  equations for  $(B, \Phi, a_x)$
\begin{align} 
\partial_v B &=\dot{B} + \frac{z^2}{2}F_z\partial_z B\, , \label{eq:teom1}
\\
\partial_v \Phi & =\dot{\Phi} + \frac{z^2}{2}F_z\partial_z \Phi \, ,\label{eq:teom2}
\\
\partial_v a_x &=\dot{a}_x + \frac{z^2}{2}F_z\partial_z a_x\, .\label{eq:teom3}
\end{align}
Knowing their time derivative at a given time, we can time-evolve  $(B, \Phi, a_x)$ to the next time step and  we can repeat the above procedure to solve for all the fields on that time step. This way we can iterate the algorithm to the time we want.

To solve the equations (\ref{eq:eom1}\,--\,\ref{eq:eom8}), we use  the Chebyshev spectral method and introduce a Chebyshev grid $z_i$ in the $z$-direction. 
This way fields are replaced by vectors,  $X(z) \rightarrow X_i = X(z_i)$,  derivative operators become matrices acting on the vectors and the differential equations translate into sets of coupled equations for the different field components $X_i$. 
More precisely,   the equation for  $\Sigma$  is  non-linear and becomes a set of non-linear coupled algebraic equations for the coefficients $\Sigma_i$, which we can collectively denote  by
\beq
f_j(\Sigma_i) = 0 \, . 
\eeq
The index $j$ counts the number of components in the equation, which  is the same as the number of variables $\Sigma_i$. We solve this set of non-linear equations using the Newton-Raphson method. This method finds an approximate solution to $f_j = 0$ as follows. First we start from a guess solution $\Sigma^{(0)}_i$. Then, use the updating routine
\beq
\Sigma^{(1)}_i = \Sigma^{(0)}_i - (J^{-1})_{ij}f_j(\Sigma_i^{(0)}),\label{eq:newton}
\eeq
where $J$ is a Jacobian matrix
\beq
J_{ij} = \frac{\partial f_i}{\partial\Sigma_j}.
\eeq
Now $\Sigma^{(1)}$ should provide a vector which is closer to the solution of $f_j = 0$ than our original guess. Repeating the algorithm by taking $\Sigma^{(1)}$ as a new guess and using (\ref{eq:newton}), we again get closer to the correct solution. Iterating this algorithm many times, one should converge to the solution of $f_j = 0$. In practice the number of iterations needed  depends on how good the initial guess was. In our numerical
algorithm  we use the solution from the previous time step as the initial guess. This way we only need a few (typically 3) iterations to solve the equation of motion to the desired accuracy (around $10^{-13}$ accuracy). 
The rest of the equations (\ref{eq:eom2}\,-\,\ref{eq:eom8}) are  all linear in the unknown variables and can be  straightforwardly solved by standard matrix inversion methods. We have used the numpy.linalg.solve and numpy.linalg.inv functions, which are included in the Python Numpy package and are based on the  LAPACK library.  

At the practical level, when solving  (\ref{eq:eom1}\,-\,\ref{eq:eom8}),  in order to simplify the numerics, we also find it convenient to subtract or rescale the near boundary behavior of some of  the fields. In particular, we work with the set of regularized fields  $X_{r}$ defined as follows 
\begin{align} \label{eq:asysub}
&F_z = \frac{1}{z^2}(1 - \frac{1}{2}k^2 z^2 + z^2 F_{z, r}) \, ,  & &   \dot{\Phi} = -\frac{3}{2}\dot{\Phi}_r   \, ,  \nonumber \\
&\Sigma = \frac{1}{z}( 1 + z^2 \Sigma_r) \, ,  & & \dot{\Sigma} = \frac{1}{2z^2}-\frac{k^2}{4} + z\dot{\Sigma}_r\,  ,\nonumber \\
& B= z^2 B_r  \, , &\qquad & \dot{B} = -\frac{3}{2}z^2\dot{B}_r \, , \\
& \Phi = z^2 \Phi_r \,, & &  \dot{a}_x = -\frac{1}{2}\dot{a}_{x, r}\, . \nonumber 
\end{align}
The functions $F_x,\,a_x,\,a_v$ are left intact.
Note that in these redefinitions, $\dot{\Phi}_r$ is not the dot derivative acting on $\Phi_r$, but a new variable defined through the equation in (\ref{eq:asysub}). The same applies for all the other dotted fields. 

In addition to (\ref{eq:eom1}\,-\,\ref{eq:eom8}), there are other components of Einstein's and Maxwell's equations that do not vanish identically for our ansatz. These are in principle redundant with (\ref{eq:eom1}\,-\,\ref{eq:eom8}), but in practice they are useful for testing our numerical solutions. 

To  evolve  $(B, \Phi, a_x)$   we use the fourth order explicit Runge-Kutta method. The time domain is divided in discrete time-steps $v_n$ and the value of a field at step $n+1$, $X(v_{n+1})$, is obtained as the value at step  $X(v_n)$, plus the weighted average of four different time increments  determined  in terms 
of (\ref{eq:teom1}\,-\,\ref{eq:teom3}). 

Using this algorithm, we can solve the full numerical problem modelling the pump probe experiment. 
In practice we have found it computationally faster to separate the problem of determining the probe
conductivity as a separate problem. Thus, we use the above algorithm to solve for the spacetime corresponding
to the system subject to the pump electric field. To obtain the probe conductivity, we
linearize the equations of motion around the numerically known background spacetime and solve them using
a similar numerical procedure as above. The main advantage of this procedure is that when solving the 
linearized equations of motion, we do not need to use the Newton-Raphson method, but all the equations
are solved using linear algebra, which is faster. This becomes particularly useful when we consider several
probe ``experiments" for the same pump pulse. We have checked the numerical accuracy of solving the linearized system
by comparing the results to those obtained using the (slower) full code for both pump and probe parts.
Further checks are provided by testing the system on the equilibrium states, in particular  by comparing
the numerically obtained conductivity with  the analytic formula for the DC conductivity. These agree to a very
good accuracy (in the cases we have tested they agree up to $10^{-7} \%$ accuracy).

\subsection{Numerical error estimate}
\begin{figure}[ht]
\begin{center}
\includegraphics[scale=0.6]{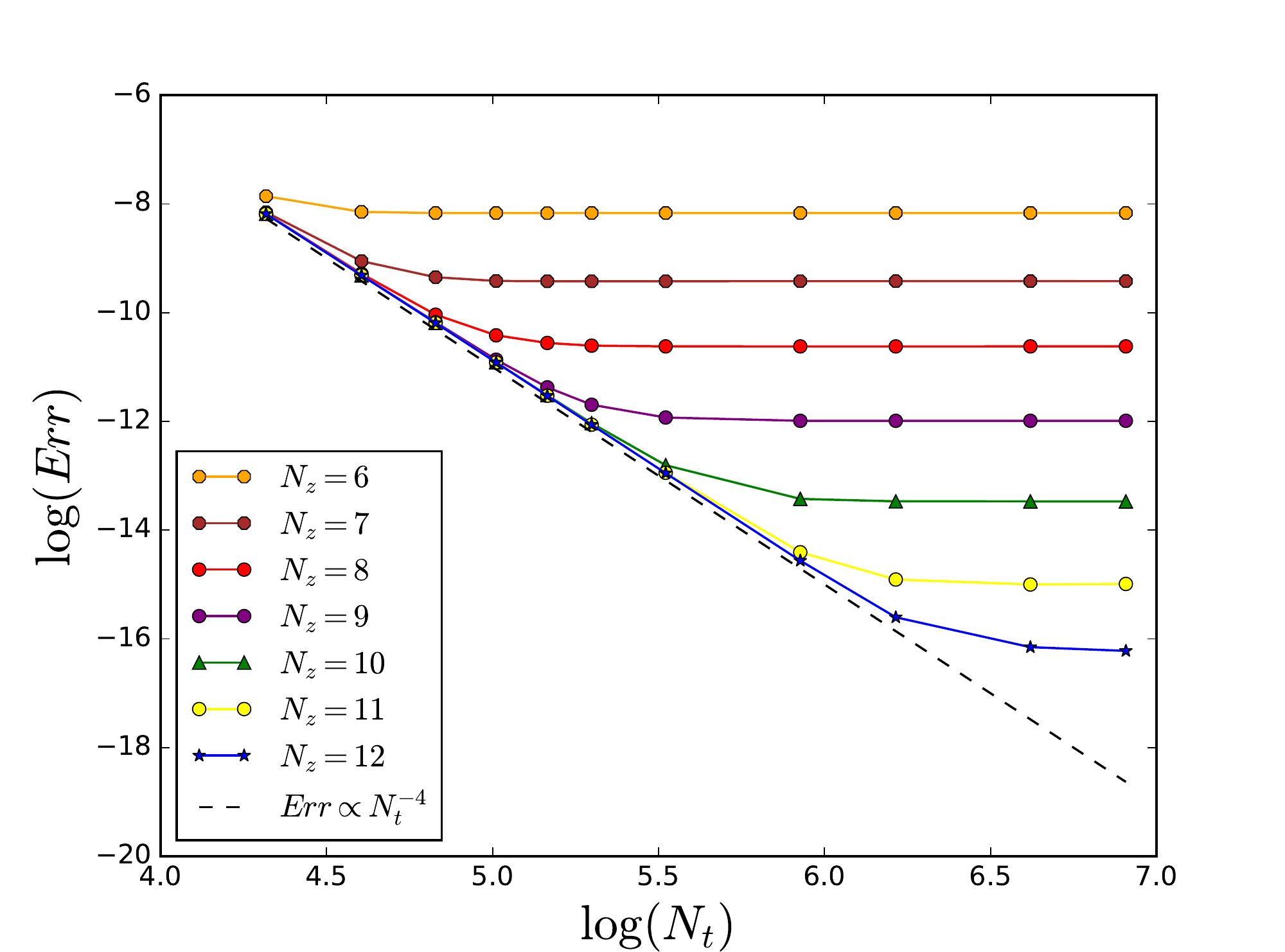}
\caption{\label{fig:error} Numerical error as a function of the number of timesteps $N_t$. The
different curves correspond to different numbers $N_z$ of spatial lattice points. Here we study a Gaussian pulse
$E_x(t) = A \cos(\omega_{P} t) \exp\left( -\frac{(t - t_0)^2}{(\Delta t)^2}\right)$ with the choice of parameters: $A = 0.5$, $t_0 = 3$, $\Delta t_P = 1$, $\omega_P = \pi/2$, $\rho = 0.5$, $k = 1.0$, $m_{I} = 0.5$, where $m_{I}$ is the initial mass of the black hole. Here we have chosen a shorter pulse to keep the computational time shorter. The spatial size of the computational domain is $z\in [0, 1]$ while the timelike size is $v\in [0, 10 ]$.}
\end{center}
\end{figure}

There are two sources for the numerical error in our procedure. The first one arises from discretizing the $z$ coordinate,
and the second one arises from discretizing the time coordinate. As a measure of the numerical error we use the remaining
three redundant equations of motion. For an exact solution, these equations would be automatically solved. Denoting the 
equations as Eq$_i=0$ where $i = 1,2,3$, we consider the following quantity
\beq
\textrm{Err} = \sqrt{\sum_{i=1}^3 \textrm{max}_{z,v}\(\textrm{Eq}_i\)^2}.
\eeq
We evaluate the equations on the spacetime grid using fourth order finite differences for approximating time derivatives
and then find the maximum value of $|\textrm{Eq}_i|$ within the grid. Finally we take the root mean square of the 
maximum error of the three equations. This error measure is displayed in Fig.~\ref{fig:error} as a function of the number of timelike $N_t$ and spacelike $N_z$ lattice points for fixed timelike and spacelike size of the computational domain. 

From this error measure, we find that the numerical error approximately first decays as
$\textrm{Err}\propto N_t^{-4}$ as $N_t$ is increased and then saturates to a constant value. This is expected as
there is a remaining error due to finite number of spatial lattice sites $N_z$. Increasing this number
then decreases the saturated value approximately exponentially. Thus, as both $N_t$ and $N_z$ are increased
the error is found to decrease rapidly, which gives strong evidence that the numerical calculation is converging
towards a solution of the continuum equations of motion.\footnote{Eventually as $N_t$ and $N_z$ are sufficiently large,
the error saturates again due to the finite accuracy of Python floating point numbers.} 
In practice we have found that rather small values of the spatial lattice sites such as $N_z=8$ or $N_z = 10$ are sufficient to 
give reliable results. For $N_t$ we use the highest values from those shown in Fig.~\ref{fig:error}. This is forced
by the fact that the probe pulses have to be short in order to reasonably approximate delta functions.
On the other hand the timelike computational domain has to be large in order to get a reliable
Fourier transform of the differential conductivity, without finite size effects. For example for a computational domain of
length of order $10^3$ we use $N_t$ of the order $10^{6}$. This results in a computational
time of the order of tens of hours on a laptop.


\section{Non-equilibrium background spacetimes} \label{sec:noneqbst}

To model the process of applying the pump electric field, we start from an initial state corresponding to an equilibrium black brane dual to a state at a given temperature $T_I$.  The time dependent pump electric field then takes the system out of equilibrium to a configuration captured by the ansatz \eqref{eq:ansatz} to finally reach a new equilibrium configuration at  a different  temperature $T_F$. Throughout this process we keep $k$ fixed and $\rho$ is conserved, as guaranteed by Ward's identities. 

\begin{figure}[ht]
\begin{center}
\includegraphics[width=0.49\textwidth]{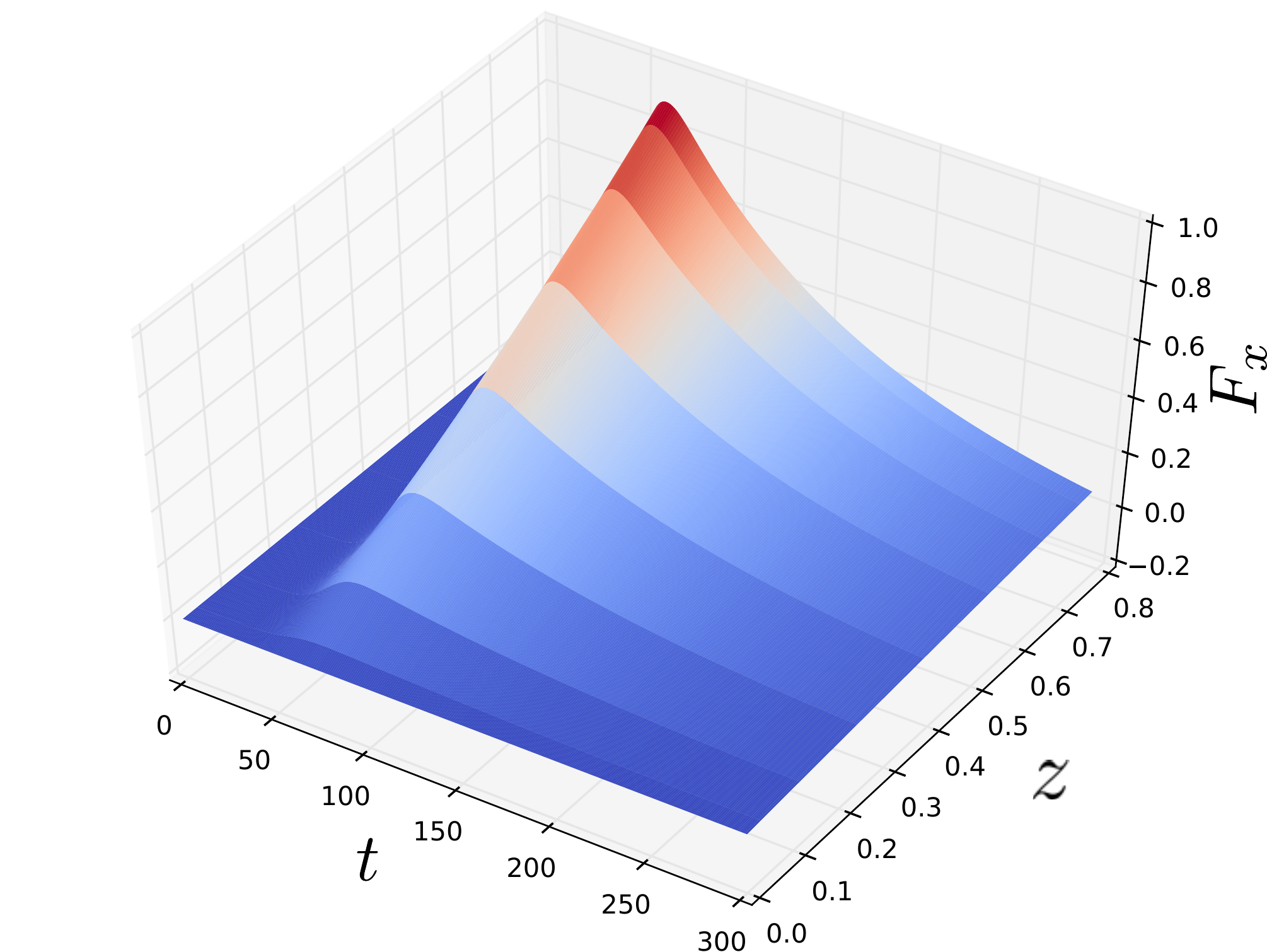} \includegraphics[width=0.49\textwidth]{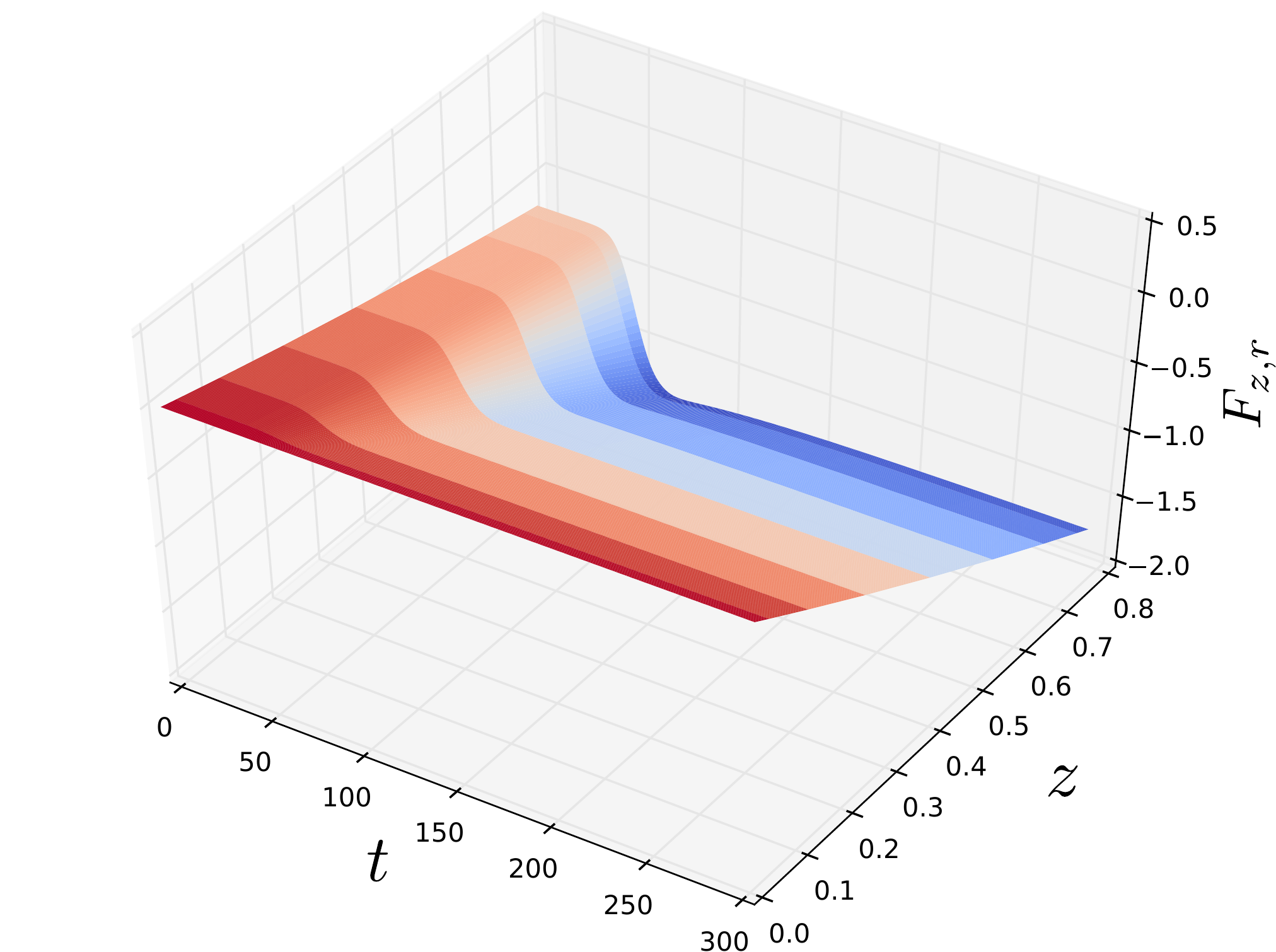}
\caption{\label{fig:bulk_metric_fns} Plots of bulk profile of the metric functions $F_{z,r}$ and $F_x$ interpolating between initial and final equilibrium state.  The parameters corresponding to the plot are: $\mu_I = 1, T_I = 0.2, T_F = 0.3, k = 0.2, 
 \omega_P = 0$.}
\end{center}
\end{figure}

In more detail, the starting equilibrium configuration in terms of the regularized fields defined in \eqref{eq:asysub} corresponds to setting
\beq
F_{z,r} = -m_I z + \frac{1}{4}\rho^2 z^2, \qquad a_{v} = \rho z  - \mu_I,
\eeq
and all the other regularized fields  in \eqref{eq:asysub} to zero. The parameters $m_I$ and $\mu_I$ are determined in terms of $T_I, k$ and $\rho$ according to the relations of Sec.~\ref{sec:model}. The specific form we use for the pump field is given by
\beq\label{eq:pump_pulse_form}
E_x(t) = A \cos(\omega_{P} t) e^{-\frac{(t - t_0)^2}{(\Delta t)^2}} \frac{1 - \tanh \frac{t - t_0 - 3\Delta t}{\delta}}{2} \, . 
\eeq
This represents a Gaussian wavepacket with central frequency $\omega_{P}$ and width $\Delta t$, centered at $t_0$ and cut off by a smoothed step function at $t_{\rm end}\equiv t_0+3\Delta t$ (from which time onwards we consider the pumping to have finished). Throughout this section, we choose the parameters $t_0 = 50$, $\Delta t = 15$ and $\delta = 0.01$. The pulse amplitude $A$ is instead tuned in order to obtain the desired increase in temperature. At the time of the pulse, the metric functions start time-evolving, exciting all the rescaled fields defined in  \eqref{eq:asysub}. Notice that in particular, this will give a nontrivial expectation value for the boundary operators associated to the bulk fields. More specifically, according to our ansatz, the current $J_{\mu}$ associated to the bulk gauge field, the operator $O$ associated to the scalar excitation $\Phi$ and the non-isotropic stress-energy tensor $T_{\mu}$ associated to the bulk metric will acquire a time dependent expectation value.  At late times they will all  settle to new equilibrium values with
\beq
F_{z,r} = -m_F z + \frac{1}{4}\rho^2 z^2, \qquad a_{v} = \rho z  - \mu_F,
\eeq
and again all other rescaled fields vanishing. The final mass parameter of the black hole will increase throughout the process, $m_F > m_I$, consistently with the fact that energy has been pumped into the system.  As an example, Fig.~\ref{fig:bulk_metric_fns} shows plots of some metric function components obtained from the numerical solution.

To obtain boundary theory expectation values from the bulk solution, one has to perform the corresponding holographic renormalization procedure \cite{Andrade:2013gsa}. The resulting one point functions are given in terms of asymptotics of the 
bulk fields as \cite{Withers:2016lft}
\begin{align} 
\eps = \langle T_{tt}\rangle &= -2 F_{z,r}'  \, ,\nonumber
\\
\langle T_{tx}\rangle &= 3 F_x' \, ,\nonumber
\\
p_x-p_y=\langle (T_{xx} - T_{yy})\rangle &= 12 B_r'  \, ,
\\
\langle \mathcal{O}\rangle &= 3 \Phi_r' \, ,\nonumber
\\
\rho=\langle J_t\rangle &= a_{v}'   \, ,\nonumber
\\
\langle J_x\rangle &= E_x(t) + a_ {x}' \, , \nonumber
\end{align}
where the different bulk function appearing on the r.h.s.\ are all evaluated at the AdS boundary $z\to0$.

\subsection{Vanishing pulse frequency} \label{sec:0pulsefreq}

\begin{figure}[h!]
\begin{center}
\includegraphics[width=0.32\textwidth]{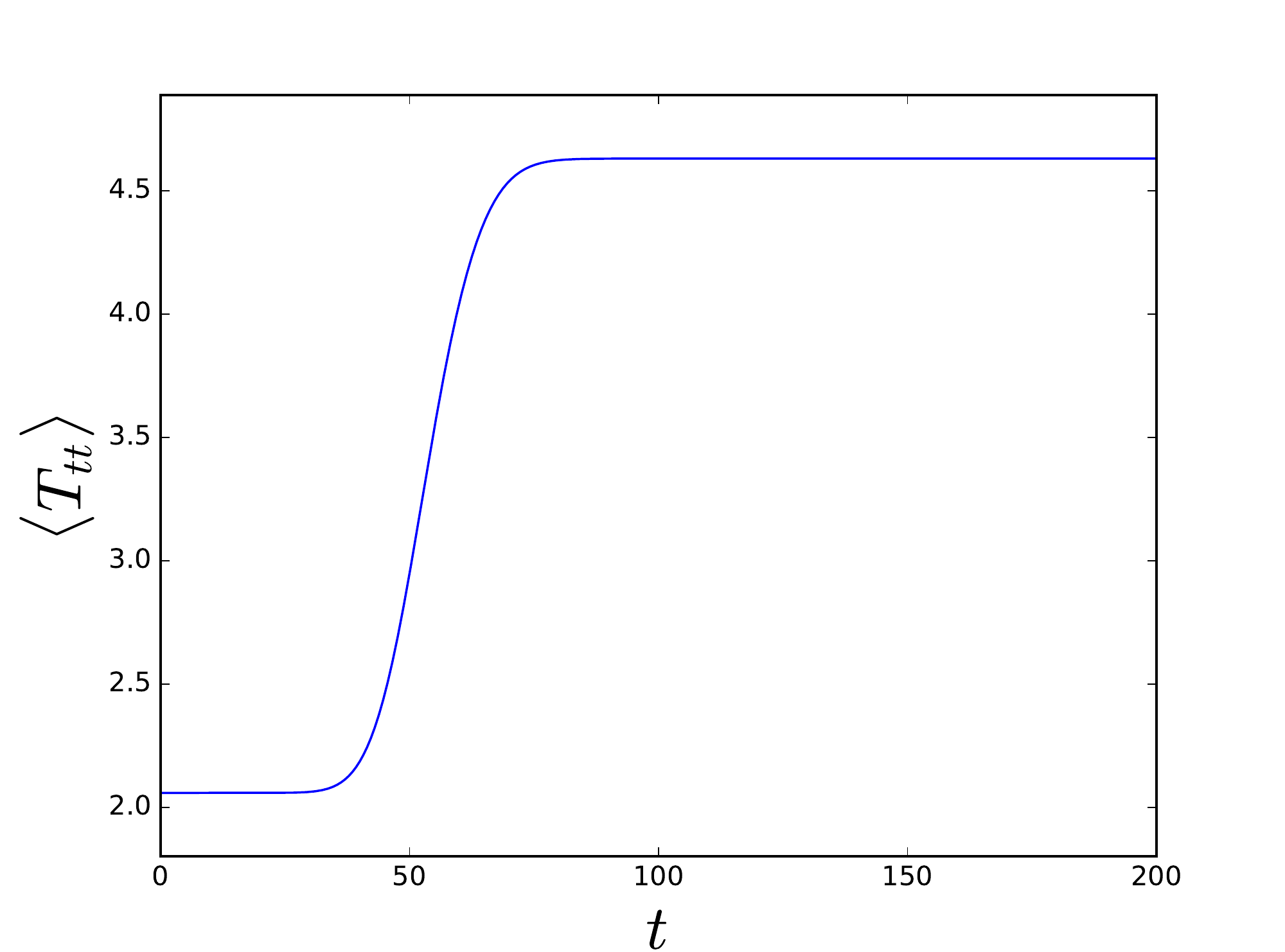} \includegraphics[width=0.32\textwidth]{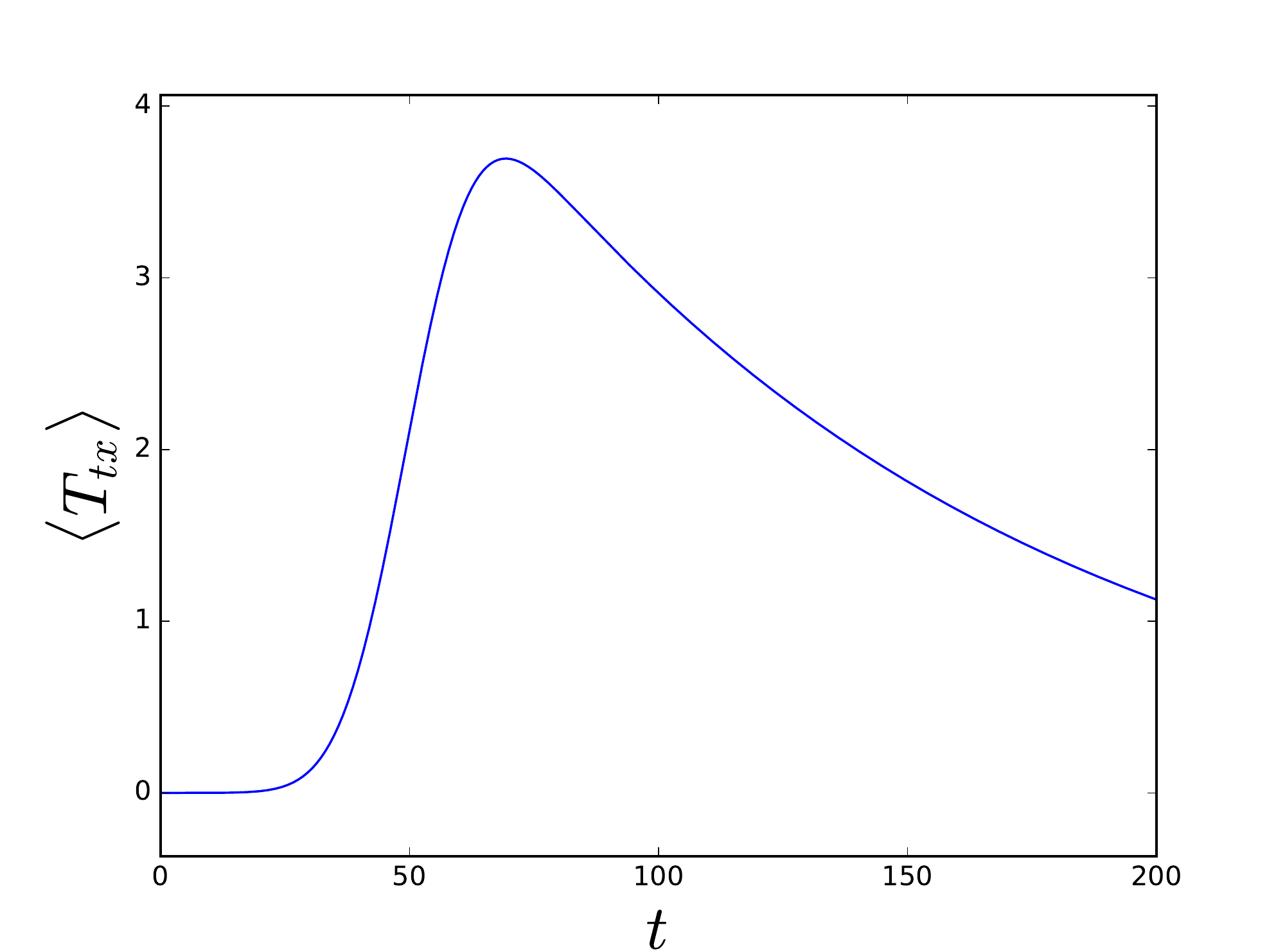}
\includegraphics[width=0.32\textwidth]{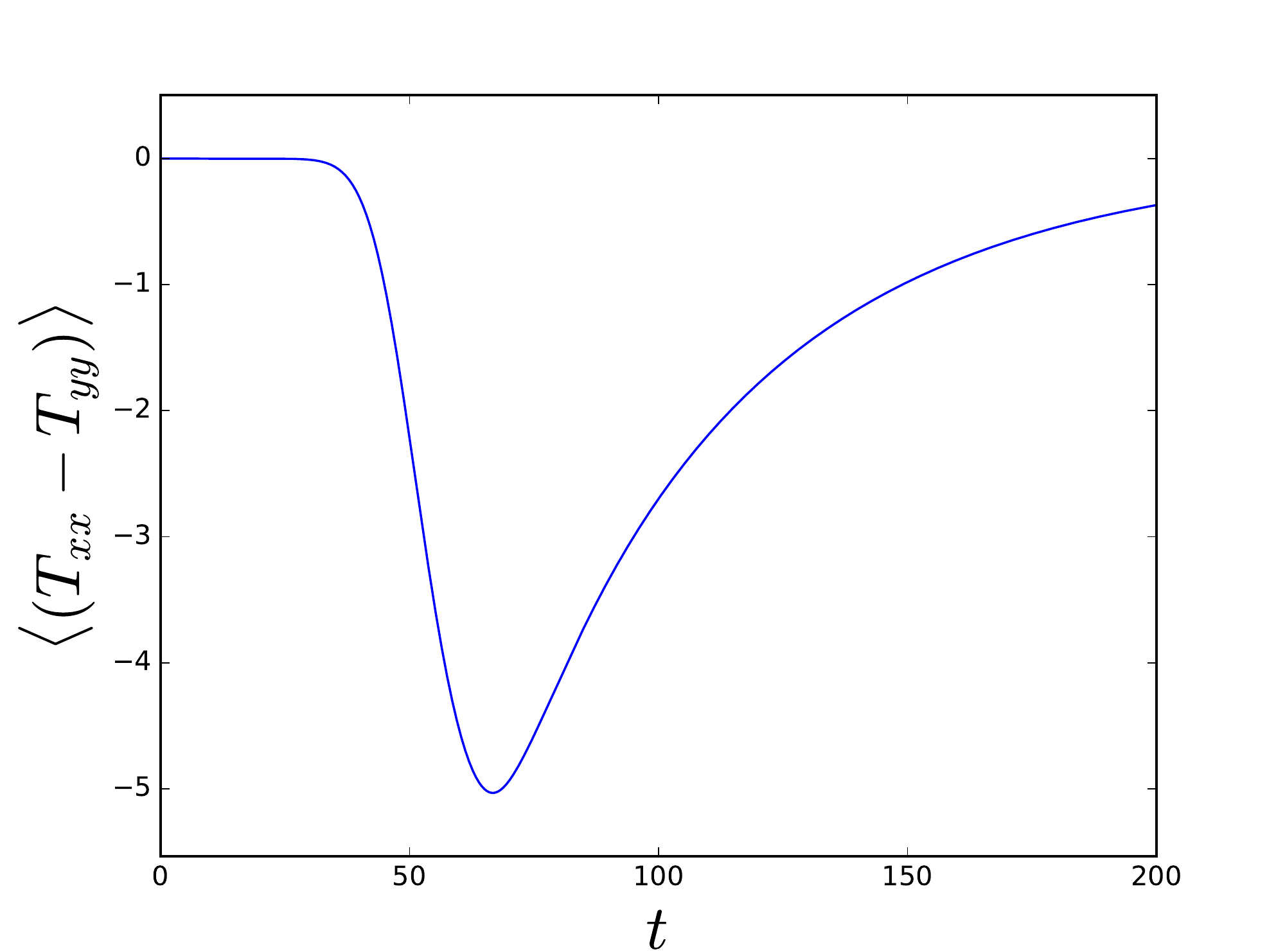}\\
 \includegraphics[width=0.32\textwidth]{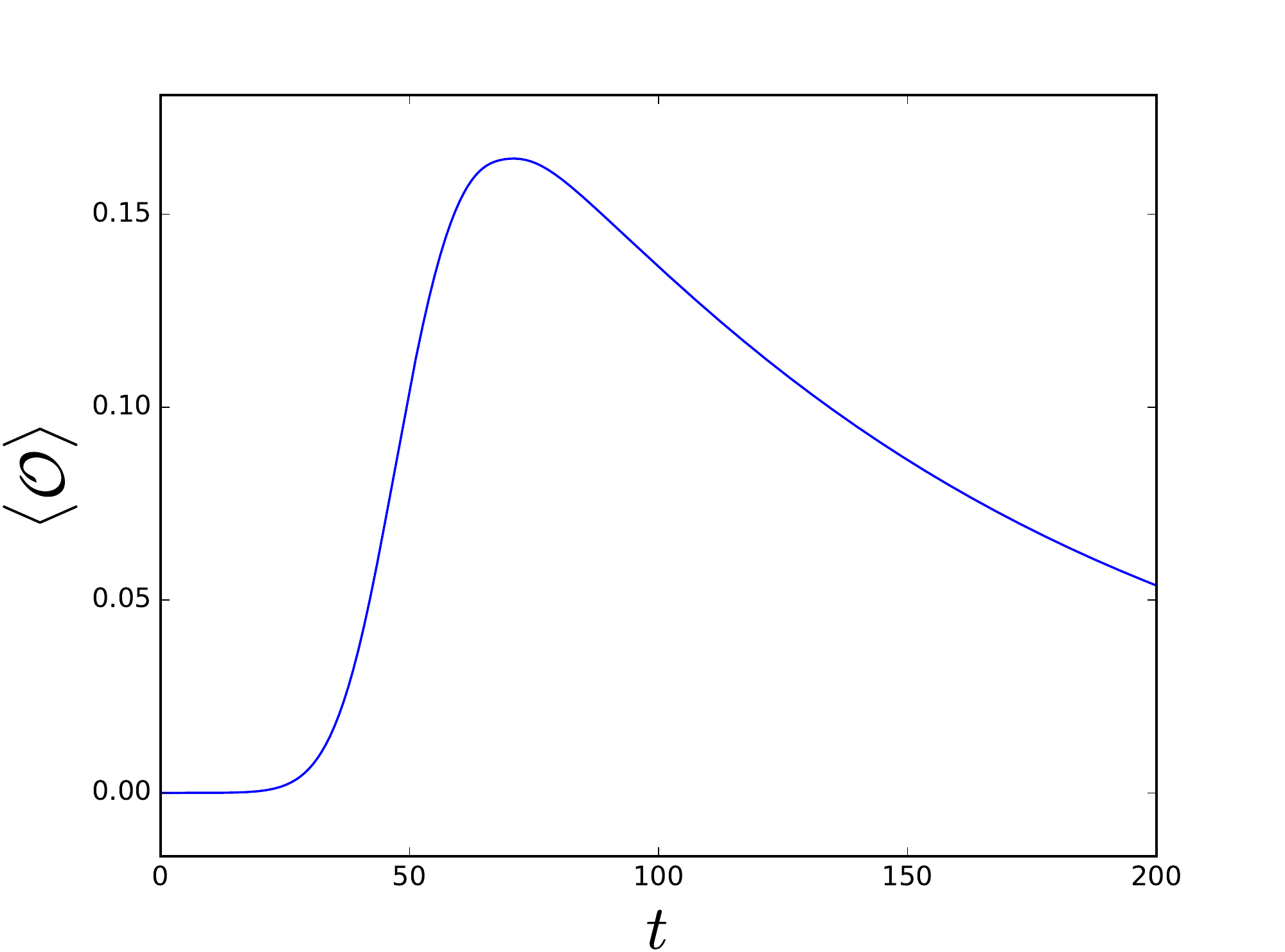} \includegraphics[width=0.32\textwidth]{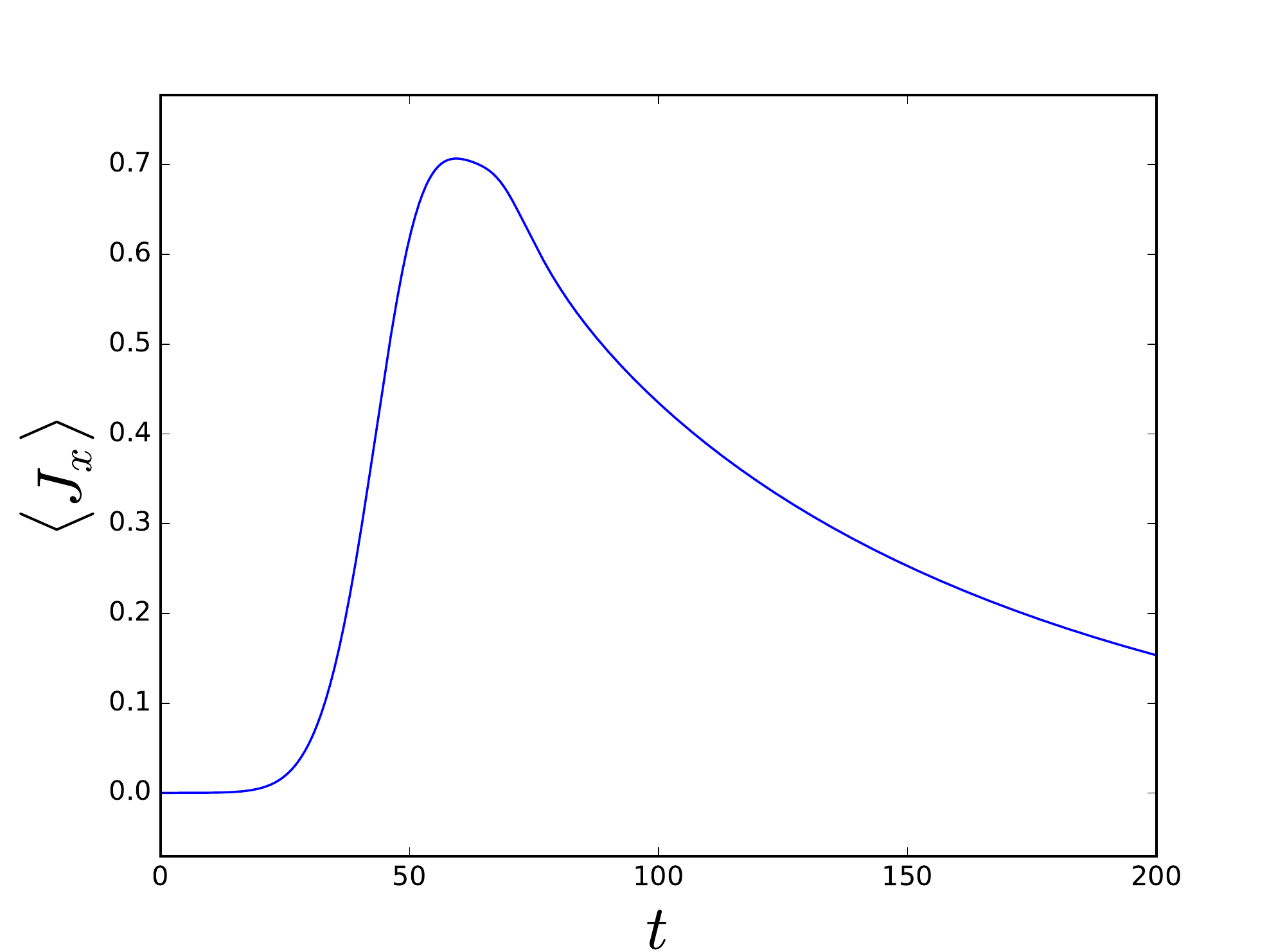}
\caption{\label{fig:1pt1} Time dependence of the expectation value of one point functions the dual field theory. The parameters considered here  are the same of Fig.~\ref{fig:bulk_metric_fns}. }
\end{center}
\end{figure}

We start by considering the particular case where the pump field is not oscillating, $\w_P =0$. Fig.~\ref{fig:1pt1} shows the boundary theory expectation values for the same type of time dependent state represented in Fig.~\ref{fig:bulk_metric_fns}. 
In particular one can observe that the pump electric field $E_{x}(t)$ induces an electric current and a 
momentum current in the field theory. Furthermore, there is a substantial pressure 
anisotropy induced and, in the case represented here,  the energy density increases by more than a factor of two.

\begin{figure}[ht]
\begin{center}
\includegraphics[width=0.45\textwidth]{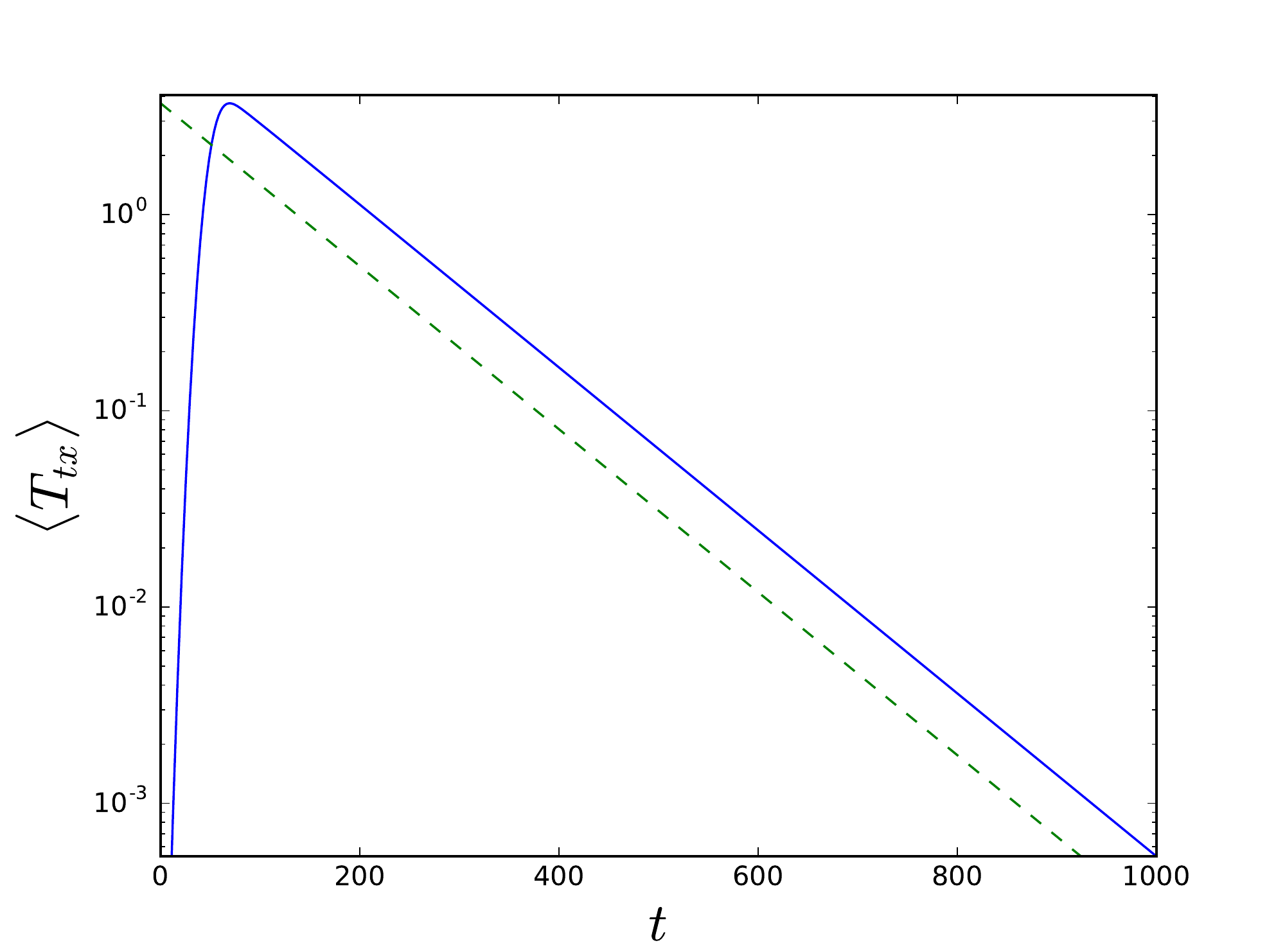}
\includegraphics[width=0.45\textwidth]{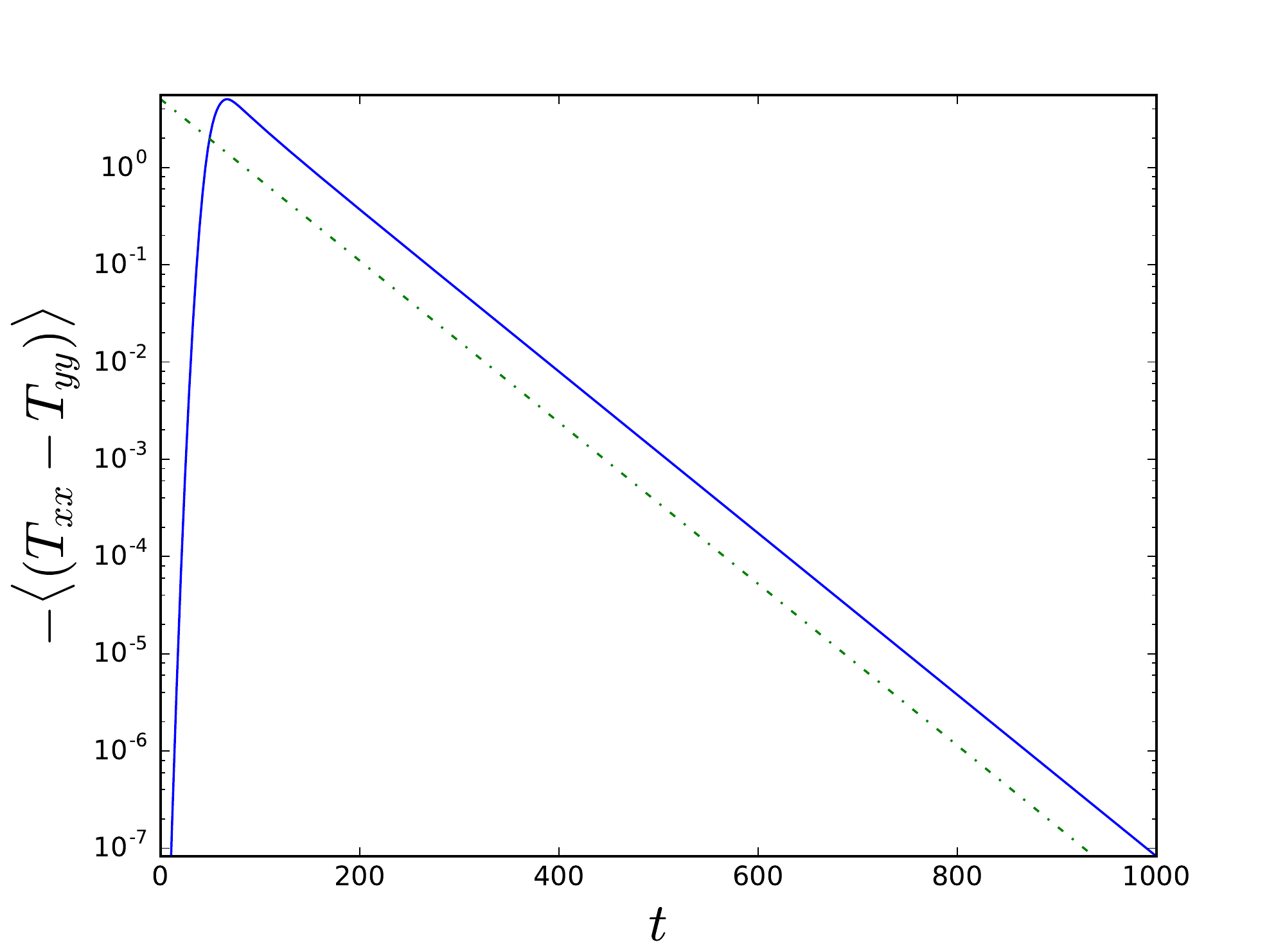}
\includegraphics[width=0.45\textwidth]{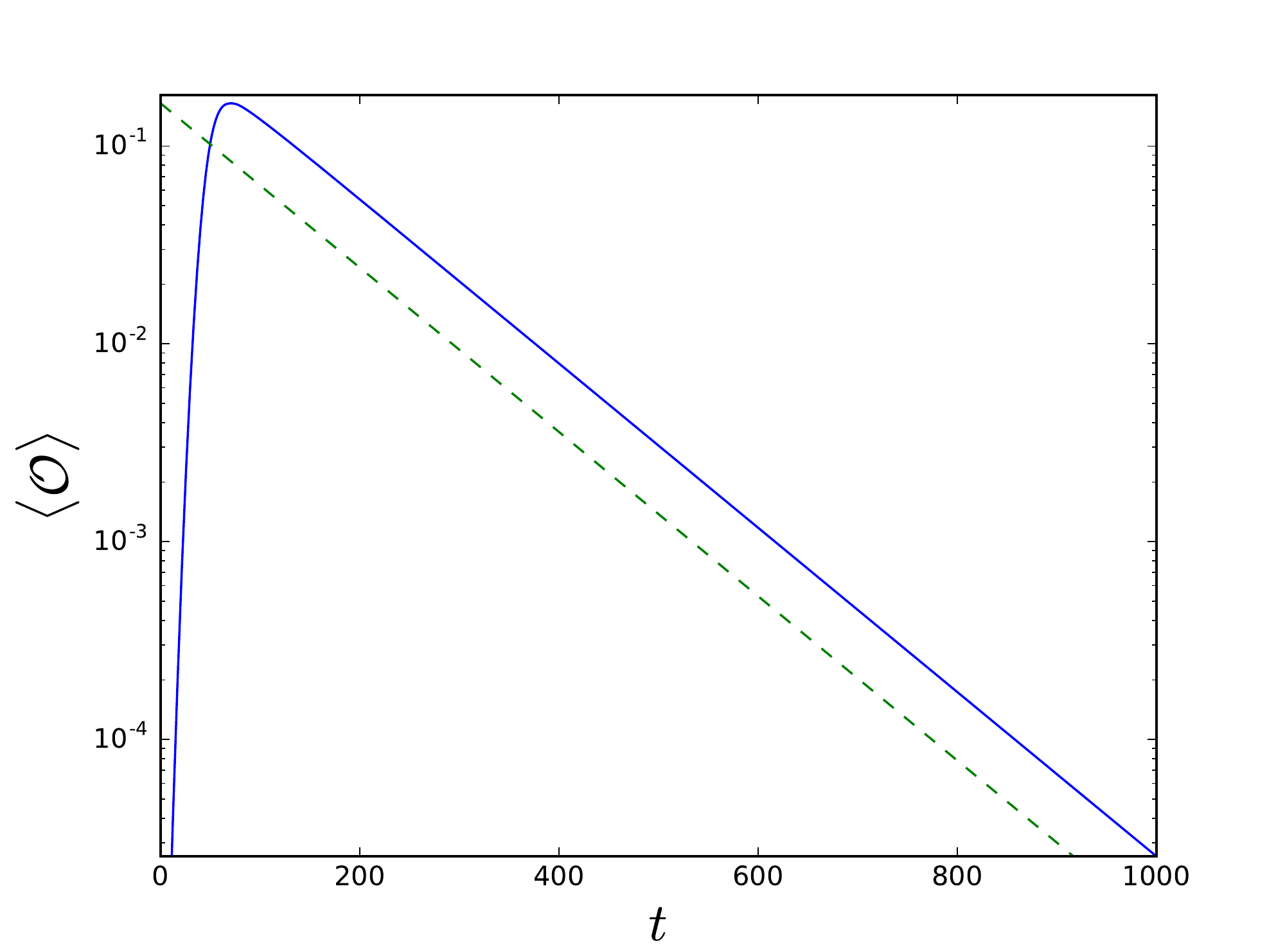}
\includegraphics[width=0.45\textwidth]{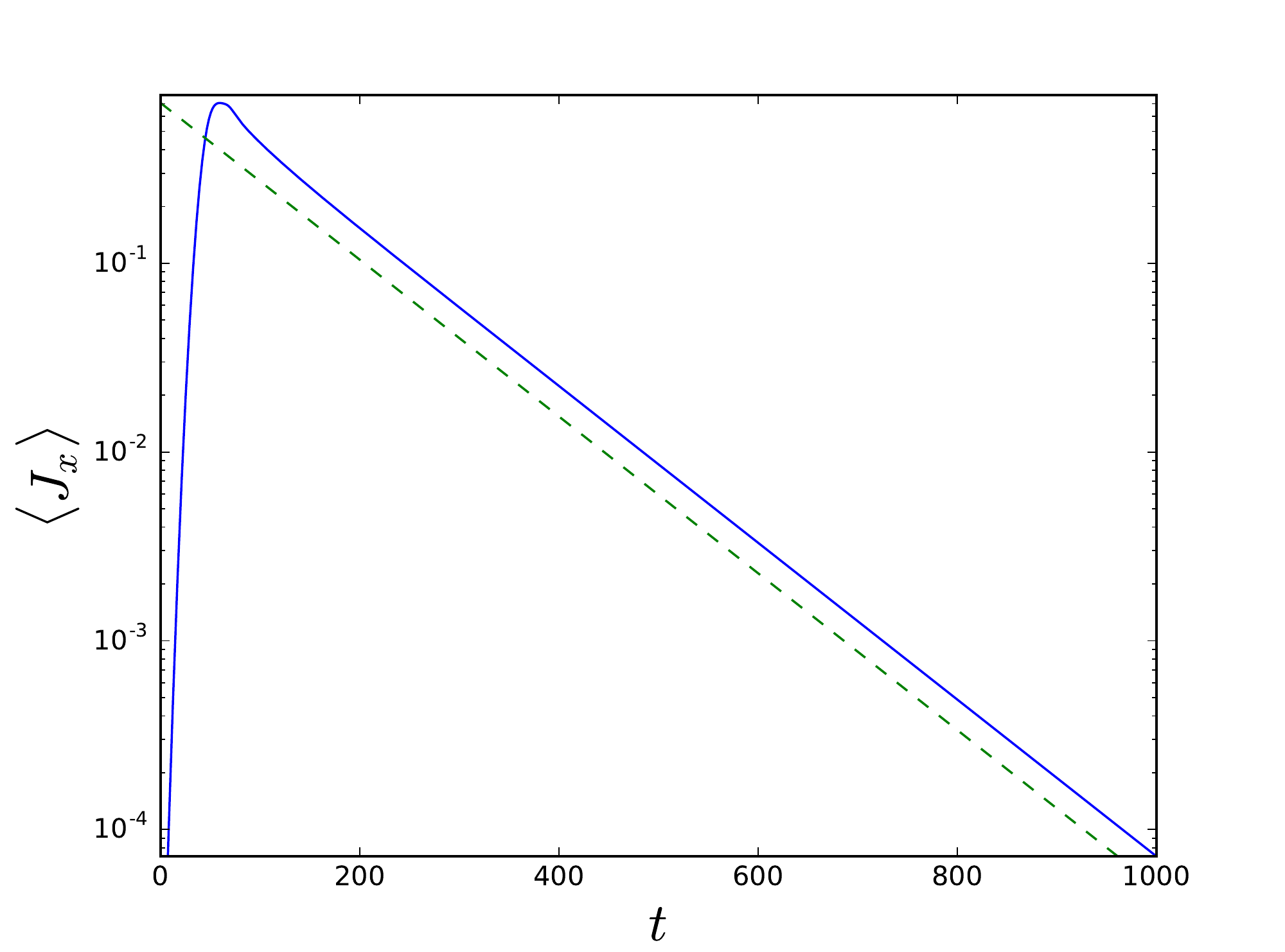}
\caption{\label{fig:1ptlog} Logarithmic plots of expectation values $\langle T_{tx}\rangle,  \langle (T_{xx} - T_{yy})\rangle, \langle \mathcal{O}\rangle,$ and $\langle J_{x}\rangle$. The dashed line corresponds to $e^{-\omega_i t}$ while the dot-dashed line corresponds to $e^{-2\omega_i t}$.}
\end{center}
\end{figure}

All one point functions, except for the energy density, seem to have a relaxation time far longer than the time scale of the pump field.  Inspecting the logarithmic plots in Fig.~\ref{fig:1ptlog} one finds that
they decay towards equilibrium exponentially in time. The rates of the exponentials are consistent with
\beq
\langle T_{tx}\rangle \propto e^{-\omega_i t}, \quad
\langle (T_{xx} - T_{yy})\rangle \propto e^{-2\omega_i t}, \quad
\langle \mathcal{O}\rangle \propto e^{-\omega_i t}, \quad
\langle J_{x}\rangle \propto e^{-\omega_i t},
\eeq
where $\omega_* = -i\omega_i$ is the, purely imaginary, lowest quasinormal mode in the vector channel. It is important to note that the pressure anisotropy is decaying with double the rate of the other expectation values.

We are able to provide an explanation for this, working under the reasonable assumption -- supported by the numerical calculations -- that the deviations from thermality are sufficiently small at late times, so that the equations of motion can be expanded in powers of the deviations.  
For this, consider an  expansion around the final thermal black brane configuration  
\begin{align} \label{eq:fluctdef}
&F_z  = \frac{1}{z^2} f(z) + \delta F_z \, ,& &  F_x   = \delta F_x,   \, ,  \nonumber \\
&\Sigma = \frac{1}{z} + \delta \Sigma  \, , & &a_x  = \delta a_x  \,  ,\nonumber \\
& a_v  = -\mu + \rho z + \delta a_v \, ,& & \Phi  = \delta \Phi   \, , \\
&B   = \delta B     \nonumber 
\end{align}
with $f(z)$ being the equilibrium metric function defined in \eqref{eq:eqsol} and $\delta X$ indicating  fluctuations. 
At the linear level in the deviations, the equations of motion decouple into two sets. One set describes the vector fluctuations 
\begin{align} \label{eq:linerizedV}
& z^2 \partial_z\( \frac{f}{z^2} \del_z\delta\Phi\)  - 2 z \partial_z\( \frac{1}{z} \del_v\delta\Phi\) + kz^2\partial_z\delta F_x  = 0\, ,
\nonumber \\
 & \partial_z\( f \del_z  \delta a_x  \) - 2\partial_z\partial_v\delta a_x 
+ \rho \del_z\(   z^2\delta F_x\)  = 0 \, ,
\\
& \del_z \( \frac{1}{z^2} \del_z \(z^2  \delta F_x\)\) + \rho \del_z\delta a_x - \frac{k}{z^2} \partial_z\delta \Phi = 0 \, .  \nonumber 
\end{align}
The other set describes tensor fluctuations (also often called scalar fluctuations).  Imposing AdS boundary and initial conditions, on finds that $\delta \Sigma$, $\delta a_v$ and $\delta F_z$ vanish, after which the remaining tensor fluctuation $\delta B$ is governed by
\beq
z^2 \partial_z\( \frac{f}{z^2} \del_z\delta B\)  - 2 z \partial_z\( \frac{1}{z} \del_v\delta B\)  - k^2\delta B = 0.
\eeq
At late times, the vector fluctuations decay with a rate set by the lowest vector quasinormal mode. At linear order, the $\delta B$ field is decoupled from the vector fluctuations and therefore remains zero. If we go to quadratic order in the fluctuations, however, the two sectors are no longer decoupled.  
In particular, the $B$ field equation of motion is now sourced by terms quadratic in the vector sector fields $\delta \Phi, \delta a_x$ and  $\delta F_x$. Setting the linearized tensor fluctuations to zero and indicating with $\delta \delta B$ the quadratic fluctuation for $B$, from the linear combination of Einstein's equations $E_{xx} -E_{yy}$ one gets 
\begin{align}
z^2 \partial_z\( \frac{f}{z^2} \del_z\delta\delta B\)  - 2 z \partial_z\( \frac{1}{z} \del_v\delta\delta B\)  - k^2\delta\delta B = \frac{1}{z}\(\del_z\(z^2 \delta F_x \)\)^2 + z \del_{z} \delta a_x \(f  \del_{z} \delta a_x - \del_v  \delta a_x  \)  \, .
\end{align}
From this we can argue that the decay rate of the source term sets the decay rate of the $B$ field, which is thus twice the decay rate of the  vector perturbations. That is, twice the imaginary part of the lowest vector quasinormal mode. This explains the factor of two in the decay rate we see from the numerics in Fig.~\ref{fig:1ptlog}.

\subsection{Increasing the pulse frequency }

So far we have studied the case of an approximately Gaussian pump electric field (\ref{eq:pump_pulse_form}) with a vanishing mean frequency $\omega_P = 0$. This has reproduced the by now standard story that the late time relaxation of the 
black brane solution is dominated by the lowest quasinormal mode that gets excited. A slight subtlety was that some
of the metric components relax with a rate given by twice the lowest quasinormal mode from a different sector.
Next, we will study how this picture changes as we increase the mean frequency of the pump pulse.

\begin{figure}[ht]
\begin{center}
\includegraphics[width=0.32\textwidth]{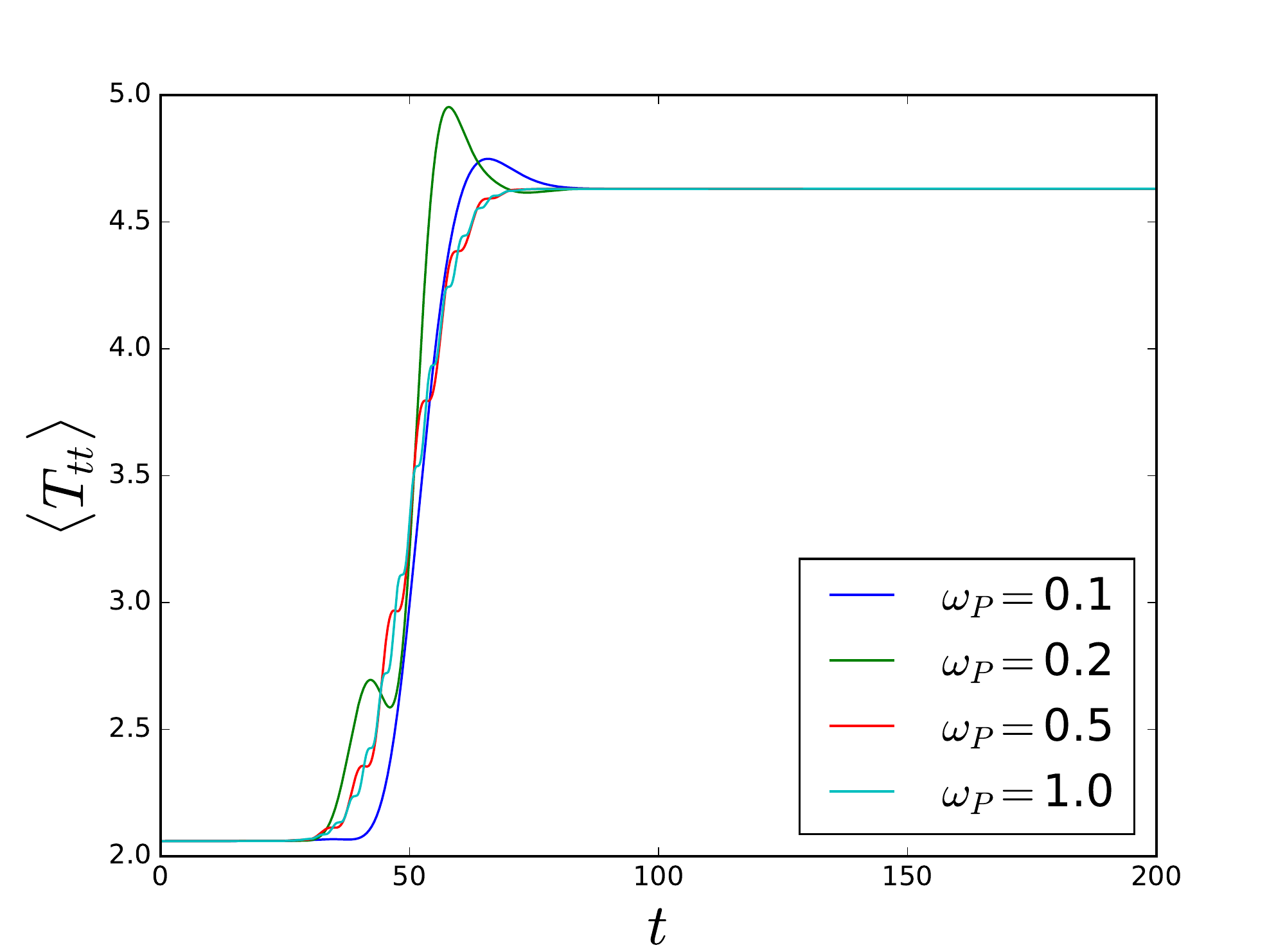} \includegraphics[width=0.32\textwidth]{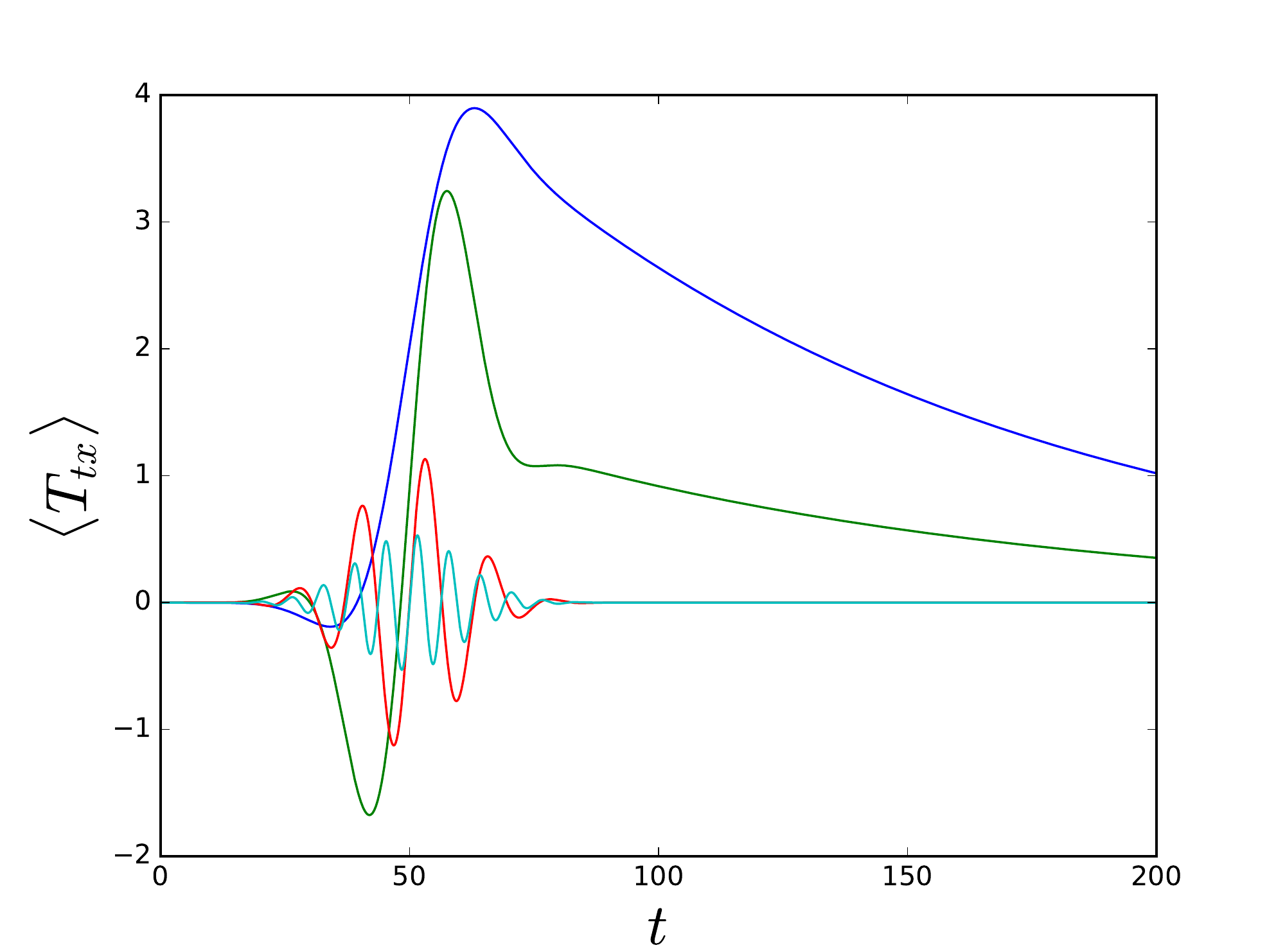}
\includegraphics[width=0.32\textwidth]{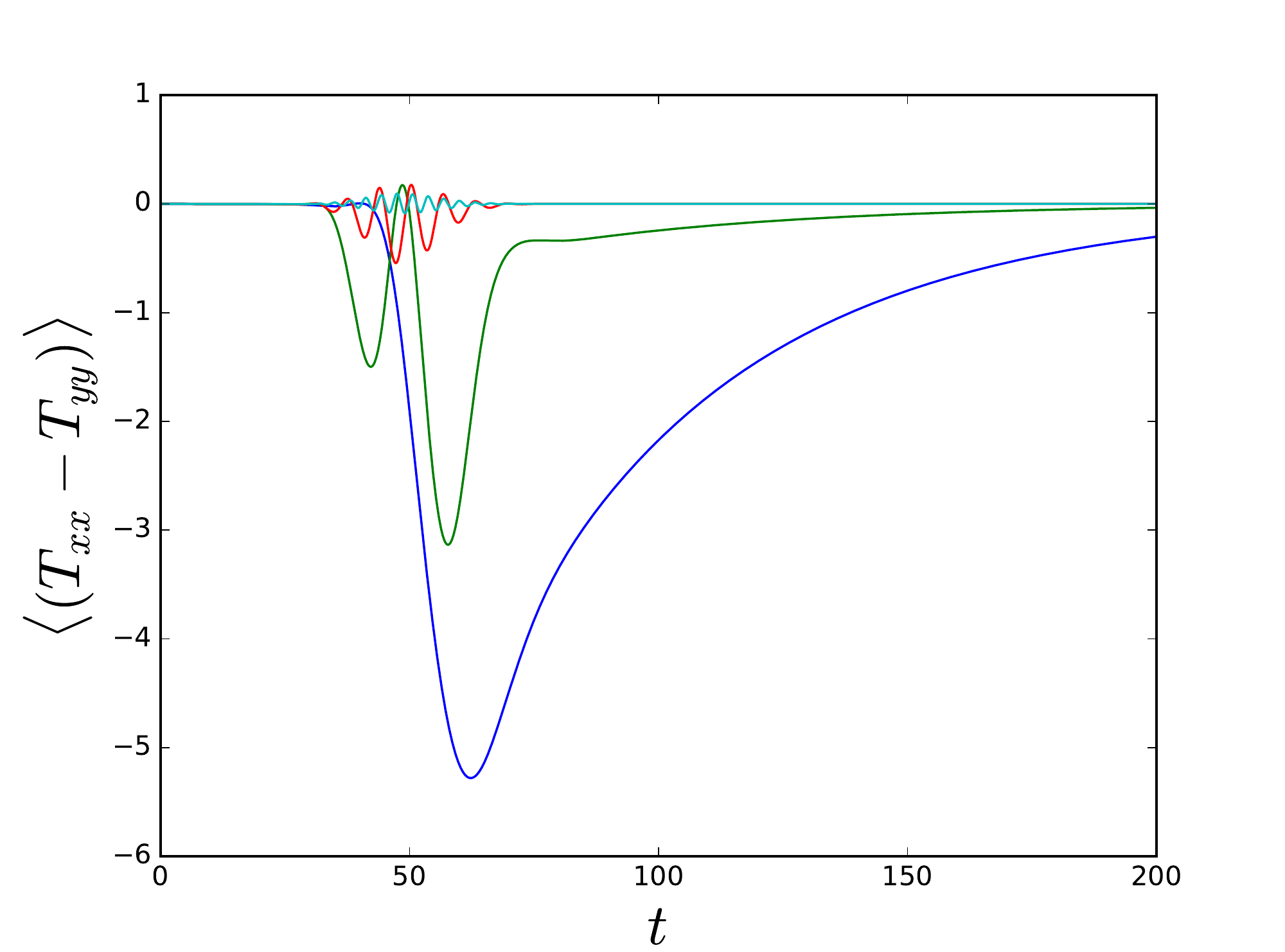}\\
 \includegraphics[width=0.32\textwidth]{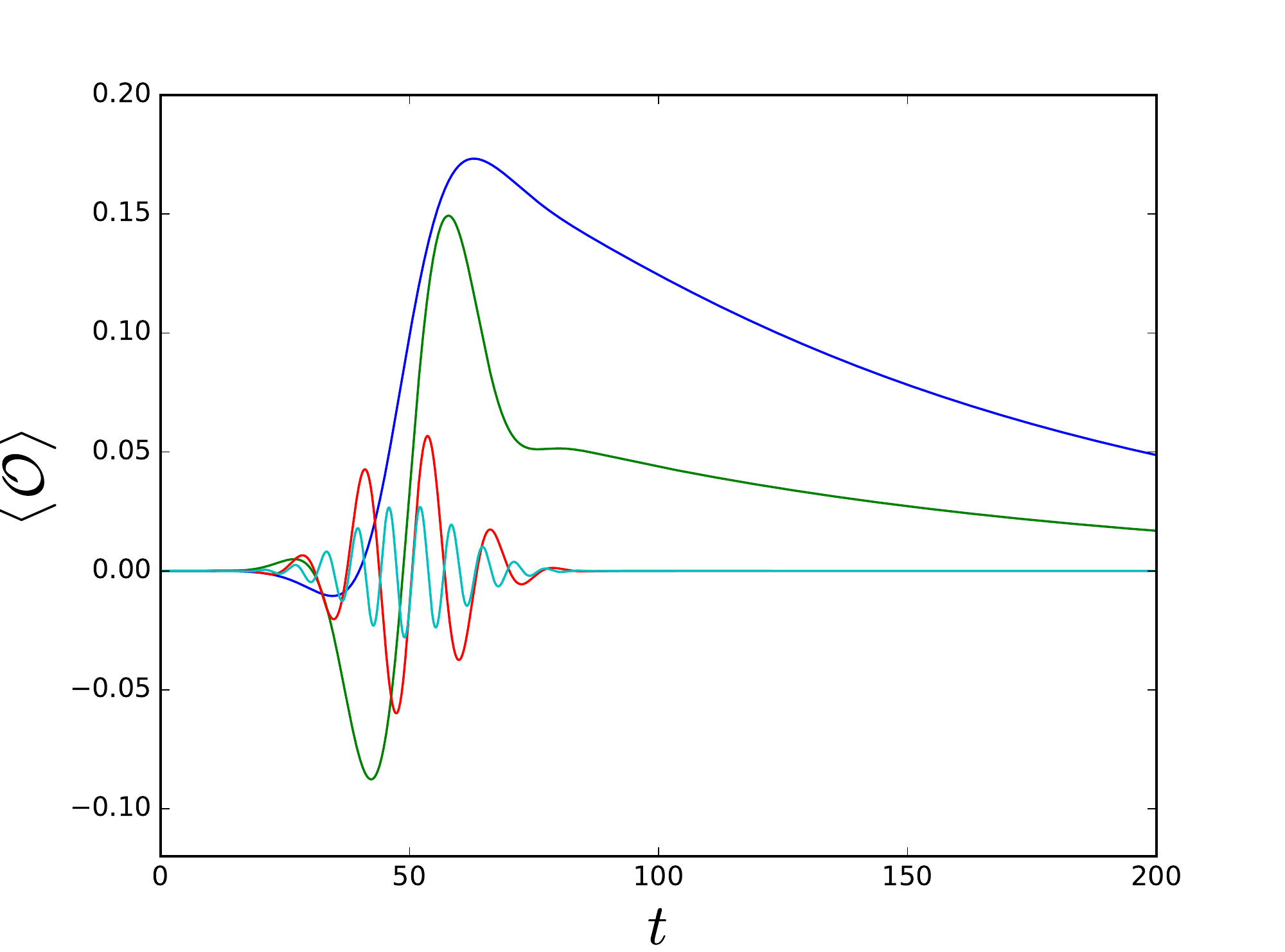} \includegraphics[width=0.32\textwidth]{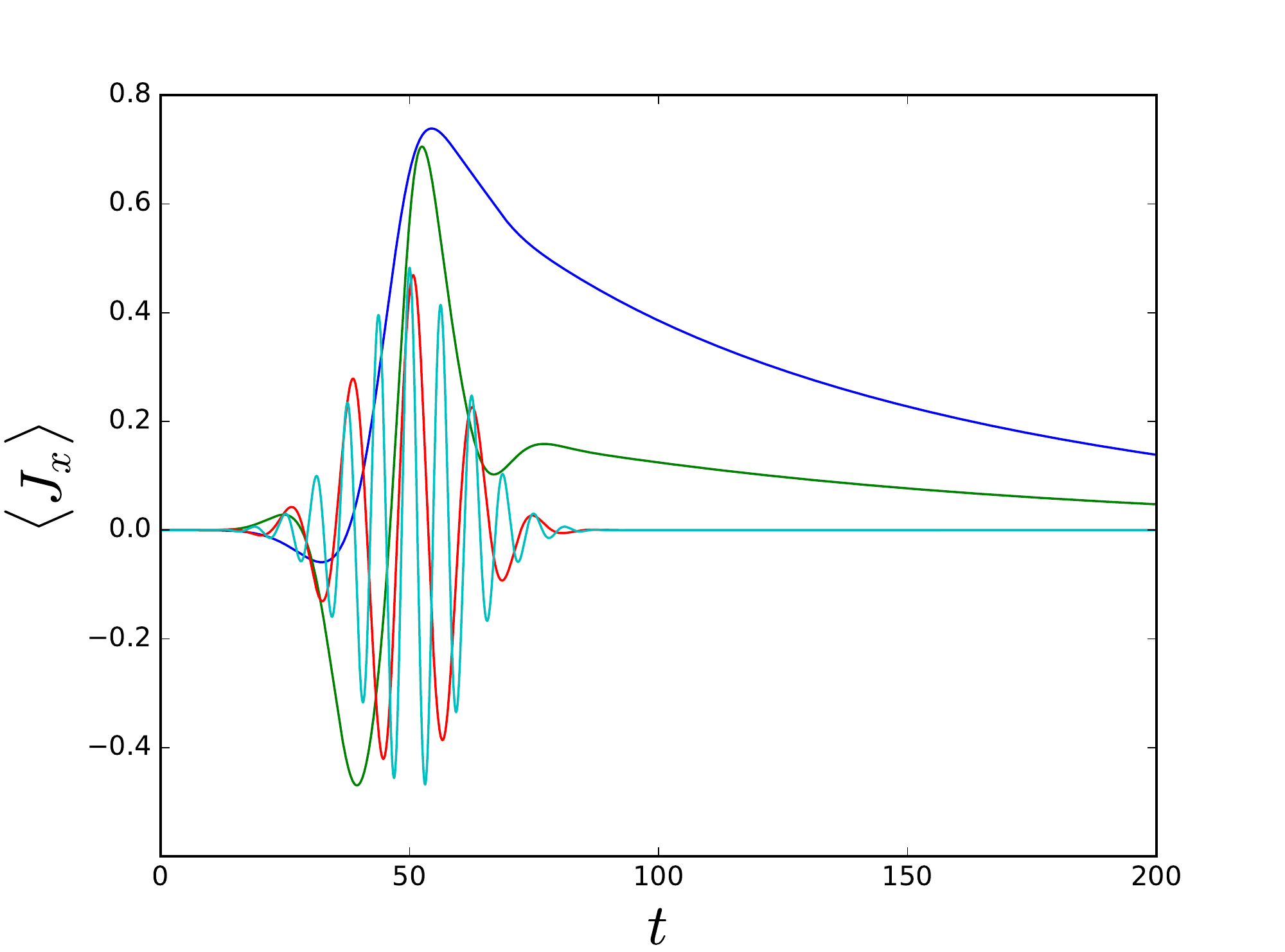}
\caption{\label{fig:1pts_omega} Plots of expectation values of one point functions for  different pump frequencies $\omega_P=(0.1, 0.2, 0.5, 1.0)$. Again   $\mu_I = 1, T_I = 0.2, T_F = 0.3, k = 0.2.$ The plots show the exponential QNM approach to equilibrium, with decreasing amplitudes for increasing $\w_P$.}
\end{center}
\end{figure}

In Fig.~\ref{fig:1pts_omega} we show the one point functions for increasing values of $\omega_P = (0.1, 0.2, 0.5, 1.0)$.
The one point functions still exhibit the late time quasinormal mode tails. The magnitudes of the tails
are decreasing rapidly with increasing $\omega_P$. Already at $\omega_P = 0.5$ the tail becomes invisible 
by the eye. So for practical purposes the quasinormal mode tail has disappeared. 
Furthermore, the magnitudes of $\langle T_{tx}\rangle, \langle (T_{xx} - T_{yy}) \rangle, \langle \mathcal{O} \rangle$ are decreasing with increasing $\omega_P$, while the magnitudes of $\langle T_{tt}\rangle$ and $\langle J_x\rangle$ stay fixed. 

We note that, strictly speaking, even if the leading QNM has only infinitesimal amplitude, one could still choose to refer to its decay constant as the decay time. If the amplitude of the QNM is below the experimental resolution, however, then it is not measurable, and in that sense irrelevant. In this paper, we therefore refer to thermalization as instantaneous or very fast if the slower decay modes have zero or negligible amplitude.

These observations suggest that the spacetime could be approximated with the Vaidya spacetime, with an appropriately  chosen time dependent mass function,  at large enough $\omega_P$.  In the rest of this section we provide evidence in support of this claim.  First we will show that  working  in the limit of a large pulse frequency the leading order solution is exactly   of the Vaidya form. We obtain this result working analytically  in the large frequency expansion. Next, we analyze  the amplitude associated to the quasinormal mode decay described above to show how this is determined by the relation between the power spectrum of the pump pulse and the quasinormal mode frequency.

 \subsection{Large frequency solution}

In the regime where the pump frequency is very large compared to the other parameters of the gravitational background the bulk solution can be studied analytically.\footnote{A related but distinct situation where an analytical treatment is also possible and the resulting geometry takes the Vaidya form was considered in \cite{Buchel:2013gba} in the context of abrupt holographic quenches.} We assume the electric field is of the simple oscillating form 
\be
E_x(t) = \cos(\w_P t)\Omega(t)\, ,  \label{eq:pulsesimple}
\ee
 where the enveloping   function $\Omega(t)$ is assumed to have compact support and slow variation compared to the cosine.
Using the knowledge obtained from the numerical solution and inspecting the equations of motions, we formulate an ansatz for the $1/\w_{P}$ expansion of each field and for the type of time dependence (rapidly or slowly varying) for each term in the expansion, and proceed to solve the resulting system of equations order by order.  The details of the analysis are reported in Appendix \ref{app:largew}. 

For the different fields and metric components,  the leading correction to the unperturbed solution  induced  by the rapidly varying source  $E_x$  takes the form
\begin{align} 
&F_z =\frac{1}{z^2} \left(1 - \frac{1}{2}k^2 z^2 - m z^3 + \frac{1}{4}\rho^2 z^4 \right)+ F^{(0)}_{z} + \dots& \qquad & \Sigma = \frac{1}{z}+  \frac{1}{\w_{P}^5} \Sigma^{(5)}+ \dots&  \nonumber  \\
&F_x = \frac{1}{\w_{P}}   F^{(1)}_{x}+ \dots & \qquad & \beta =   \frac{1}{\w_{P}^3} \beta^{(3)}+ \dots & \\
&a_v = - \mu + \rho z +  \frac{1}{\w_{P}^3} a^{(3)}_{v} + \dots& \qquad & a_x =    \frac{1}{\w_{P}^2}   a^{(2)}_{x}  + \dots \nonumber \\
&\Phi  = \frac{1}{\w_P^2}  \Phi^{(2)}+ \dots\, .\nonumber  
\end{align}
At leading order in the frequency expansion only $F_z$ gets corrected by\footnote{Strictly speaking here and in the expression \eqref{eq:m(v)} below we are  assuming that on the r.h.s.\ we are consistently taking  only the leading contribution from the integral $ \frac{1}{2}\int^v_{-\infty} dv' E_x(v')^2 $, which in general will also have subleading terms in $1/\w_P$ (see Appendix \ref{app:largew}).}
\beq
F^{(0)}_{z}(z,v) =  -\frac{z}{2}\int^v_{-\infty} dv' E_x(v')^2   \, .
\eeq

This directly shows that the response to the rapidly oscillating electric field  takes at leading order the  Vaidya spacetime form
\begin{align}
ds^2 = \frac{1}{z^2} \Big[  -  \left(1 - \frac{1}{2}k^2 z^2 - M(v) z^3 + \frac{1}{4}\rho^2 z^4 \right) dv^2    - 2 dv dz + dx^2 + dy^2 \Big] \, ,
 \end{align}
with  the mass function $M(v)$  given by the background $m$  value plus the contribution coming from   $F^{(0)}_{z}$, that is     
\beq
M(v) =  m+\frac{1}{2}\int^v_{-\infty} dv' E_x(v')^2  \label{eq:m(v)}\, .   
\eeq
 The first correction to the Vaidya form of the geometry comes from the  $F_x$  component of the metric  at order $1/\w_P$ 
\be 
F_{x}(z,v) = \frac{1}{3} \rho  z \int_{-\infty}^{v} d v'~  E_x (v') + O(\w_P^{-2}) \,  \label{eq:FXleading} \, .
\ee
In the limiting approximation where one can   treat the function $\Omega(t)$ as a constant under the integral, we would simply have $F_{x} \approx  \rho  z \sin(\w_P v ) \Omega(v) / (3 \omega_P)$. 
Notice however that in our case for those times $v$ where $E_x(v)$ has no support, that is times where the pump pulse has been turned off,  the  suppression of this correction   is even stronger. In fact, with a choice  of $E_x(v)$ of the form \eqref{eq:pulsesimple}  and for  $\Omega$ any smooth function,  $F_{x}$  is suppressed more strongly than any inverse power of $\omega_P$, as follows from basic Fourier analysis.

At order $1/\w_P^2$ also $\Phi$ and  $a_x$  get their leading contribution from the pulse, which we report here for comparison with the result obtained from the numerics
\begin{align}
&\Phi(z ,  v)   =  \frac{ z^3    \rho k }{12}  \int^{v}_{-\infty}  \int^{v'}_{-\infty}E_x(v'') dv'' dv'  + O(w_P^{-3}) \, , \\
&a_{x}(z , v)  = \frac{1}{6} z^3 \rho^2   \int^{v}_{-\infty}  \int^{v'}_{-\infty}E_x(v'') dv'' dv' + O(w_P^{-3}) \, , 
\end{align}
while the order $1/\w_{P}^3$   gives the leading corrections to the  background values of $a_v$ and $B$  
\begin{align}
 a_{v}(z, v ) &  = - \mu + \rho z +   \frac{\r^3 z^6}{36}\(  \int^{v}_{-\infty}E_x(v') dv'\) \(   \int^{v}_{-\infty}  \int^{v'}_{-\infty}E_x(v'') dv'' dv'\)  + O(w_P^{-4})\, , \\
B(z,  v ) &=  - \frac{\r^2 z^5}{16}\(  \int^{v}_{-\infty}E_x(v') dv'\) \(  \int^{v}_{-\infty}  \int^{v'}_{-\infty}E_x(v'') dv'' dv'\) + O(w_P^{-4}) \, . 
\end{align}

\begin{figure}[ht]
\begin{center}
\includegraphics[width=0.49\textwidth]{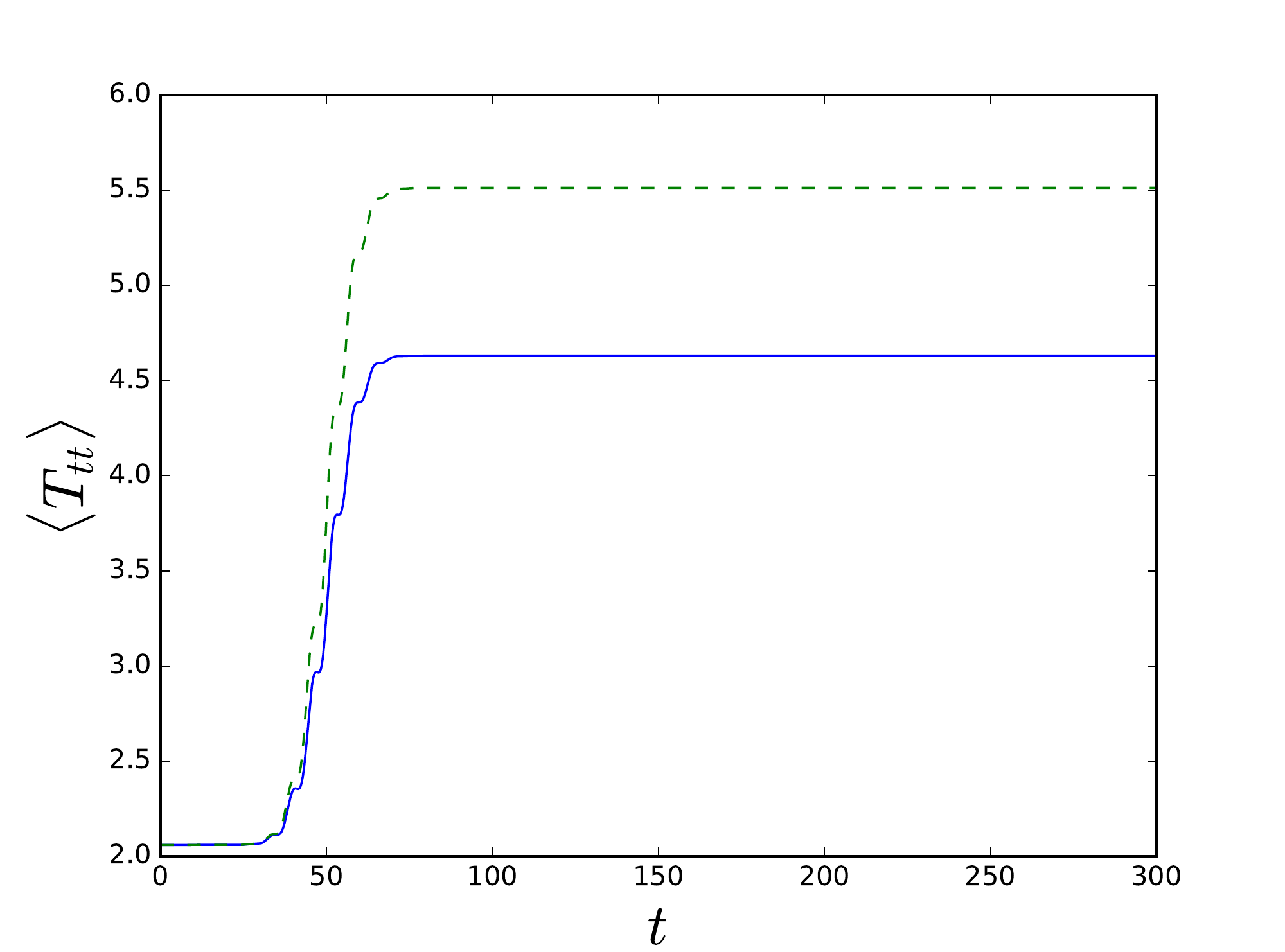} \hfill
\includegraphics[width=0.49\textwidth]{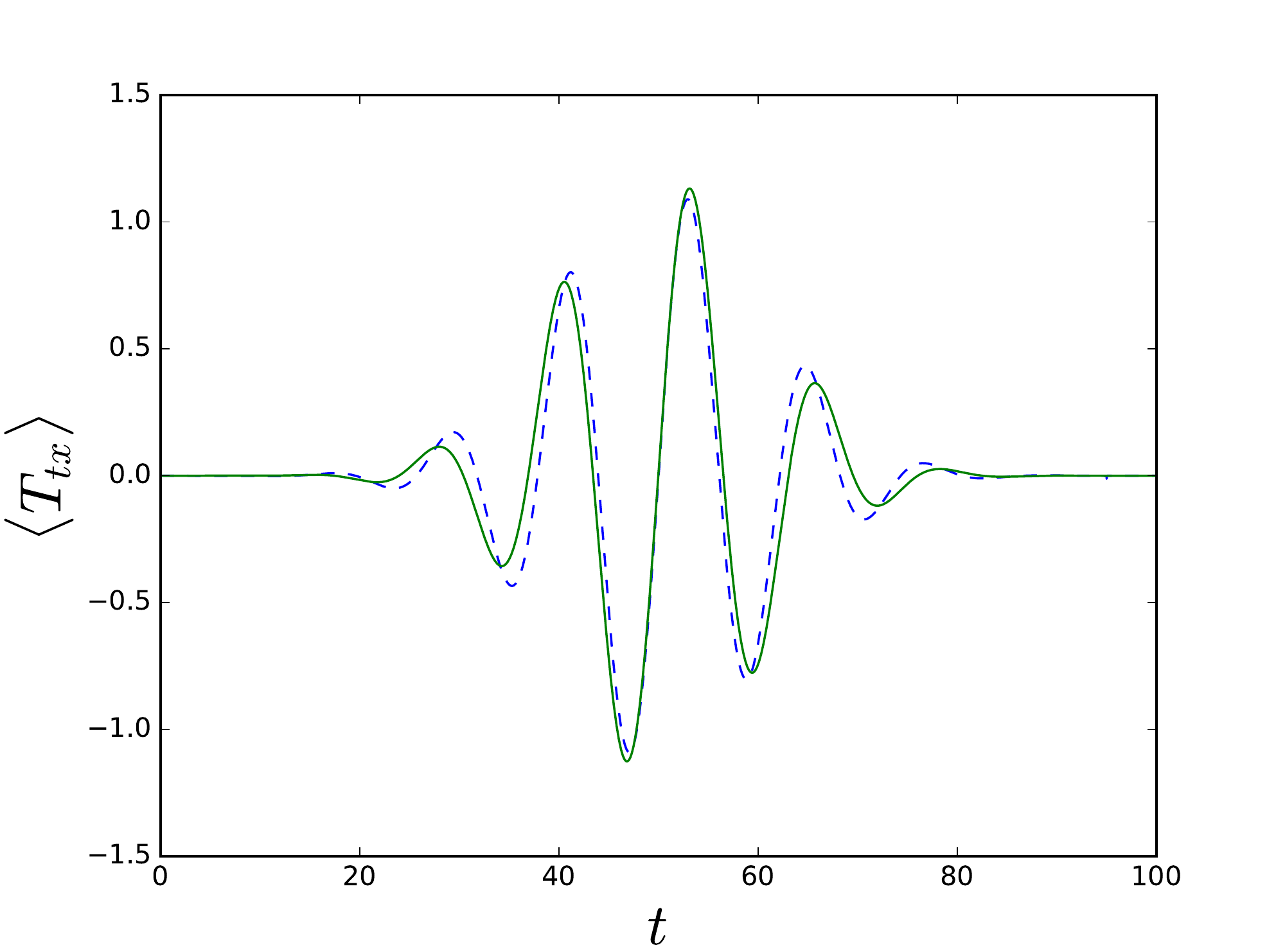}
\includegraphics[width=0.49\textwidth]{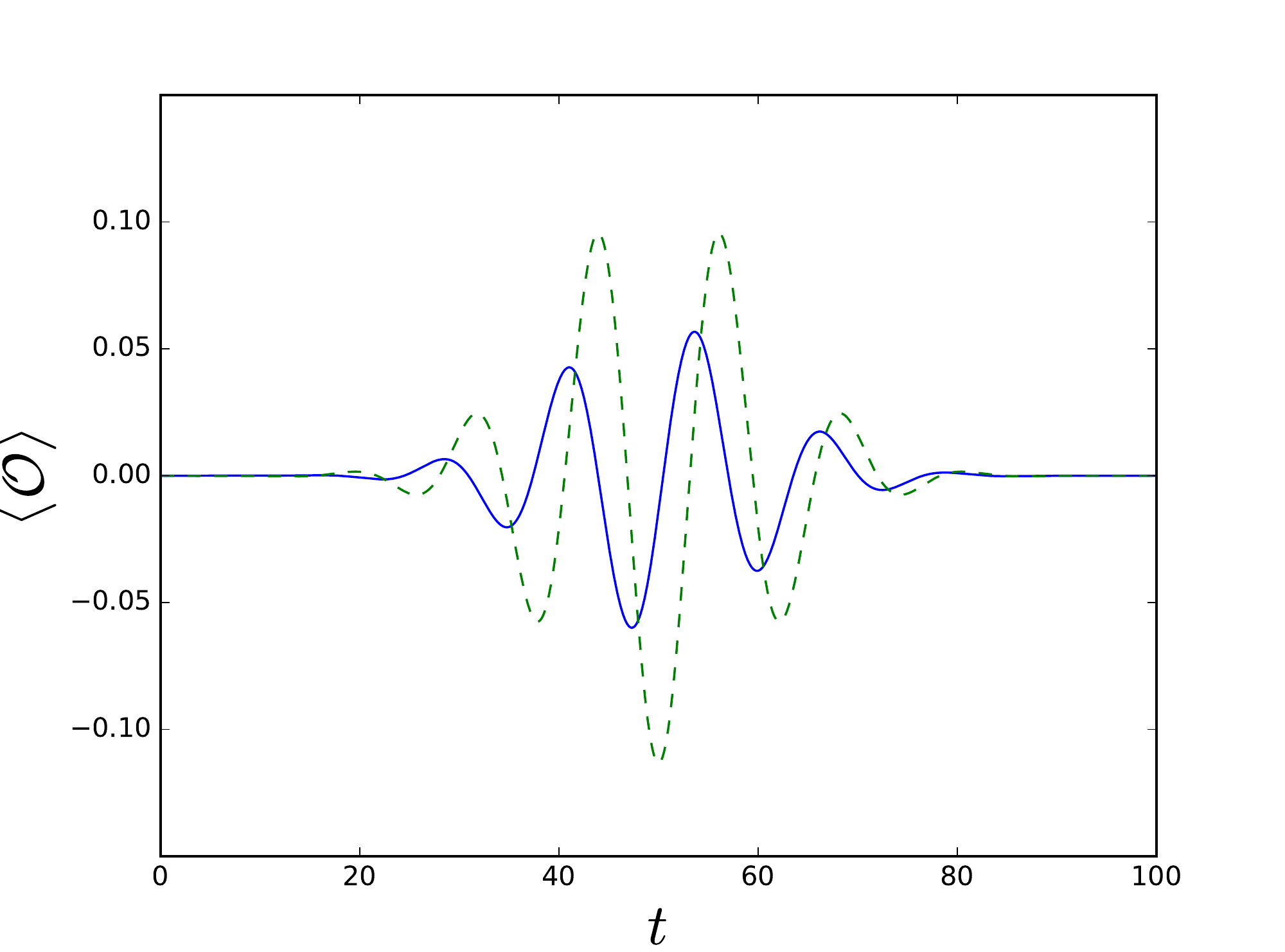} \hfill
\includegraphics[width=0.49\textwidth]{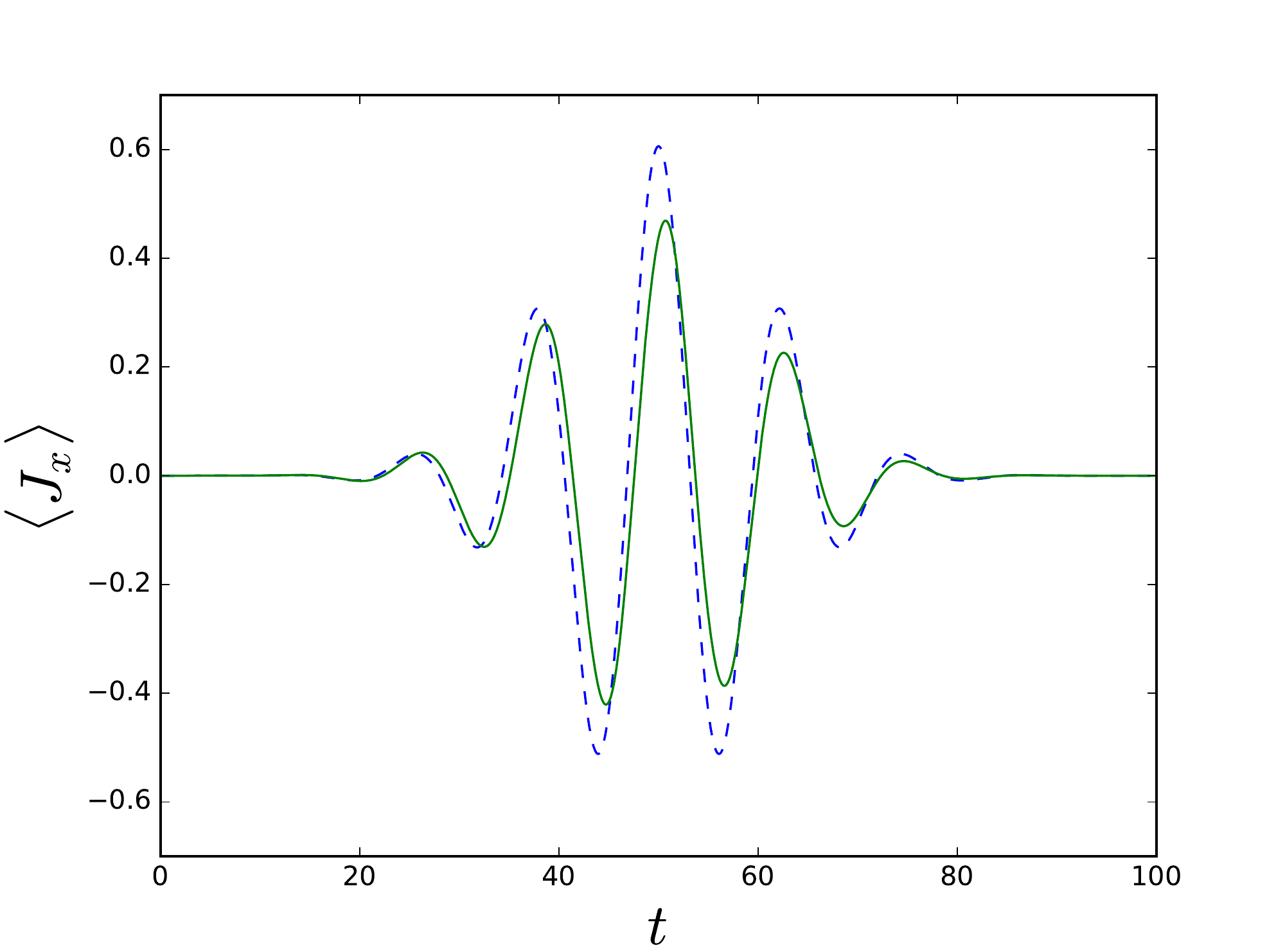}
\caption{\label{fig:vaidya_1pt} One point functions from the full numerical solution (solid) compared to the analytic large $\omega_P$ approximation at leading order (dashed), for $\omega_P = 0.5$.}
\end{center}
\end{figure}

\begin{figure}[ht]
\begin{center}
\includegraphics[width=0.48\textwidth]{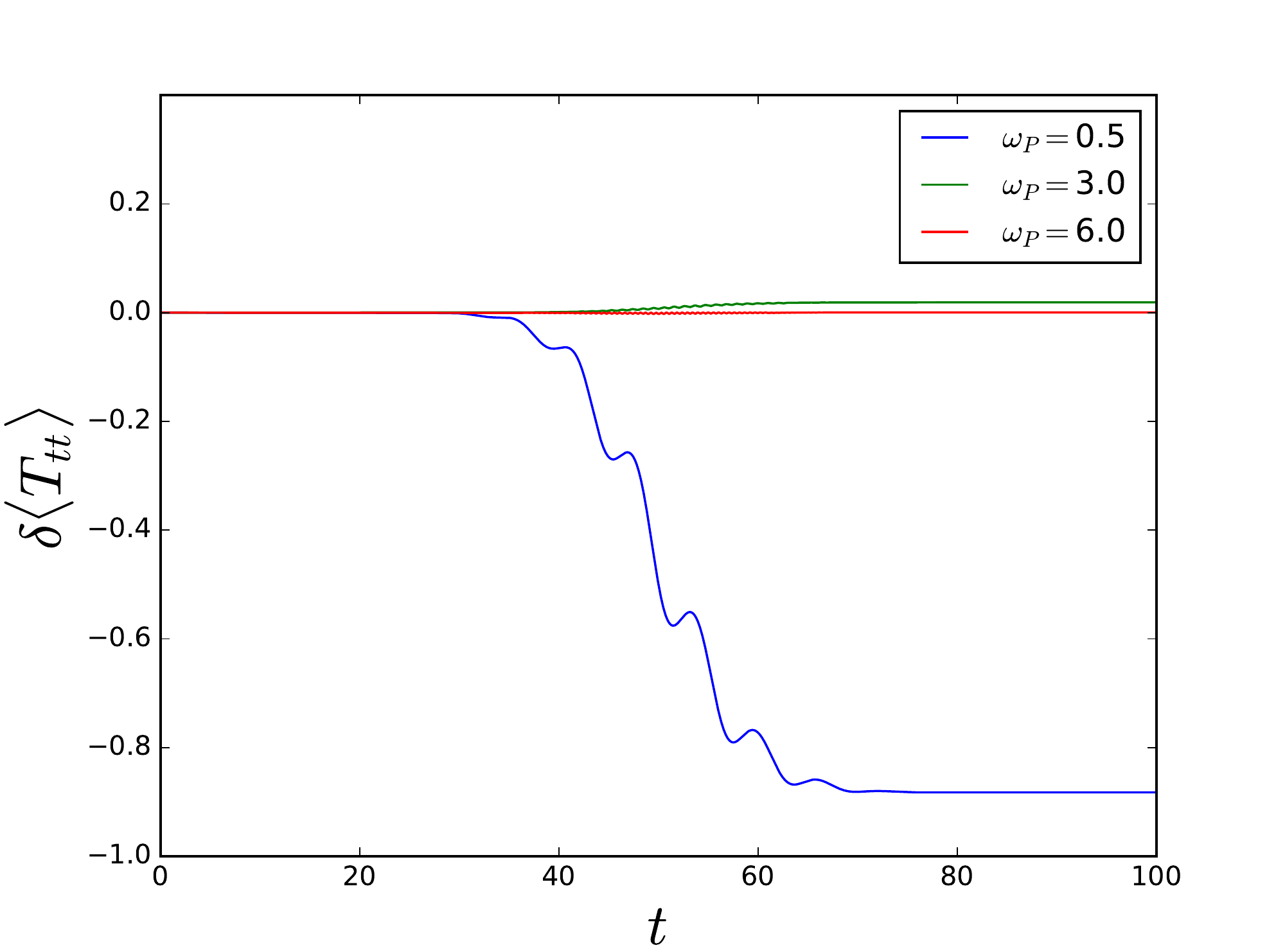}\hfill
\includegraphics[width=0.48\textwidth]{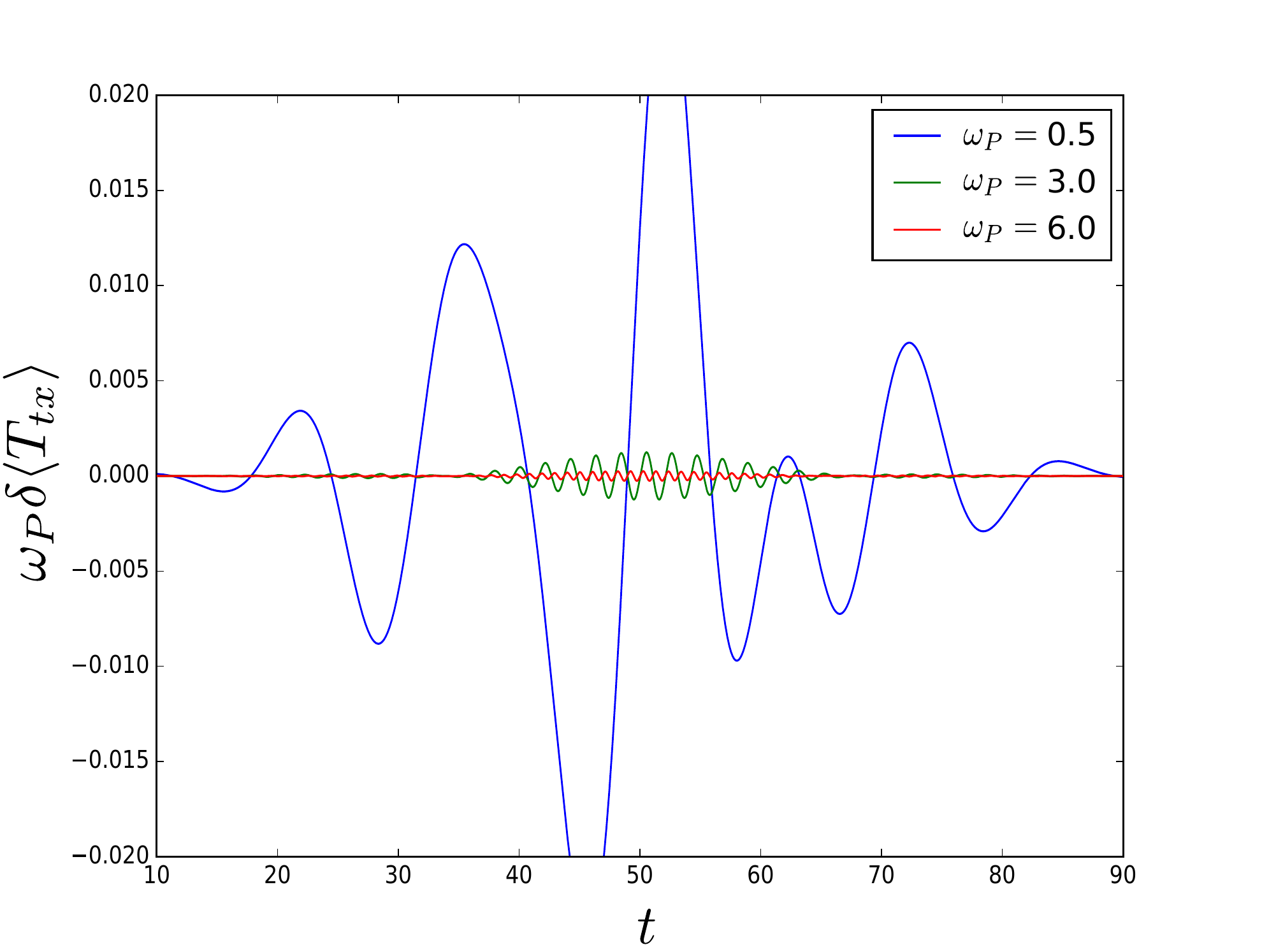}
\includegraphics[width=0.47\textwidth]{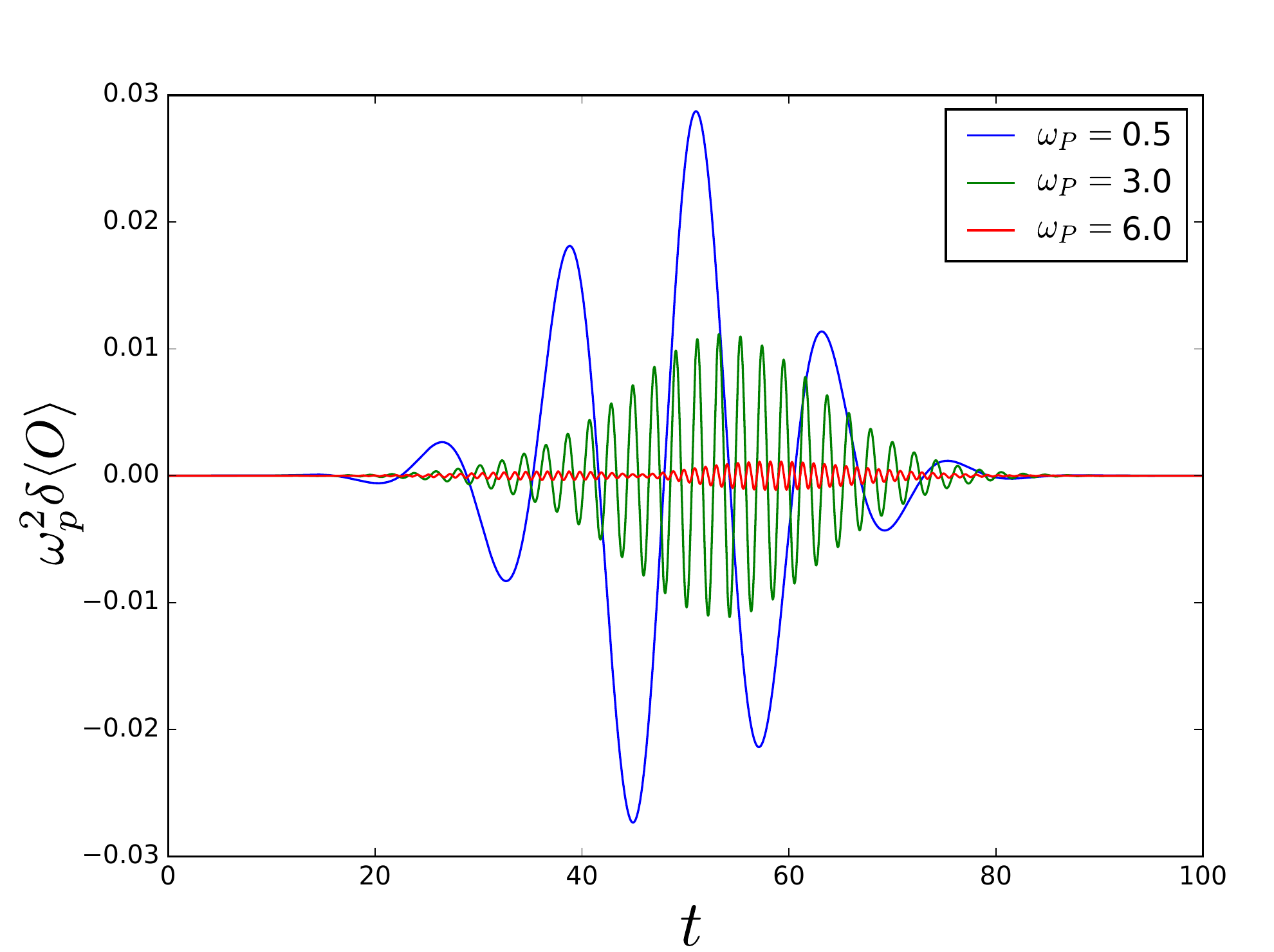}\hfill
\includegraphics[width=0.48\textwidth]{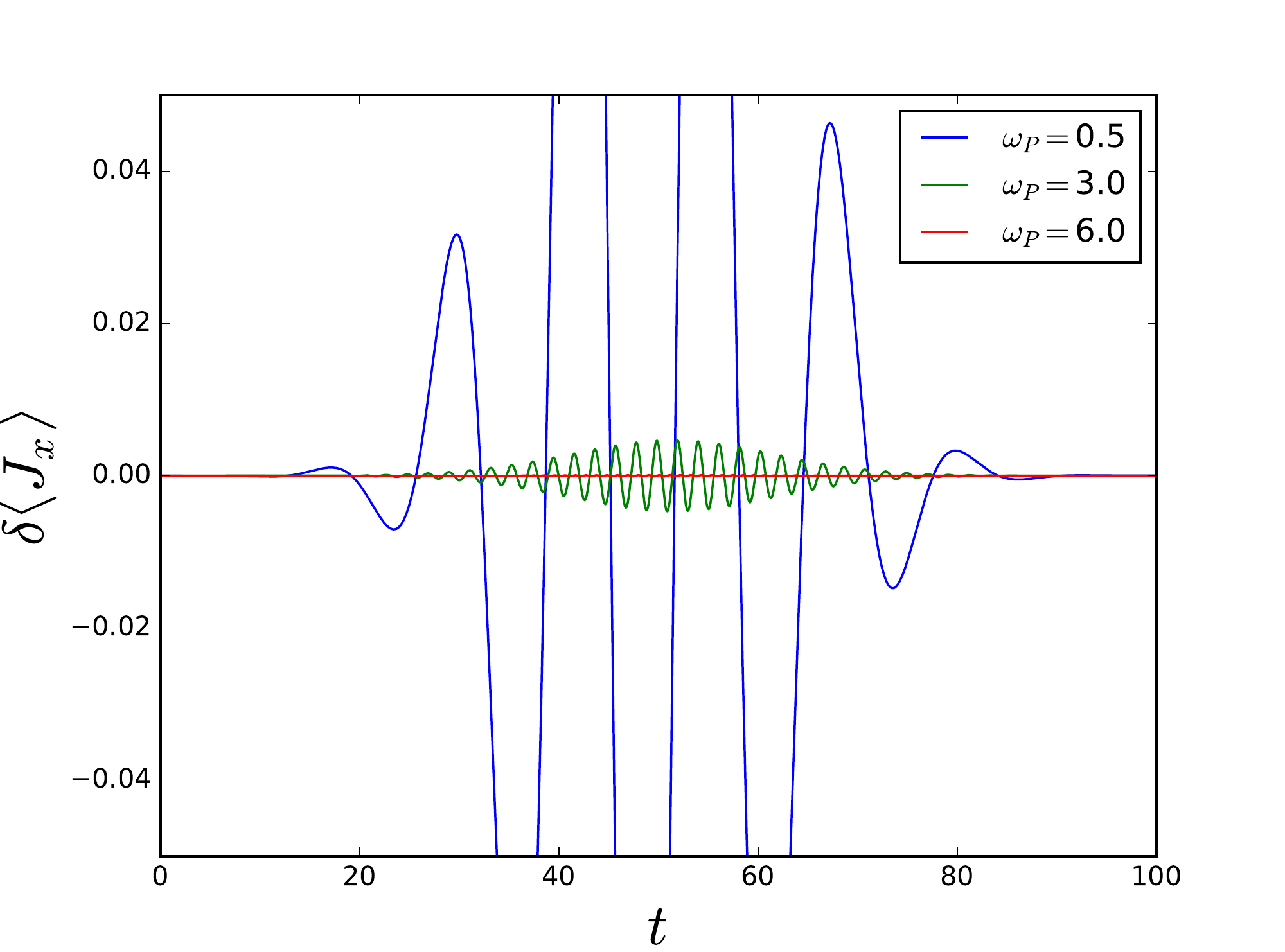}
\caption{\label{fig:delta_1pt} Difference between the expectation value obtained from numerics and from the large $\omega_P$ analytic solution. The differences are seen to decrease as  $\omega_P$ increases.}
\end{center}
\end{figure}

In Fig.~\ref{fig:vaidya_1pt} we show the one point functions obtained from the full numerical solution (solid curves) together with the large $\omega_P$ analytic solution. In this example we have a fairly small value of $\omega_P=0.5$. Thus,
the two results do not agree quantitatively very precisely, although the qualitative form of the solutions is already very similar.
Furthermore for $T_{tx}$, the difference between the full solution and the approximate analytic one is already 
surprisingly small.
To test the convergence of the approximate analytic solution to the full numerical solution at large $\omega_P$, we define the subtracted one point functions
\begin{align}
\delta \langle T_{tt}(t)\rangle &= \langle T_{tt}(t)\rangle - 2 m - \int^{t}_{-\infty} dt' E_x(t')^2,
\\
\delta \langle T_{tx}(t)\rangle &= \langle T_{tx}(t)\rangle - \rho \int_{-\infty}^t dt'  E_x(t'),
\\
\delta \langle \mathcal{O}(t)\rangle & = \langle \mathcal{O}(t)\rangle - \frac{\rho k}{4}\int^{t}_{-\infty} dt'\int^{t'}_{-\infty} dt'' E_x(t'').
\end{align}
These are plotted in Fig.~\ref{fig:delta_1pt}, where we have multiplied them  with appropriate powers of $\omega_P$ in order to make the
corresponding expectation values order one in the large $\omega_P$ limit. 
It can be seen that the full numerical solution is converging well to the approximate analytic one.


\subsection{Estimating the size of non-thermality from linear response theory}

In this subsection, we estimate the size of the quasinormal mode contributions to the one point
functions (and thus, to the gravitational background solution) from linear response theory. That is,
we assume that the electric field $E_x$ is small and we evaluate at linearized level its effect on one point functions. This assumption is clearly not a priori valid for our setup, but the final result we obtain this way is surprisingly close to the exact numerical results.

Linear response theory tells us that the leading contribution to the expectation value
of an operator $\chi(t)$ due to the presence of an external electric field is given by\footnote{Here for simplicity we work in the equivalent gauge where the electric field is generated  by $A_x$, that is $E_x = -\partial_t A_x$. In writing the linear response  \eqref{eq:lin_resp} we are also assuming that the expectation value of the operator $\chi(t)$ is zero when the electric field is absent, which is the case for the operators we will be considering.} 
\beq
\langle\chi(t)\rangle = \int_{-\infty}^{t} dt' G_R^{\chi,J_x}(t,t') A_x(t') \label{eq:lin_resp} \, ,
\eeq
where 
\be
G_R^{\chi,J_x}(t,t') = - i\theta(t-t')\int d^2x' \langle [\chi(t, x), J_x(t',x')]\rangle \, . \label{eq:GR}
\ee
The expectation value is taken in the final equilibrium thermal state, and in writing  \eqref{eq:GR}  we have taken into consideration the fact that we are considering a spatially homogeneous configuration and consequently $G_R^{\chi,J_x}$ is independent of the spatial position.  We will be considering the operator $\chi$  to be one of the vector sector $T_{tx}, J_x, \mathcal{O}$. 

The retarded correlator can be expanded at late times (i.e.\ large $|t-t'|$) in terms of quasinormal modes  
\beq
G_R^{\chi,J_x}(t,t') = \theta(t-t')\sum_n g_n e^{-i\omega_n (t-t')} \, ,
\eeq
where $\omega_n$ are the quasinormal modes frequencies shared by the correlators of the vector
sector operators 
and $g_n$ are the residues of the quasinormal mode poles in the Fourier transformed correlator. 

In the situation we are interested in, $A_x$ vanishes after the pulse is  turned off. For late enough times $t$ compared to the time where the source has been turned off, we can reliably substitute  the quasinormal mode expansion inside the integral (\ref{eq:lin_resp}). This way we obtain the late time expression 
\beq
\langle\chi(t)\rangle = \sum_{n} g_n e^{-i\omega_n t}\int_{-\infty}^{t} dt' A_x(t') e^{i\omega_n t'} \, .
\eeq
Thus, $\chi(t)$ decays at late times with a rate set by the quasinormal modes and an amplitude set by the integral involving $A_x$. When $t$ is large enough to be
outside the support of $A_x$,  the time integral with upper limit is formally equal to the integral over the entire temporal domain. Integrating by parts we can express this in terms of the electric field
\beq
\int_{-\infty}^{\infty} dt' A_x(t') e^{i\omega_n t'} = -\frac{i}{\omega_n}\int_{-\infty}^{\infty} dt' E_x(t') e^{i\omega_n t'}\, ,\label{eq:QNM_size}
\eeq
which gives for the  linear response of the operator $\chi$ at late enough times where the electric pump field has been turned off 
\beq \label{eq:QNMdecomposition}
\langle\chi(t)\rangle = - i \sum_{n} \frac{g_n}{\w_n} e^{-i\omega_n t}  \int_{-\infty}^{\infty} dt' E_x(t') e^{i\omega_n t'} \, .
\eeq
This shows how the amplitude associated to each quasinormal mode contribution is determined by the Fourier transform of the pulse electric field evaluated at the frequency of the quasinormal mode itself.
At the times we focus on in our analysis, the lowest lying QNM with purely imaginary frequency generically dominates over the others. There can be in principle cases where other QNMs, with frequency having a non-vanishing real part, may be in resonance with the pump pulse and compete with the leading QNM. Nonetheless, these contributions would decay fast in time and quickly become negligible.

\begin{figure}[t]
\begin{center}
\includegraphics[width=0.48\textwidth]{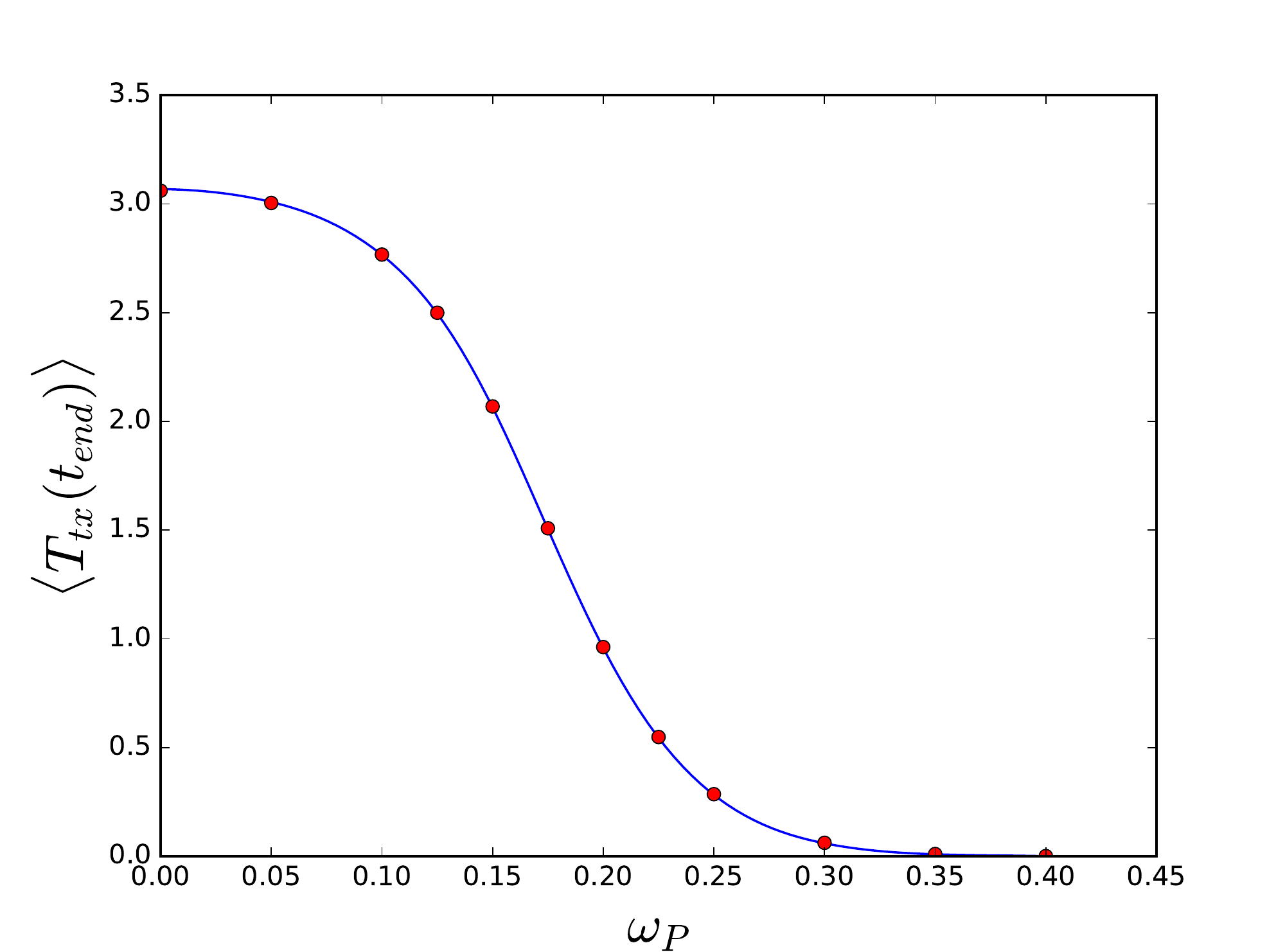}
\includegraphics[width=0.48\textwidth]{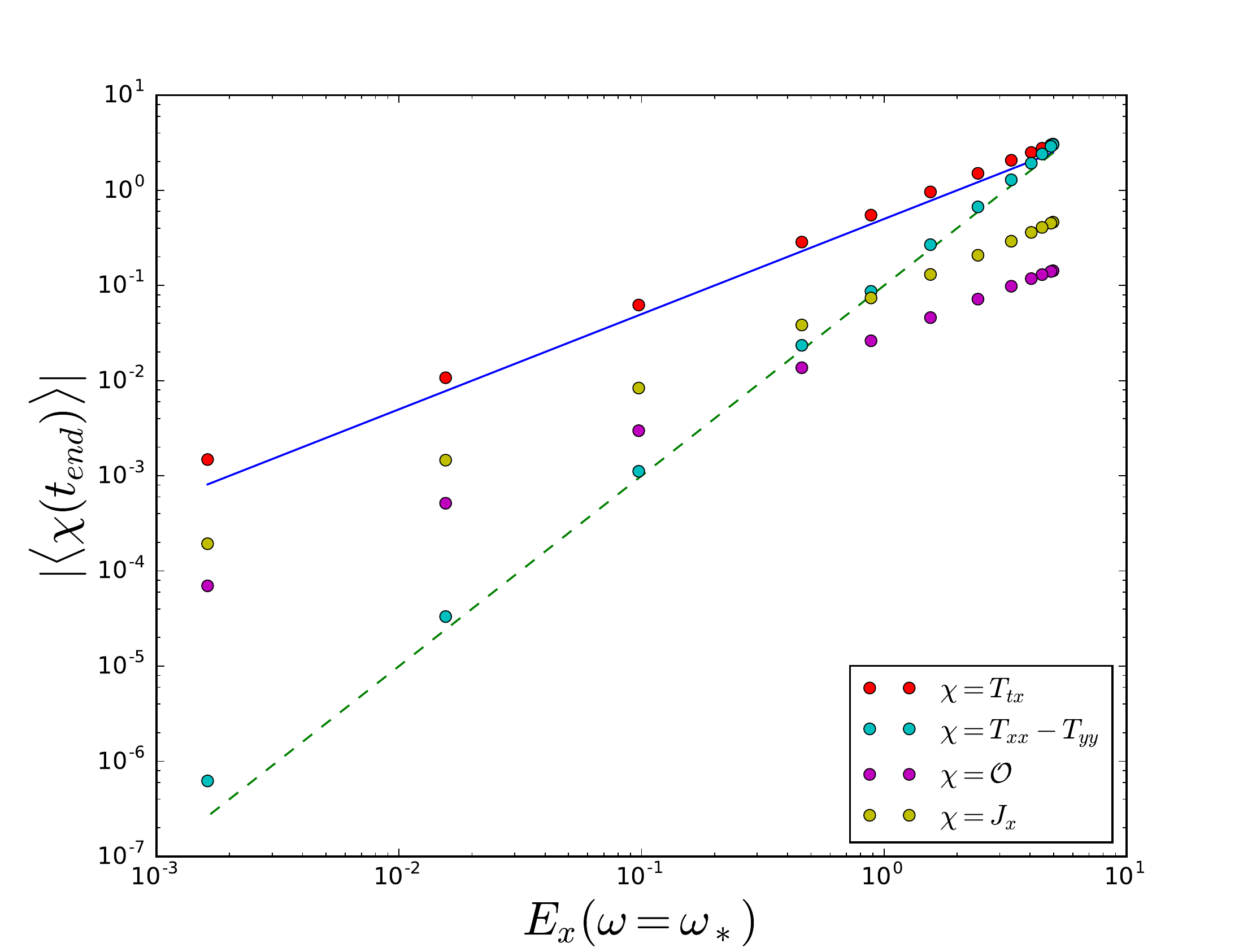}
\caption{\label{fig:QNM_comp} Left: The red dots are data points obtained from the full numerical solution. The blue
curve is a fit to the functional form (\ref{eq:QNM_size}) with a single fitting parameter (the overall scale). The 
root-mean-square error of the fit is $RMSE\approx 0.0036$. Right: Different one point functions evaluated at $t=t_{end}$
plotted as functions of the Fourier transformed electric field (\ref{eq:QNM_size}). The blue solid line corresponds to
a linear relation, while the green dashed line corresponds to a quadratic relation.}
\end{center}
\end{figure}

In Fig.~\ref{fig:QNM_comp} we plot the expectation value $\langle T_{tx}\rangle$ at the time $t=t_{end}$ when
the pump pulse turns off. In the large $\omega_P$ approximation the expectation value is immediately zero at this
time. In the numerics we see significant deviations of $\langle T_{tx}\rangle$ from zero for small $\omega_P$. 
The deviation mainly arises from the lowest quasinormal mode contribution, whose size is given by
the Fourier transform (\ref{eq:QNM_size}) in the linear response approximation. 
The blue curve in   Fig.~\ref{fig:QNM_comp} 
is obtained by fitting (\ref{eq:QNM_size}) with a constant coefficient in front as a fitting parameter.
As the figure shows, the linear response curve fits the numerical data very well with a root-mean-square error\footnote{The root-mean-square error is the positive root of $RMSE^2 = MSE = n^{-1}\sum_{i=1}^n(y_i - y_i^{fit})^2$.} of $RMSE\approx 0.0036$. Similarly, also the other one point functions  $\langle J_x\rangle$ and  $\langle\mathcal{O}\rangle$ can be fitted with the form (\ref{eq:QNM_size}). An exception is
the one point function $\langle(T_{xx} - T_{yy})\rangle$ which is better fitted to $(E_x(\omega = \omega_*))^2$.
In the right part of Fig.~\ref{fig:QNM_comp}, we show a loglog plot of the one point functions versus the Fourier transformed
electric field. For the vector sector operators, the relation is approximately linear while for $\langle(T_{xx} - T_{yy})\rangle$
the relation is approximately quadratic. The quadratic approximation appears for the same reason as the factor of two 
in the decay of $\langle(T_{xx} - T_{yy})\rangle$, as the corresponding field is sourced by the squares of the vector sector
fields.


\section{Driven oscillator toy model} \label{sec:toy}

As we will show in this section, two signatures in the evolution of our holographic setup can be captured and explained by a
simple toy model given in terms of a driven damped harmonic oscillator. These are the instantaneous relaxation at large driving frequency $\omega_P\rightarrow \infty$ and the smallness of the quasinormal mode contributions when 
$\omega_P$ is separated from the real parts of the quasinormal mode frequencies.  The equation of motion of the driven 
oscillator is given by
\beq
\ddot{x} = -\omega_0^2 x - \kappa \dot{x} + f(t) \, ,\label{eq:osc}
\eeq
where $\omega_0$ is the undamped oscillation frequency and $\kappa$ the damping strength.  
We will choose the driving force to have the form
\beq
f(t) = \cos(\omega_P t) \Omega(t), \label{eq:time_decomp}
\eeq
where $\Omega(t)$ varies much more slowly than the cosine.
Without a driving force,
displacements of $x$ decay back to zero exponentially in time as  
\beq
x(t) = A_{-} e^{ -i \omega_- t} + A_{+} e^{- i \omega_+ t}\, , \label{eq:OQNM}
\eeq
where  
\beq
\omega_{\pm} =   -\frac{i}{2}(\kappa \pm \nu)\, , \quad \nu = \sqrt{\kappa^2 - 4\omega_0^2} \, . 
\eeq
These complex frequencies represent the analogue of the quasinormal mode frequencies in our system. 
The driven system   \eqref{eq:osc}  is solved explicitly  as
\be
x(t) =  A_-(t)e^{- i \omega_- t }  + A_+(t)  e^{-i \omega_+ t } \, , \label{eq:Osol}
\ee
where the time dependent amplitudes associated to each mode are 
\be\label{Apm}
A_{\pm}(t) =  \mp\frac{1}{\nu}\int^{t }_{-\infty}  dt'  e^{i   \omega_\pm t' } f(t' ) \, .
\ee
In writing the solution  we have assumed that $x(t)$ vanishes at early times before the driving force has been turned on.

Let us consider the large $\omega_P$ limit of the solution. The basic intuition is that in this case $x(t)$ oscillates fast in which case the equation of motion can be approximated as $\ddot{x}(t) \approx f(t)$, with the approximate solution $x(t) \approx -\cos(\omega_P t)\Omega(t)/\omega_P^2$. 
We can see this from the exact solution using integration by parts twice,\footnote{Integrating $\cos(\omega_P t)$ and taking derivatives of the rest.} leading to
\beq
A_{\pm}(t) = \mp\frac{1}{ \nu\omega_P} e^{i\omega_{\pm} t}\Omega(t) \sin(\omega_P t) \mp \frac{1}{ \nu\omega_P^2}(i\omega_{\pm} \Omega(t)+ \Omega'(t))e^{i\omega_{\pm} t}\cos(\omega_P t) + O(\omega_P^{-3}).
\eeq
Substituting into (\ref{eq:Osol}), we arrive at
\beq
x(t) = -\frac{1}{\omega_P^2}\cos(\omega_P t) \Omega(t) + O(\omega_P^{-3}) = -\frac{f(t)}{\omega_P^2} + O(\omega_P^{-3}).\label{eq:largewsol}
\eeq
Thus, we see that the oscillator follows the driving force instantaneously and, in particular, it relaxes instantaneously as the force vanishes. This behavior is similar to the instantaneous thermalization of  Vaidya spacetimes where one point functions relax to their thermal values as soon as the boundary source turns off. Notice also that the series expansion in $1/\omega_P$ can be computed to arbitrary orders by repeatedly integrating by parts. This way the instantaneous behavior is seen to hold to all orders in the $1/\omega_P$ expansion.

The next question we want to address is that what happened to the quasinormal mode contributions and whether a large $\omega_P$ is a necessary condition to have instantaneous thermalization.

At late times, that is for times after the driving function has been turned off,  it is apparent that the solution written in the form \eqref{eq:Osol} takes  the ``quasinormal mode''  decay form \eqref{eq:OQNM}. In fact,  $A_{\pm}(t)$  become time independent  as long as one considers values of $t$ where  the driving force has been turned off, and therefore at late times give the amplitude associated to the quasinormal modes. 
For concreteness, let us assume that  $\Omega(t)$ has support on a compact region so that the driving function is turned off after some time $t_{\rm end}$.   For all those times $t>t_{\rm end}$ where the driving force has been turned off,  one can formally replace the integrals in  \eqref{Apm} with integrals over the entire time range, that is 
\beq
A_{\pm} = \mp\frac{1}{\nu}\int^{\infty}_{-\infty}  dt'  e^{i   \omega_\pm t' } f(t' ) = \mp\frac{1}{2\nu}\(\hat{\Omega}( \omega_\pm+ \omega_P ) + \hat{\Omega}(\omega_\pm-\omega_P)\),
\eeq
where we denote the Fourier transforms as
\beq
\hat{\Omega}(\omega) = \int_{-\infty}^{\infty} dt' e^{i\omega t}\Omega(t) \, .
\eeq
For  any  smooth choice of $\Omega(t)$, the coefficients $\hat\Omega$ of the quasinormal modes will be suppressed more strongly than any inverse power of the arguments $  \omega_\pm \pm \w_P$. 

To build some more explicit intuition, we can specialize to an example close to our holographic calculation by choosing a Gaussian envelope
\beq
\Omega(t) =  e^{-\frac{t^2}{\Delta t^2} } \, .    \label{eq:gaussOmega} 
\ee
Strictly speaking with this choice the forcing pulse is never turned off. However, for all practical purposes, at large enough times compared to the Gaussian width we can consider the driving force to be vanishing. With this choice, the  amplitudes of QNMs take the form 
\be
\begin{aligned} \label{eq:modeampl}
  A_{\pm} = \mp\frac{ \sqrt{\pi}  \Delta t }{{2\nu}} \( e^{-\frac{( \omega_{\pm} + \w_P)^2 \Delta t^2}{4} } +  e^{-\frac{( \omega_{\pm}- \omega_P)^2 \Delta t^2}{4} } \).
 \end{aligned}
\ee
 Notice  that, modulo a factor $2\nu$, these are nothing else than the Fourier transform of the driving force  %
\beq
\hat f(p) = \frac{\sqrt{\pi} \Delta t }{2}  \(e^{-\frac{\Delta t^2 (p+\w_P)^2}{4}}  + e^{-\frac{\Delta t^2 (p -\w_P)^2}{4}} \) \ 
\eeq
evaluated at the  QNM frequencies $p =  \omega_{\pm }$.  
 This shows explicitly how the amplitudes  associated to QNMs depend on the overlap between the spectrum of the driving force and  the real part of the QNM frequencies:  If the driving frequency $\w_P$  coincides with the real part of a QNM frequency there will be no Gaussian suppression of the amplitude of the corresponding mode. Conversely, for ${\rm Re}( \omega_{\pm} \pm \w_P)\neq 0$ the QNM amplitude is exponentially suppressed in the square of this combination.

Let us  further  specialize to  the over-damped case,  where  $\kappa > 2m$.  This exactly mimics our holographic setup  in the fact that the relevant QNM frequencies are purely imaginary. Since $\w_{\pm}$ have no real part,  the amplitudes of the QNM excitations are exponentially suppressed in $\w_P$ as 
\begin{align}
 A_{\pm}   =\mp \frac{1}{\nu} \sqrt{\pi}  \Delta t \cos(\frac{\omega_P |\omega_{\pm}| \Delta t^2}{4}) e^{\frac{  |\omega_{\pm}|^{2} \Delta t^2} {4} } e^{-\frac{  \omega^2_P \Delta t^2}{4} }    \,  . 
\end{align}
Thus, for large $\omega_P \Delta t$ the quasinormal mode contributions are very strongly suppressed.


\section{Out of equilibrium conductivity \label{sec:conduct}}

Next, we want to study the conductivity properties of the non-equilibrium states we have prepared with a pump pulse.
For this purpose one introduces a smaller ``probe" electric field $\delta E_x(t)$. This electric field
induces a change in the current $\delta \langle J_x(t)\rangle$. This way we can define a real-time
conductivity called differential conductivity $\sigma(t,t')$ through the relation
\beq
\delta \langle J_x(t)\rangle = \int_{-\infty}^t dt'\sigma(t, t') \delta E_x(t')\, .\label{eq:difcond}
\eeq 
Although there is no standard definition of frequency space conductivity out of equilibrium, we will
follow \cite{Lenarcic2014} and define
\beq 
\sigma(\omega, t) = \int_t^{t_m}dt' e^{i\omega(t' - t)}\sigma(t', t)\, ,\label{eq:cond}
\eeq
where $t_m$ is the time at which the experiment ends. The conductivity defined this way can
be related to the current two point function as discussed in Appendix \ref{sec:curcurcor}. In thermal equilibrium,
the conductivity (\ref{eq:cond}) approaches the standard definition of optical conductivity at frequency $\omega$,
and spatial momentum $k=0$ as the observation time $t_m$ is sent to infinity. This is also shown in Appendix
\ref{sec:curcurcor}.

In practice we calculate the conductivity in two steps. First, we calculate the differential conductivity
appearing in (\ref{eq:difcond}), and then perform a Fourier transform to obtain (\ref{eq:cond}). 
By choosing a probe field $\delta E_x(t) = \epsilon \delta(t - t_0)$, (\ref{eq:difcond}) becomes
\beq
 \delta \langle J_x(t)\rangle = \epsilon \sigma(t, t_0) \, .
\eeq
This way we obtain  the differential conductivity directly from the knowledge of  $\delta \langle J_x(t)\rangle$. In the numerical implementation, we actually use a smoothed version of a delta function, which we choose to be a Gaussian
\beq
\delta E_x(t) = \epsilon \frac{1}{\sqrt{2\pi}\delta t} e^{-\frac{(t-t_0)^2}{2(\delta t)^2}}\, .
\eeq
In the limit $\delta t\rightarrow 0$, this approaches a delta function, while in practice we keep $\delta t$ non-vanishing but small enough that it does not affect our results considerably. 
Smoothing out the delta function over a small scale $\delta t$ affects the conductivity at frequencies $\omega \propto 1/\delta t$ and larger, while  the main interesting time-dependence in the conductivity is seen for frequency $\omega = O(1)$. In our analysis, we have taken $\delta t = 0.05$, which has only small (order $1\%$) effect on the conductivity in the regime of interest.

\subsection{Numerical results}

\begin{figure}[t]
\begin{center}
\includegraphics[width=1 \textwidth]{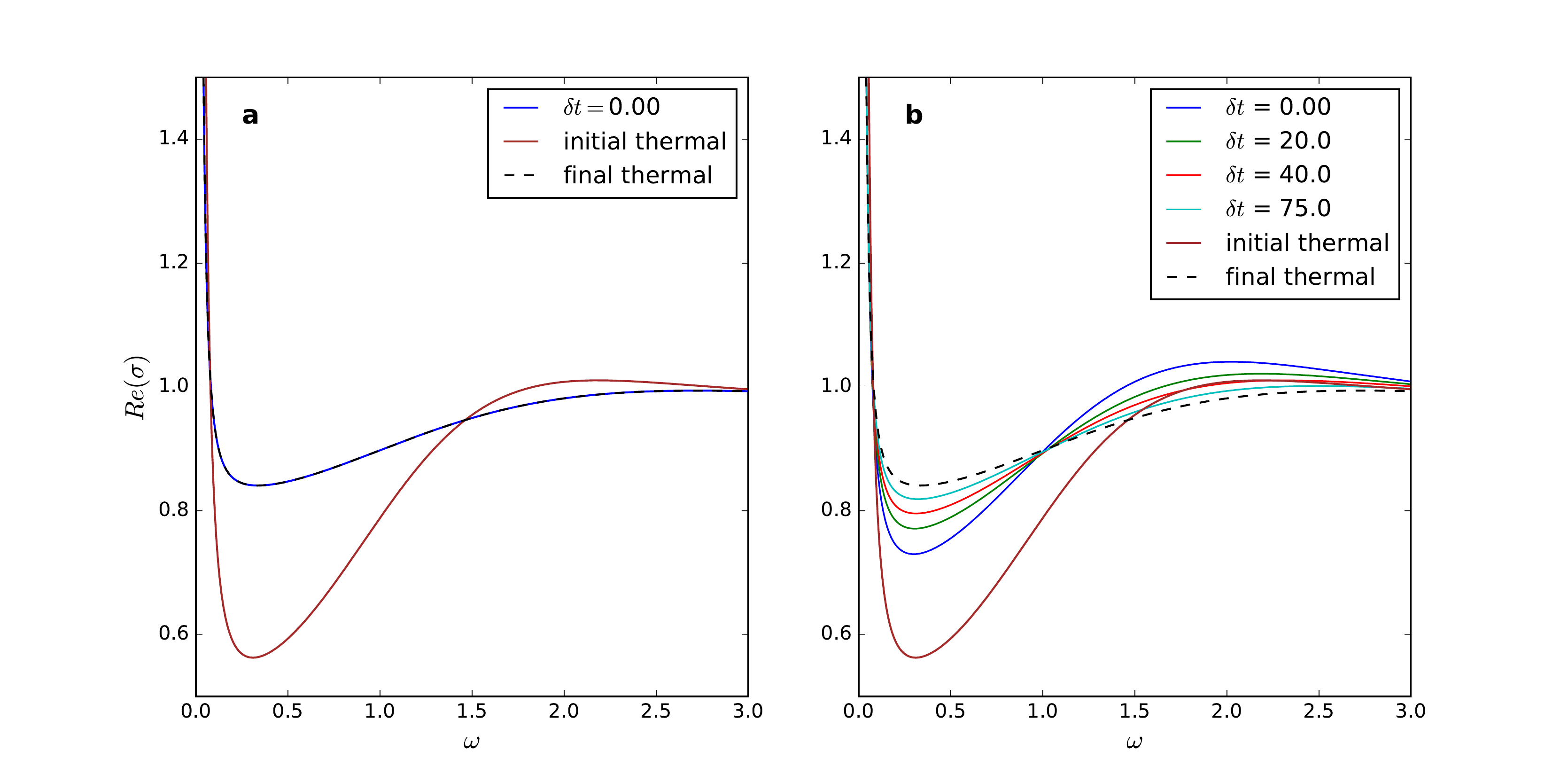}
\caption{\label{fig:sigma_omega_non_eq} The optical conductivity at different times and in the initial and the final thermal equilibrium states. Left: The pump frequency is $\omega_P = 0.5$, and the optical conductivity right after the pump pulse has ended coincides with the final equilibrium one. Right: The pump frequency is $\omega_P=0$,  and the optical conductivity is seen to interpolate in time between the initial and final equilibrium values.}
\end{center}
\end{figure}

First, we consider the pump pulse profile (\ref{eq:pump_pulse_form}) with different values of the pump frequency $\omega_P$. Fig.~\ref{fig:sigma_omega_non_eq} shows the optical conductivity as a function of $\omega$ for different values of times. The time $\delta t$ is measured from the time $t_m = 3\Delta t + t_0$ at which the pump pulse practically turns off (due to the smoothed theta function in (\ref{eq:pump_pulse_form})), 
\beq
\delta t = t - t_m.
\eeq
Fig.~\ref{fig:sigma_omega_non_eq}a shows the conductivity for the large frequency pump pulse which appears thermal immediately at time $\delta t=0$. This is consistent with the results of the previous section which showed that the larger $\omega_P$, the closer to the Vaidya spacetime we get. 
On the other hand Fig.~\ref{fig:sigma_omega_non_eq}b shows the conductivity for $\omega_P=0$, in which case the conductivity deviates significantly from the final thermalized conductivity for all times displayed. 

\begin{figure}[t]
\begin{center}
\includegraphics[width=0.47 \textwidth]{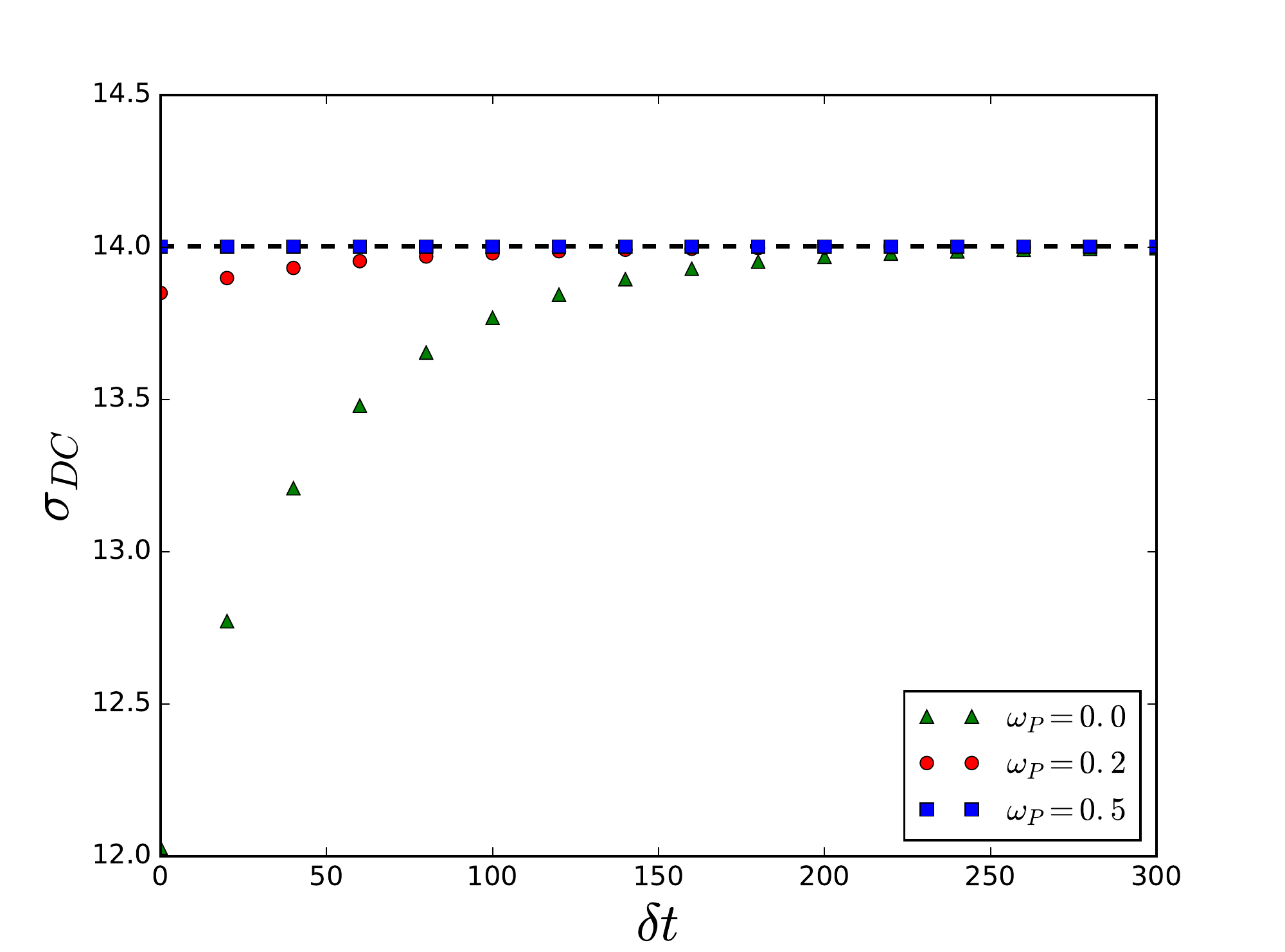}  
\includegraphics[width=0.47 \textwidth]{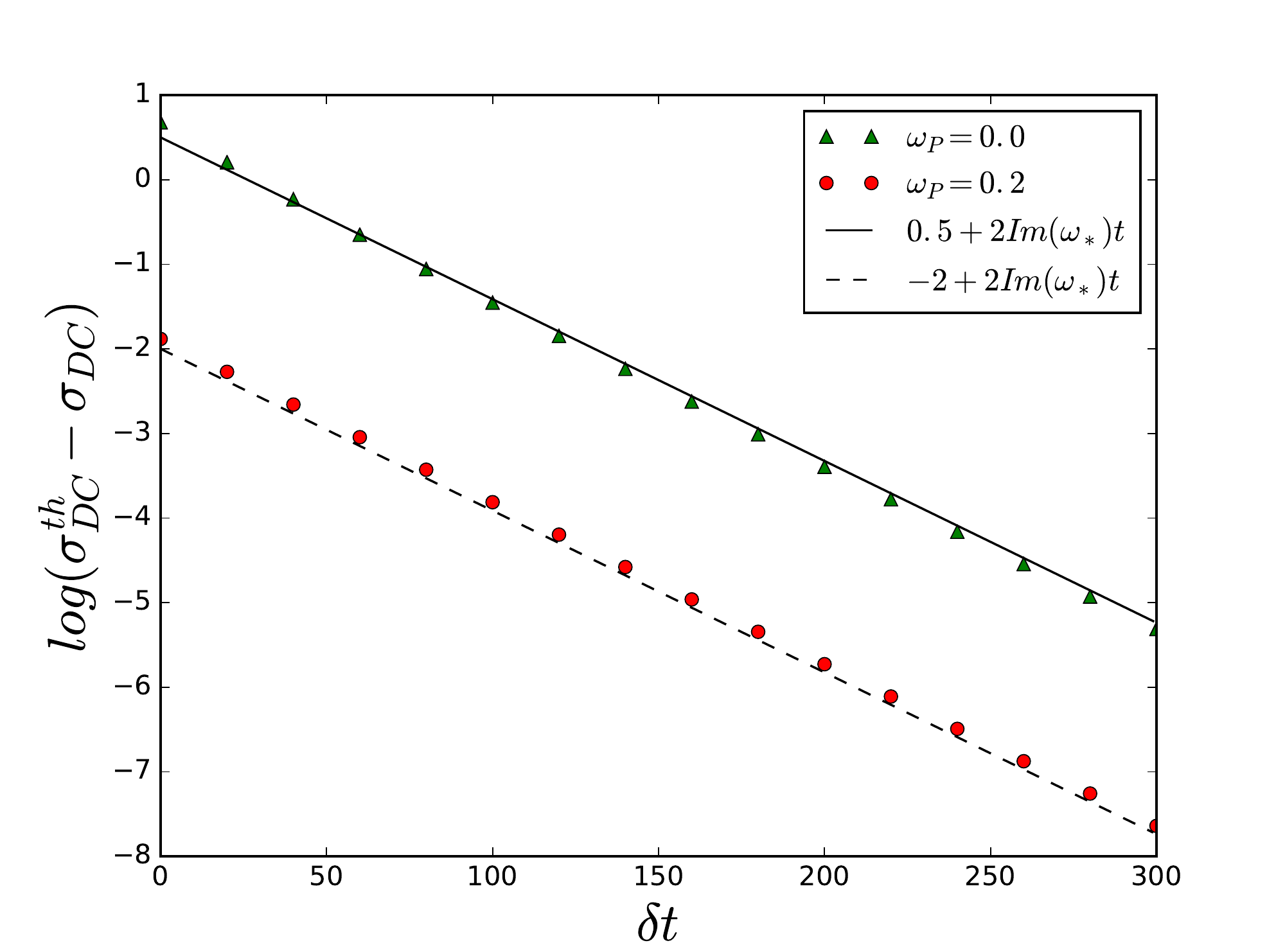}
\caption{\label{fig:sigma_dc_non_eq} The DC conductivity as a function of time.}
\end{center}
\end{figure}

Next, we study how the conductivity approaches its final thermalized value. We will focus on the DC conductivity, that is, the optical conductivity $\sigma(\omega)$ at $\omega=0$.  Fig~\ref{fig:sigma_dc_non_eq} shows $\sigma_{DC}$ as a  function of time. 
In the previous section we saw that the background spacetime approaches a static black hole with a rate set by the lowest quasinormal mode. Thus, we might expect that the conductivity approaches its final thermalized value with the same rate.
This is where we find a slightly surprising result. The conductivity approaches its thermalized value with a rate $2 \textrm{Im}(\omega_*)$. We will come back to this factor of two in the next subsection. Thus, the DC conductivity is consistent with the approximate form
\beq
\sigma_{DC}(\delta t) = \sigma_{DC}^{th} + C e^{-2 \textrm{Im}(\omega_*) \delta t},\label{eq:approx_decay}
\eeq
where $\sigma_{DC}^{th}$ is the final thermalized value of the DC conductivity. The coefficient $C$ quantifies 
how far out of equilibrium the conductivity gets. Defining
\beq
\delta \sigma_{DC}= \sigma_{DC}^{th} - \sigma_{DC}(\delta t = 0)\, ,
\eeq
with the approximate form (\ref{eq:approx_decay}) we have $\delta \sigma_{DC}=-C$. 

\begin{figure}[t]
\begin{center}
\includegraphics[width=0.8 \textwidth]{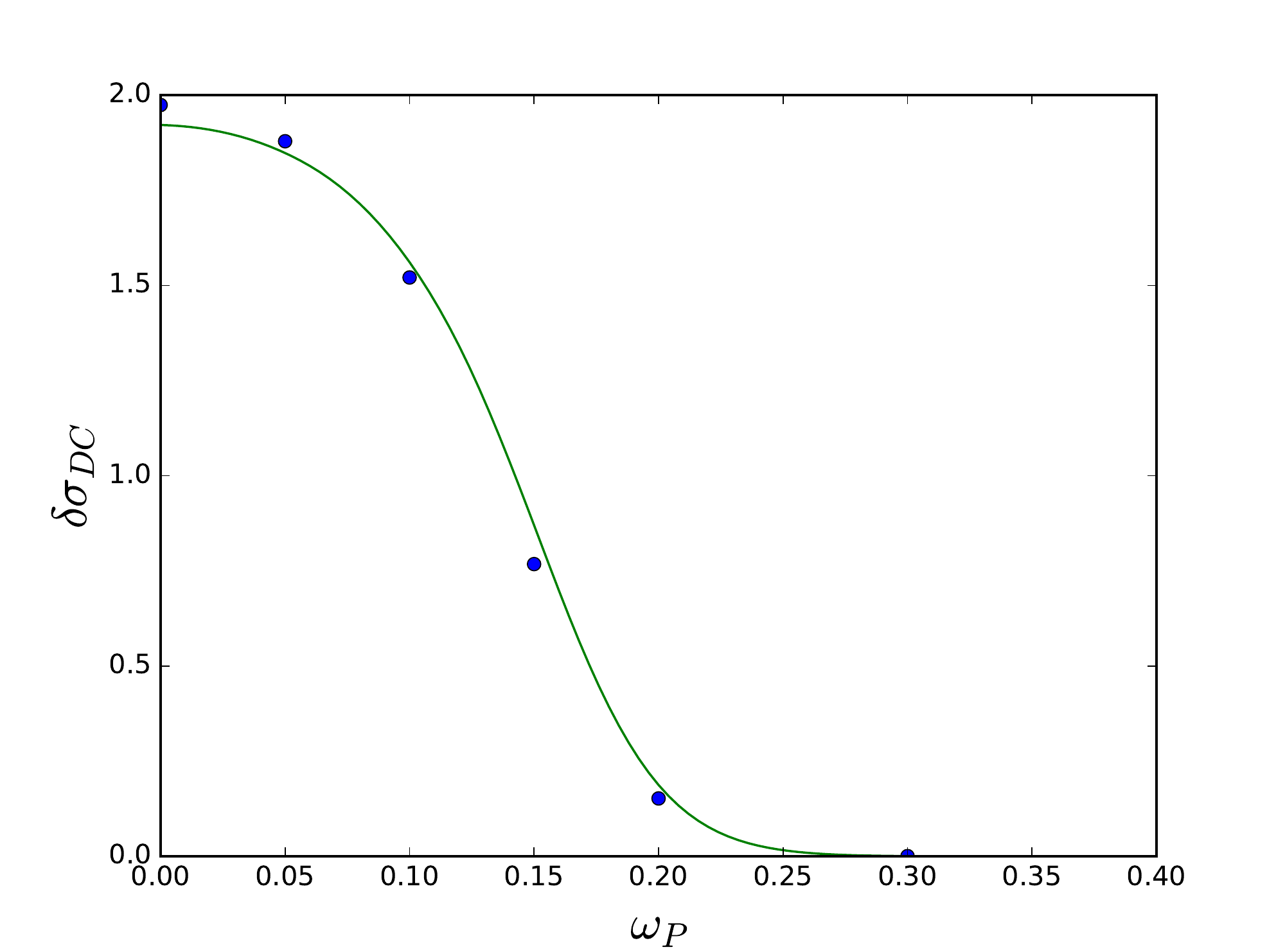}
\caption{\label{fig:C_vs_omega_P} The maximum deviation of the DC conductivity from its thermalized
value as a function of the pump frequency $\omega_P$. The solid green curve is a one parameter fit 
to a form $(E_x(\omega_*))^2$ }
\end{center}
\end{figure}

In Fig.~\ref{fig:C_vs_omega_P} we show how $\delta \sigma_{DC}$ behaves as a function of $\omega_P$.
As before, the initial and final temperatures are kept fixed while varying $\omega_P$.
Clearly the largest deviation appears as $\omega_P=0$. Recalling that the amplitude of the deviation of the 
background spacetime from the equilibrium one was well approximated by the Fourier transformed electric field
$E_x(\omega_*)$, we are motivated to also attempt a similar fit to the deviation of the conductivity from its 
equilibrium value. Fitting $\delta \sigma_{DC}$ with $E_x(\omega_*)$ does not give a good fit but instead
$E_x(\omega_*)^2$ does. A fit to $E_x(\omega_*)^2$  is shown in Fig.~\ref{fig:C_vs_omega_P} as the solid
green curve. This is a one parameter fit with the overall amplitude being the only parameter. Thus, our results 
for the time dependence and $\omega_P$ dependence of the non-equilibrium conductivity can be summarized in
\beq
\sigma_{DC}(\delta t) = \sigma_{DC}^{th} + \gamma E_x(\omega_*)^2 e^{-2 \textrm{Im}(\omega_*) \delta t},\label{approx_decay_2}
\eeq
with a constant coefficient $\gamma\approx -0.0798$. 

\begin{figure}[t]
\begin{center}
\includegraphics[width=0.8 \textwidth]{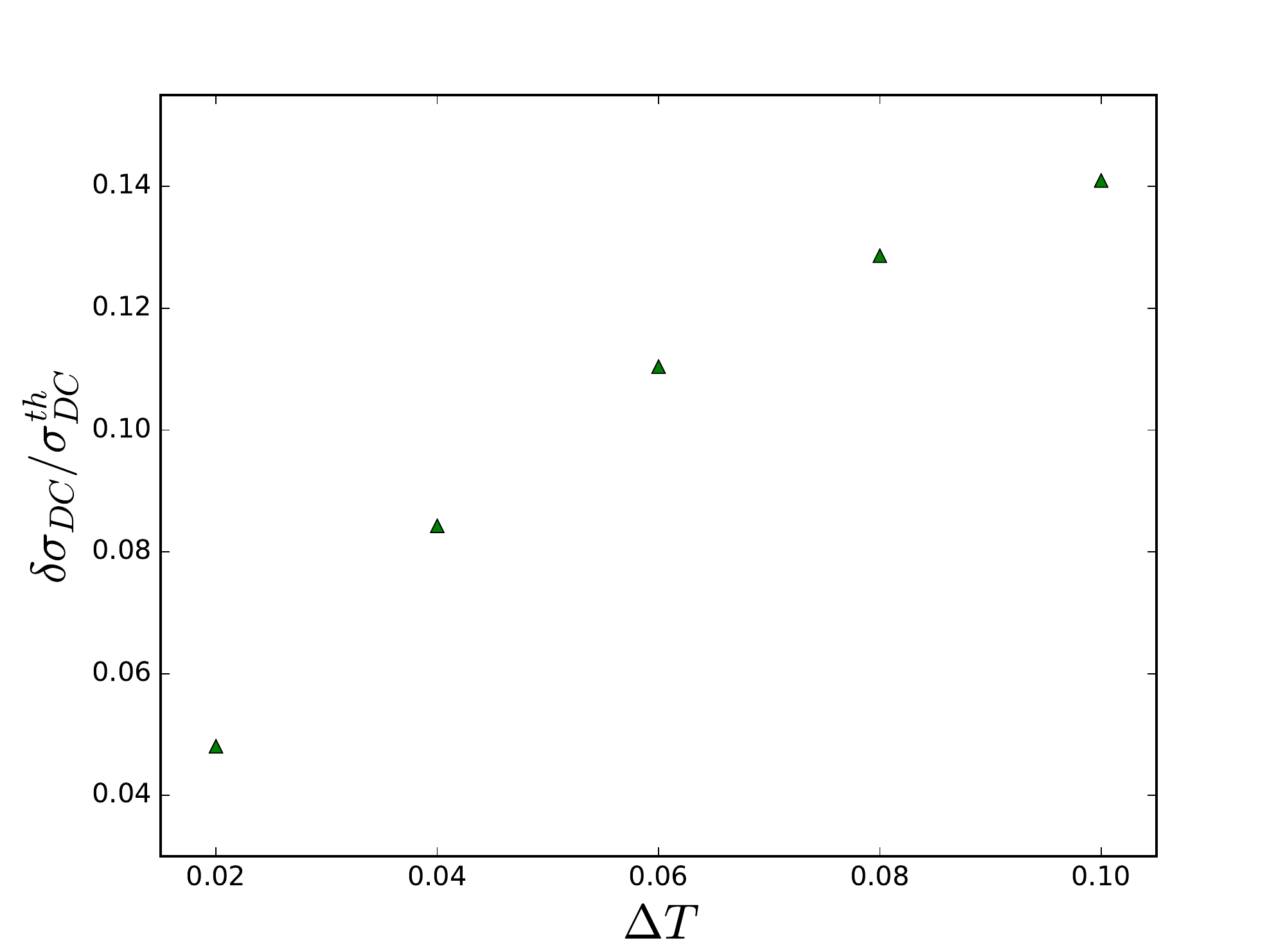}
\caption{\label{fig:dsigma_vs_dT} The maximum deviation of the DC conductivity from its thermalized
value as a function of the temperature difference $\Delta T = T_f - T_i$.}
\end{center}
\end{figure}

Finally we study how the
deviation from thermality behaves as we change the difference between the initial and final temperatures. 
We have chosen to keep the final temperature $T_f$ fixed and to vary 
the difference $\Delta T = T_f-T_i$ by changing the initial temperature $T_i$. A plot of $\delta \sigma_{DC}$ as a function of $\Delta T$ is shown in 
Fig.~\ref{fig:dsigma_vs_dT}.
This confirms the intuition that the more we increase the magnitude of the pumping electric field,
the further from equilibrium the conductivity deviates (in units of the final thermalized conductivity $\sigma_{DC}^{th}$).

\subsection{A symmetry argument for the thermalization rate}

We have given numerical evidence that the conductivity thermalizes with a rate $e^{-2i\omega_* t}$
where $\omega_*$ is the lowest vector quasinormal mode. The conductivity deviates from thermality
simply because the background spacetime deviates from a static black hole. Thus, it is not surprising
that the thermalization time scale of the conductivity is related to that of the background spacetime. What is somewhat surprising instead is the factor of $2$ relating the two rates. Here we provide a symmetry argument for it.

Let us start considering the  symmetries of the bulk action  \eqref{eq:action} of the holographic model we are studying.  These include the subgroup  $SO(2) \times SO(2)$, where the first factor represents the rotations $M$ acting on the spatial coordinates $x^{i}=(x,y)$ common to the boundary, and the second factor the global rotations $R$ that act non-trivially only on the scalar duplet $\phi_{I} = (\phi_1,\phi_2)$, rotating the two fields into one another.

The equilibrium solution, and more specifically the scalar field configuration \eqref{eq:scalarsk}
\beq
\phi_I = k \delta_{I i} x^i \, , 
\eeq
explicitly breaks this symmetry as  $SO(2) \times SO(2) \to  \sores$, with the residual $SO(2)$ being  the subgroup that leaves the scalar field configuration invariant,
\be
\phi_{I}(x^i) \to R_{I}^{~J}\phi_{J}(M^{k}_{~i}x^i) =  \phi_{I}(x^i) \, , 
\ee
that is
\be
 R_{I}^{~J}\delta_{Jk}M^{k}_{~i} = \delta_{Ii} \, . 
\ee
The bulk metric and the bulk gauge field configurations \eqref{eq:eqsol} do not break any of the original $SO(2) \times SO(2)$ symmetry, as they are isotropic (and homogeneous) in the boundary spatial coordinates and they do not transform under the global $SO(2)$ symmetry.  
Hence the residual $SO(2)$ is automatically preserved by the bulk metric and gauge fields, and the entire equilibrium solution is a scalar of $ \sores$.

One can conveniently organize  deviations from the equilibrium background solution according to representations of the  $\sores$. 
Starting with the ansatz  \eqref{eq:ansatz} for the non-equilibrium solution, and  employing the definitions given in  \eqref{eq:fluctdef}, we can split the fluctuations as  
\begin{align}
&\d F_z, \d\Sigma, \d a_v && \textrm{scalar},\nonumber
\\
 &\d F_x, \d a_x, \d \Phi & &\textrm{vector},\nonumber
\\
 &\d B &  &\textrm{symmetric traceless tensor}.\nonumber
\end{align}
As we described, at sufficiently late times the spacetime is close to thermal equilibrium and one can to a good approximation expand the equations of motion in perturbations around the equilibrium spacetime. 

At the linearized level the three type of perturbations completely decouple from each other. This can be immediately understood by thinking about the equation of motions in terms of the $\sores$ symmetry. In general, the linearized equations for the fluctuations can be schematically written as
\begin{align} \label{eq:lin_eqs_schem}
\mathcal{L}^{(1)}_{SS} \d S + \mathcal{L}^{(1)}_{SV} \d V +\mathcal{L}^{(1)}_{ST} \d T = 0 ,  \nonumber \\
\mathcal{L}^{(1)}_ {VS}\d S+ \mathcal{L}^{(1)}_{VV} \d V +\mathcal{L}^{(1)}_{VT} \d  T = 0, \\
\mathcal{L}^{(1)}_ {TS}\d S+ \mathcal{L}^{(1)}_{VS} \d S +\mathcal{L}^{(1)}_{TT} \d T = 0 .  \nonumber 
\end{align}
Here  $\d S , \d V,\d T$  collectively indicate fluctuations belonging the scalar, vector and traceless tensor sector respectively. $\mathcal{L}^{(1)}_{XY}$ indicates operator constructed from the equilibrium solution and derivative operators, which  acting on linearized fluctuations establish a map from the sector $Y$ to the sector $X$. Since we only consider spatially homogeneous background fields and fluctuations, if follows that the operators $\mathcal{L}$  effectively do not contain any derivative operator in the boundary spatial coordinates -- or rather these have a trivial effect. This, together with the fact that the equilibrium solution completely belongs to the singlet representation of $\sores$, implies that the  operators $\mathcal{L}^{(1)}_{XY}$ are diagonal in $XY$. Thus the three sector decouples completely 
\begin{align}  
\mathcal{L}^{(1)}_{SS} \d S = 0 ,  \qquad   \mathcal{L}^{(1)}_{VV} \d V   = 0,  \qquad  \mathcal{L}^{(1)}_{TT} \d T = 0 \, ,   
\end{align}
as we explicitly discussed in Sec.~\ref{sec:0pulsefreq}.  To linearized level the only non-vanishing perturbations we considered were the vector fluctuations, which decayed towards equilibrium with a rate $q = e^{-i\omega_* t}$ set by the lowest vector quasinormal mode  $\omega_*$.

Working at the quadratic level in the fluctuations, the equations of motion can now be written according to the structure  given by  
$\sores$ in the form 
\begin{align} \label{eq:2nd_eqs_schem}
 \mathcal{L}^{(2)}_{SS} \d\d S + \mathcal{L}^{(2)}_{SV} \d\d V +\mathcal{L}^{(2)}_{ST} \d\d T= \mathcal{J}_{S} ,  \nonumber \\
\mathcal{L}^{(2)}_ {VS}\d\d S + \mathcal{L}^{(2)}_{VV} \d\d V +\mathcal{L}^{(2)}_{VT} \d\d  T = \mathcal{J}_{V}, \\
\mathcal{L}^{(2)}_ {TS}\d\d S+ \mathcal{L}^{(2)}_{VS} \d\d S +\mathcal{L}^{(2)}_{TT} \d\d T = \mathcal{J}_{T}.  \nonumber 
\end{align}
Now  $\d\d X $  indicates quadratic fluctuations and, in a similar way as above, $\mathcal{L}^{(2)}_{XY}$ indicates an operator constructed from the equilibrium solution and derivative operators. The sources $\mathcal{J}_{X}$ collect terms that are quadratic in the linearized fluctuations  $\d S , \d V, \d T$. Using the same symmetry argument as above, together with the fact the we only have vectorial perturbations at the linear order,  we can conclude that the equations for the quadratic fluctuations relevant to our case take the form
\begin{align} 
 \mathcal{L}^{(2)}_{SS} \d\d S   = \mathcal{J}_{S} , \qquad     \mathcal{L}^{(2)}_{VV} \d\d V   = 0 , \qquad \mathcal{L}^{(2)}_{TT} \d\d T =  \mathcal{J}_{T} \, . 
\end{align}
Again, this consistently reproduces what we observed in Sec.~\ref{sec:0pulsefreq}.
In particular, as the leading excitations of the scalar and tensor sector are sourced by expressions that are quadratic in  $\d V \propto q$, we have  that $\d\d S, \d\d T \propto  \d V^2 \propto q^2$.
 
All in all, this analysis teaches us that the non-equilibrium solution written in a perturbative expansion has the schematic structure
\be \label{eq:backstruct}
\textrm{Background} \approx S + \d V +  \d\d S+ \d\d T +O(q^3) \, ,
\ee
with $S \propto q^0$   indicating the equilibrium, scalar sector, solution.  

Given this, we can proceed to analyze the linearized perturbations that compute the optical conductivity in the background  \eqref{eq:backstruct}. 
We indicate these new sets of linearized fluctuations  collectively as $\tilde \d s , \tilde  \d v, \tilde \d t$ to distinguish them from the background ones.  Similarly we use $\tilde{ \mathcal{L}}$  to indicate the operators that act on these fluctuations in the equations of motion, which take an analogous form as  \eqref{eq:lin_eqs_schem}.

We are interested in how much these fluctuations deviate from the form they would take around the background equilibrium, $O(q^0)$, solution. For this we organize our computation in an expansion in $q$. Making  this perturbative structure  explicit, we have up to second order in $q$ included
\begin{align}
&\tilde{ \mathcal{L}}_{ss} \approx \tilde{\mathcal{L}}_{ss, 0} +  \tilde{\mathcal{L}}^{S}_{ss,2}\d\d S  , 
\quad &\tilde{ \mathcal{L}}_{sv} &\approx  \tilde{\mathcal{L}}^{V}_{sv,1}\d V   \, ,  
 \quad &\tilde{ \mathcal{L}}_{st} &\approx   \tilde{\mathcal{L}}^{T}_{st,2}\d\d T \, ,\nonumber 
\\
&\tilde{ \mathcal{L}}_{vv} \approx  \tilde{\mathcal{L}}_{vv, 0} +  \tilde{\mathcal{L}}^{S}_{vv,2}\d\d S   +  \tilde{\mathcal{L}}^{T}_{vv,2}\d\d T    \, , 
& \tilde{ \mathcal{L}}_{vs} &\approx    \tilde{\mathcal{L}}^{V}_{vs,1}\d V    \, ,  
& \tilde{ \mathcal{L}}_{vt} &\approx     \tilde{\mathcal{L}}^{V}_{vt,1}\d V   \, ,   
\\
&\tilde{ \mathcal{L}}_{tt} \approx \tilde{\mathcal{L}}_{tt, 0} +  \tilde{\mathcal{L}}^{S}_{tt,2}\d\d S   \, ,
& \tilde{ \mathcal{L}}_{ts}& \approx   \tilde{\mathcal{L}}^{T}_{ts,2}\d\d T   \, , 
&\tilde{ \mathcal{L}}_{tv} &\approx     \tilde{\mathcal{L}}^{V}_{tv,1}\d V   \, . \nonumber 
\end{align}
 Similarly we write the fluctuations as  $\tilde \d s = \tilde\d s_0 + \tilde \d s_1 + \dots $, with $ \tilde \d s_i$ being order $q^i$,  and analogously  for the other sectors.

Solving the resulting equations order by order in $q$, the zeroth order problem corresponds to the study of linearized fluctuations around the thermal equilibrium value. All three sectors remain decoupled and to compute the optical conductivity one only excites the vector sector,  so only   $\tilde\d v_0$ is non-vanishing. 

At the next order in $q$  we have  
\be
\tilde{ \mathcal{L}}_{XY,0} \tilde \d Y_1 +  \tilde{ \mathcal{L}}_{XY,1} \tilde \d Y_0   = 0 ,  
\ee
and using the fact that $\tilde\d s_0 =\tilde\d t_0 =0$ and the explicit form of  $\tilde{ \mathcal{L}}_{XY,1}$ that can be read form above, it is easy to see that the equation for $\tilde \d v_1$ reduces to 
\be
\tilde{ \mathcal{L}}_{vv,0}  \tilde \d v_1  = 0 \, .  
\ee
Thus, there is no order $q$ contribution to the optical conductivity. 

At the next order instead 
\be
\tilde{ \mathcal{L}}_{XY,0} \tilde \d Y_2 +  \tilde{ \mathcal{L}}_{XY,1} \tilde \d Y_1 +\tilde{ \mathcal{L}}_{XY,2} \tilde \d Y_0   = 0 ,  
\ee
and generically all the sectors get sourced by the lower orders solution. Concentrating on $\tilde  \d v_2$, which is the relevant sector for the optical conductivity,  
\begin{align}
\tilde{ \mathcal{L}}_{vv,0} \tilde \d v_2  &= -   \tilde{ \mathcal{L}}_{vs,1} \tilde \d s_1  - \tilde{ \mathcal{L}}_{vt,1} \tilde \d t_1 - \tilde{ \mathcal{L}}_{vv,2} \tilde \d v_0  \nonumber \\
&= -    \tilde{\mathcal{L}}^{V}_{vs,1}\d V   \tilde \d s_1  -  \tilde{\mathcal{L}}^{V}_{vt,1}\d V \tilde \d t_1 - (\tilde{\mathcal{L}}^{S}_{vv,2}\d\d S   +  \tilde{\mathcal{L}}^{T}_{vv,2}\d\d T) \tilde \d v_0  \,. 
\end{align}
We therefore conclude that for the vector fluctuations the leading  deviation from their thermal value is of order $q^2$. That is, we showed that the rate of thermalization of the optical conductivity corresponds to $e^{-2i\omega_* t}$.


\section{Discussion} \label{sec:conjecture}
 
In this paper we  have analyzed the pattern of conductivity thermalization in a minimal holographic setting that includes finite charge density and a mechanism of weak momentum dissipation. We have shown that, when quenched by a laser pulse with a significant DC component, the equilibration time of the conductivity is given by half the lowest-lying imaginary bulk quasinormal mode: $\tau = - 1/ (2\Im  \omega_*)$. The appearance of the QNM governing momentum relaxation can be understood from the fact that the DC component of the electric field sets the charges of a finite-density system in motion, thus inducing finite momentum densities in the system. We have also provided a symmetry argument for the factor of two. If the mean frequency of the wave packet is large, the pulse lacks resonance with the corresponding QNM, and the thermalization is effectively instantaneous. The latter result is surprising from different points of view and gives rise to a number of questions.

The instantaneous thermalization of the conductivity is surely not intuitive from the boundary perspective, but  is somewhat natural from the bulk point of view once we are given a   background dynamics of the Vaidya form.  However, 
when going to finite density,  a priori  we would  not have expected the bulk dynamics to remain (even approximately) of the instantaneously thermalizing Vaidya form as in the zero density case \cite{Horowitz:2013mia,Bardoux:2012aw}. 
It is then interesting to ask to what extent this result is generic and, especially, whether it still holds in non-relativistic systems. A natural option would be to consider Einstein-Dilaton-Maxwell holographic models that explicitly exhibit Lifshitz scaling and hyperscaling violation, which have also been generalized to include momentum dissipation \cite{Cremonini:2016avj}. However, a closer look reveals that they do not allow for a dynamical electric field, as this will generically result in a modification of the dilaton profile and affect the scaling exponents. On the other hand, models that are Lorentz-invariant in the UV and where the relevant scaling properties only emerge in the IR (see, e.g., \cite{Davison:2013txa,Gouteraux:2014hca,Andrade:2016tbr}) seem to be less constraining in this sense, and could represent an interesting context where to explore this question further.

From a broader perspective, our observation  poses a question about the underlying principles governing such ultrafast equilibration. One might ask to which extent this surprising prediction depends on the precise details of the UV  description of the field theory, and more  specifically what is the role played by the large $N$ limit, on which our classical gravity computation relies.
Recently it has been  shown that the causal behavior of certain observables associated with the Vaidya geometry, in the zero density case, can be explicitly recovered  from conformal field theory structures in the limit of infinite central charge \cite{Anous:2016kss}. It has also been suggested that $1/N$-corrections deflect the system from this regime \cite{Anous:2016kss}. This might suggest that the Vaidya-like response of the holographic strange metal to the laser pulse is simply an artifact of the regime of classical gravity.   
The eigenstate thermalization hypothesis provides us with another way to think of this instantaneous equilibration, relating rapid thermalization of  local observables to the dense entanglement within the full many-body quantum system \cite{Srednicki,He:2017vyf}. From this perspective, it is possible that the seemingly universal behavior of holographic systems is not an artifact of specific limits, but rather a generic manifestation of the entangled nature of those states that have holographic duals.

In the same way that the holographic predictions have been shown to enjoy some UV independence through the minimal viscosity \cite{Kovtun:2004de} and the early onset of  the hydrodynamic behavior \cite{Chesler:2008hg} that are reflected in heavy ion collisions experiments \cite{Schafer:2009dj}, it would be   interesting to look for signatures of this instantaneous thermalization in condensed matter systems. This is the perspective we highlighted in the companion paper \cite{Bagrov:2017tqn}. 


\section*{Acknowledgements}
 We thank T.~Andrade, C.~Ecker, A.~Ficnar, B.~Gouteraux, M.P.~Heller, D.H.~Lee, L.~Rademaker, A.O.~Starinets, S.A.~Stricker and D.~Thompson for helpful discussions. 
This research has been supported in part by BELSPO (IAP P7/37), FWO-Vlaanderen (projects G020714N, G044016N and G006918N),
Academy of Finland (grant no 1297472), and the National Science Foundation (grant no NSF PHY-1125915). Research at Perimeter Institute is supported by the Government of Canada through Industry Canada and by the Province of Ontario through the Ministry of Research \& Innovation.
 
 \appendix 
 
 \section{Large $\w_P$ solution } \label{app:largew}
 
In order to work out the solutions in a $1/\w_P$  expansion, based on what we observed from the numerical results for the bulk solution, we formulate the following  ansatz for the expansion of the different  metric components  %
\begin{align}
&F_z(z,v) =  \frac{1}{z^2}\(1 - \frac{1}{2}k^2 z^2 - m z^3 +\frac{1}{4}\rho^2 z^4\) +\sum^{\infty}_{n=0} \w_{P}^{-n}  \(  F_{z,n}(z,v) + \w_{P}^{-1}  \tilde F_{z,n}(z, \w_{P} v)  \)  , \label{eq:first_eq_w} \\
&F_x(z,v) =\   \frac{1}{\w_{P}}\sum^{\infty}_{n=0} \w_{P}^{-n}   \tilde F_{x,n}(z,\w_{P}  v)   \ , \\
&B(z,v) =  \frac{1}{\w_{P}^3}\sum^{\infty}_{n=0} \w_{P}^{-n}   \tilde B_{n}(z,\w_{P}  v)   \ , \\
&\Sigma(z,v) = \frac{1}{z}+\frac{1}{\w_{P}^5}\sum^{\infty}_{n=0} \w_{P}^{-n} \tilde \Sigma_{n}(z \w_{P},v) \ , 
\end{align}
and gauge and scalar fields
\begin{align}
&a_v(z,v) = - \mu + \rho z    +\frac{1}{\w_{P}^3}\sum^{\infty}_{n=0}  \w_{P}^{-n} \tilde a_{v,n}(z ,\w_{P} v) \ , \\
&a_x(z,v)  = \frac{1}{\w_{P}^2}\sum^{\infty}_{n=0} \w_{P}^{-n} \tilde a_{x,n}(z,\w_{P} v) \ , \\
&\Phi(z,v) = \frac{1}{\w_{P}^2}\sum^{\infty}_{n=0}  \w_{P}^{-n} \tilde \Phi_{n}(z,\w_{P} v) \ .  \label{eq:last_eq_w}
\end{align}
The response to the time dependent pulse order by order can have contributions with or without large frequency. This is encoded in the two terms in the last sum in \eqref{eq:first_eq_w}. The functions $X_{n}$ appearing in the expansion are assumed to vary slowly in time, while   $\tilde X_{n}$   are quickly varying functions  and bring with them a factor $\omega$ every time they are hit with a time derivative. To make this manifest, we explicitly included a  ``time derivative counting factor'' $\w_{P}$ in their argument, which will eventually be set to one. Notice that when solving the resulting system of equations order by order, $X_{n}$ will enter at the same order as single time derivatives of the fields $\tilde X_{n+1}$.    
We will be working under the assumption that  in the pulse \eqref{eq:pulsesimple} the enveloping function $\Omega(t)$ has always negligible variation.

At leading order  the non-identically vanishing and linearly independent equations are the $E_{vv}, E_{vz}$ and $E_{vx}$ components of Einstein's equations, which read  
\begin{align}
& 2 \del_{v} \( F_{z,0}(z,v) + \frac{1}{\w_{P}}\tilde F_{z,1}(z, \w_{P} v) \)  = z E^2_x(v) \, , \\
&z \del_{z}F_{z,0}(z,v) - F_{z,0}(z,v) =0 \, , \\
&\frac{1}{\w_{P}}\del_{ v}\(2 \tilde F_{x,0}(z,  \w_{P} v ) + z \del_{z} \tilde F_{x,0}(z,  \w_{P} v )\) = z \rho E_x(v) \, .
\end{align} 
The equation for $\tilde F_{x,0}$ is readily solved as 
\be
\frac{1}{w}\tilde F_{x,0}(z,  v ) = \frac{1}{3} \rho  z \int_{-\infty}^{v} d v'~  E_x (v')   \, .
\ee
Similarly for  $F_z$ we have
\be 
F_{z,0}(z,v) + \frac{1}{\w_{P}}\tilde F_{z,0}(z,  v)  = - \frac{z}{2}  \int_{-\infty}^{v} d v'~  E_x^2(v') \, . \label{eq:FZleading}
\ee
Notice that the integral on the r.h.s.\ will in general give an order $\w_{P}^0$ contribution and an order $\w_{P}^{-1}$ contribution as reflected on the l.h.s.\ of the equation. That is 
\begin{align}
&F_{z,0}(z,v)  = \lim_{\w_{P}  \to \infty} - \frac{z}{2}  \int_{-\infty}^{v} d v'~  E_x^2(v') \, , \\
 &\tilde F_{z,0}(z,v)  = \lim_{\w_{P}  \to \infty} - \w_{P} \(\frac{z}{2}  \int_{-\infty}^{v} d v'~  E_x^2(v')  -F_{z,0}(z,v)\) \, .
\end{align}
For later convenience we define the   order $\w_{P}^0$  quantity 
\be
  M(v) \equiv  m+ \lim_{\w_{P}  \to \infty}   \frac{1}{2} \int_{-\infty}^{v} d v'~  E_x^2(v')   \label{eq:MM} \, . 
\ee
At the following order  the non identically  vanishing and linearly independent equations are the scalar equation for $\Phi$, Maxwell's equation for the $x$ component of the gauge field and the $E_{vx}$ component of Einstein's equations: 
\begin{align}
&  \frac{2}{\w_{P}}\del_{v}\( \t \Phi_{x,0}(z, \w_{P} v ) - \del_{z} \t \Phi_{x,0}(z, \w_{P} v )  \)  + k  z^2 \ F_{x,0}(z, \w_{P} v) =0 \ , \\
&  \frac{2}{\w_{P}}\del_{v} \del_{z} \t a_{x,0}(z, \w_{P} v ) - 3 \r z \t F_{x,0}(z, \w_{P} v) =0 \ , \\
&\frac{4}{\w_{P}}\del_{ v}\( 2 \tilde F_{x,1}(z,  \w_{P} v ) +   z \del_{z}  \tilde F_{x,1}(z,  \w_{P} v )\)   + (3 k^2 z + 2 z^3 \rho^2)   \tilde F_{x,0}(z, \w_{P} v) = 0   \, .
\end{align} 
These are immediately solved as 
\begin{align}
\frac{1}{\w_{P}^2}\t\Phi_{0}(z , v) & = \frac{ z^3    \rho k }{12}  \int^{v}_{-\infty}  \int^{v'}_{-\infty}E_x(v'') dv'' dv'\, ,  \\
\frac{1}{\w_{P}^2}\t a_{x,0}(z , v)  & = \frac{1}{6} z^3 \rho^2   \int^{v}_{-\infty}  \int^{v'}_{-\infty}E_x(v'') dv'' dv'\, ,  \\
\frac{1}{\w_{P}^2}\t F_{x,1}(z,v)  &= -\( \frac{z^2 \rho  k^2}{16}  + \frac{ z^4 \rho^3}{36} \)  \int^{v}_{-\infty}  \int^{v'}_{-\infty}E_x(v'') dv'' dv'   \, .
\end{align}
In addition, the $E_{vv}$ and $E_{vz}$ component of Einstein's equations read 
\begin{align}
& 2 \del_{v} \( F_{z,1}(z,v) + \frac{1}{\w_{P}}\tilde F_{z,1}(z, \w_{P} v) \)  =  - z^3 \r  E_x(v) \tilde F_{x,0}(z, \w_{P} v) \, , \\
&z \del_{z}F_{z,1}(z,v) - F_{z,1}(z,v) =0 \, , 
\end{align}
which imply
\begin{align}
&F_{z,1}(z,v) = 0 \ ,  \\
&\frac{1}{\w_{P}^2}\tilde F_{z,1}(z, v) = - \frac{1}{18} z^4 \rho^2 \int^{v}_{-\infty} E_x(v') \int^{v'}_{-\infty}E_x(v'') dv'' dv' \, .  
\end{align}

One can continue with this procedure, in principle, to arbitrary order. Without giving all the details we just report here the result of the next order computation, where $B$ and $a_v$ receive the first correction to their background values. These are
\begin{align}
 \frac{1}{\w_{P}^3}\tilde a_{v,0}(z, v ) &=  \frac{\r^3 z^6}{36} \int_{-\infty}^{v} \( \( \int^{v'}_{-\infty}E_x(v'') dv'' \)^2+ E_x(v') \int^{v'}_{-\infty}  \int^{v''}_{-\infty}E_x(v''') dv''' dv'' \) dv' \nonumber \\ 
&=  \frac{\r^3 z^6}{36}\(  \int^{v}_{-\infty}E_x(v') dv'\) \(   \int^{v}_{-\infty}  \int^{v'}_{-\infty}E_x(v'') dv'' dv'\),  \\
  \frac{1}{\w_{P}^3}\tilde B_{1}(z,  v ) &=  - \frac{\r^2 z^5}{16} \int_{-\infty}^{v} \( \( \int^{v'}_{-\infty}E_x(v'') dv'' \)^2+ E_x(v') \int^{v'}_{-\infty}  \int^{v''}_{-\infty}E_x(v''') dv''' dv'' \) dv' \nonumber \\ 
&= - \frac{\r^2 z^5}{16}\(  \int^{v}_{-\infty}E_x(v') dv'\) \(  \int^{v}_{-\infty}  \int^{v'}_{-\infty}E_x(v'') dv'' dv'\) .
\end{align}
The remaining fields receive the following corrections:
\begin{align}
\frac{1}{\w_{P}^3}\tilde\Phi_{1}(z, v ) &= - \frac{k \r z^4}{16}\int_{-\infty}^{v} \( k^2 +\frac{3}{2} M(v') z - \frac{2}{9} \r^2 z^2\) \int^{v'}_{-\infty}  \int^{v''}_{-\infty}E_x(v''') dv''' dv''dv'\, , \\
 \frac{1}{\w_{P}^3}\tilde a_{x,1}(z,  v ) &=  \frac{\r^2 z^2}{4}\int_{-\infty}^{v} \( 1  -\frac{5}{8} k^2 z^2  -  M(v') z^3 + \frac{7}{36} \r^2 z^4 \) \int^{v'}_{-\infty}  \int^{v''}_{-\infty}E_x(v''') dv''' dv''dv' \, ,\\
  \frac{1}{\w_{P}^3}\tilde F_{x,2}(z,  v ) &=  - \frac{\r z}{4}\int_{-\infty}^{v} \( \frac{1}{3}k^2 +   \frac{z^2}{5} \(-\frac{1}{2}k^4 + \r^2 \)  - \frac{5}{48}  z^3 k^2 M(v') - \frac{13}{168} k^2 \r^2 z^4  \right. \\
 &    \qquad \qquad \qquad \qquad\left.  -\frac{1}{8}  z^5 \rho^2 M(v')  + \frac{7}{324} \rho^4 z^6  \) \int^{v'}_{-\infty}  \int^{v''}_{-\infty}E_x(v''') dv''' dv''dv' \, , \nonumber  \\
  \frac{1}{\w_{P}^3}\tilde F_{z,3}(z,  v ) &=  \frac{\r^2 z^5}{864}  (9 k^2 + 4 z^2 \r^2) \int_{-\infty}^{v} \( \( \int^{v'}_{-\infty}E_x(v'') dv'' \)^2\right. \nonumber \\
  & \qquad \qquad \qquad \qquad \qquad \qquad \qquad\left. + E_x(v') \int^{v'}_{-\infty}  \int^{v''}_{-\infty}E_x(v''') dv''' dv'' \) dv' \nonumber \\ 
&= \frac{\r^2 z^5}{864}  (9 k^2 + 4 z^2 \r^2) \(  \int^{v}_{-\infty}E_x(v') dv'\) \(   \int^{v}_{-\infty}  \int^{v'}_{-\infty}E_x(v'') dv'' dv'\) \, , \\
\frac{1}{\w_{P}^2}F_{z,2}(z,  v ) &=  \frac{\r^2 z^4}{36}   \( \( \int^{v}_{-\infty}E_x(v') dv' \)^2 - 2   \int^{v}_{-\infty} E_x(v')  \int^{v'}_{-\infty}E_x(v'') dv'' dv' \) \, .  
 \end{align}
%


\section{Relation to current-current correlator and to equilibrium optical conductivity}\label{sec:curcurcor}

Using linear response theory, the current generated by a small external electric field
is given by
\beq
\delta J_{\mu}(x, t) = \int dt' dx'\, G_{R, \mu\nu}(x, t; x', t') A^{\nu}(x', t') \, ,
\eeq
where 
\beq
 G_{R, \mu\nu}(x, t; x', t') = -i\theta(t - t')\langle [J_{\mu}(x, t), J_{\nu}(x', t')]\rangle \, 
\eeq
is the retarded current-current correlator. Specializing to a homogeneous electric field by taking
\beq
A_i(x, t) =- \int^{t}_{t_i}dt'E_i(t') \, , 
\eeq
we obtain
\beq
\delta \langle J_i(x,t)\rangle = - \int_{t_i}^{\infty} dt'dx'\int^{t'}_{t_i} dt'' G_{R, ij}(x, t; x', t') E_j(t'') \, .
\eeq
Comparing to (\ref{eq:difcond}) we find that the differential conductivity can be written in terms of the current-current 
correlator as
\beq
\sigma(t, t'') = -\int_{t_i}^{\infty}dt'dx'\theta(t' - t'')G_{R, xx}(x, t; x', t').
\eeq
From the above formula it seems that $\sigma$ might be $x$ dependent. If the system is such that the 
$JJ$ correlator is invariant under translations, then the $x$ dependence disappears after performing the 
$x'$ integral. This is the case in the system we study. 

Next, we show that the non-equilibrium conductivity \eqref{eq:cond} reduces to the standard definition of optical conductivity 
when in thermal equilibrium and as the observation time  $t_m\rightarrow\infty$.
 This can be seen by Fourier transforming the definition of the differential conductivity (\ref{eq:difcond})
\beq
\delta \langle J_x(\omega)\rangle = \int_{-\infty}^{\infty} dt\,\delta \langle J_x(t)\rangle e^{i\omega t}
= \int_{-\infty}^{\infty} dt\, dt' e^{i\omega t}\theta(t-t')\sigma(t,t')\delta E_x(t').
\eeq
In thermal equilibrium $\sigma(t,t')$ can only depend on the difference $t-t'$ due to time translational symmetry
of the thermal density matrix, so that  $\sigma(t,t') =  \sigma(t -t',0)$. Changing the integration variable $t$ into $t'' = t-t'$, the two integrals separate
and we obtain
\beq
\delta \langle J_x(\omega)\rangle = \delta E_x(\omega) \int_{-\infty}^{\infty}dt'' \theta(t'')\sigma(t'', 0) e^{i\omega t''}.
\eeq
The second term on the righthand side can be recognized as the frequency space non-equilibrium conductity (\ref{eq:cond})
evaluated at $t=0$. Thus, we see that in thermal equilibrium it satisfies
\beq
\sigma(\omega, 0) = \frac{\delta \langle J_x(\omega)\rangle }{\delta E_x(\omega)},\label{eq:equilibrium_case}
\eeq
which is the standard definition of optical conductivity in thermal equilibrium. Since $\sigma(\omega, t)$
is independent of $t$ in thermal equilibrium,  (\ref{eq:equilibrium_case}) holds for arbitrary $t$. This
establishes the claim that the non-equilibrium conductivity reduces to the equilibrium conductivity 
in thermal equilibrium and as $t_m\rightarrow\infty$. If $t_m$ is finite, the step that fails above is the
factorization of the integrals after the change of variables $t''= t-t'$, since this changes the integration limits.
But as long as one considers values of $t\ll t_m$, the integral defining $\sigma(\omega, t)$ can be safely 
extended to infinity.




\begin{thebibliography}{99}

\bibitem{Ammon:2015wua}
  M.~Ammon and J.~Erdmenger,
 ``Gauge/gravity duality: Foundations and applications,''
Cambridge University Press 2015.
 
\bibitem{Zaanenetalbook}
  J.~Zaanen, Y.~W.~Sun, Y.~Liu and K.~Schalm,
 ``Holographic Duality in Condensed Matter Physics,''
Cambridge University Press 2015.

\bibitem{DeWolfe:2013cua}
  O.~DeWolfe, S.~S.~Gubser, C.~Rosen and D.~Teaney,
  ``Heavy ions and string theory,''
  Prog.\ Part.\ Nucl.\ Phys.\  {\bf 75} (2014) 86
  doi:10.1016/j.ppnp.2013.11.001
  [\arXiv{arXiv:1304.7794} [hep-th]].

\bibitem{Murata:2010dx}
  K.~Murata, S.~Kinoshita and N.~Tanahashi,
  ``Non-equilibrium Condensation Process in a Holographic Superconductor,''
  JHEP {\bf 1007} (2010) 050
  doi:10.1007/JHEP07(2010)050
  [\arXiv{arXiv:1005.0633} [hep-th]].
 
\bibitem{Bhaseen:2012gg}
  M.~J.~Bhaseen, J.~P.~Gauntlett, B.~D.~Simons, J.~Sonner and T.~Wiseman,
   ``Holographic Superfluids and the Dynamics of Symmetry Breaking,''
  Phys.\ Rev.\ Lett.\  {\bf 110} (2013) no.1,  015301
  doi:10.1103/PhysRevLett.110.015301
  [\arXiv{arXiv:1207.4194} [hep-th]].

\bibitem{Chesler:2014gya}
  P.~M.~Chesler, A.~M.~Garcia-Garcia and H.~Liu,
  ``Defect Formation beyond Kibble-Zurek Mechanism and Holography,''
  Phys.\ Rev.\ X {\bf 5} (2015) no.2,  021015
  doi:10.1103/PhysRevX.5.021015
  [\arXiv{arXiv:1407.1862} [hep-th]].

\bibitem{Sonner:2014tca}
  J.~Sonner, A.~del Campo and W.~H.~Zurek,
   ``Universal far-from-equilibrium Dynamics of a Holographic Superconductor,''
  Nature Commun.\  {\bf 6} (2015) 7406
  doi:10.1038/ncomms8406
  [\arXiv{arXiv:1406.2329} [hep-th]].

\bibitem{Callebaut:2014tva}
  N.~Callebaut, B.~Craps, F.~Galli, D.~C.~Thompson, J.~Vanhoof, J.~Zaanen and H.~b.~Zhang,
  ``Holographic Quenches and Fermionic Spectral Functions,''
  JHEP {\bf 1410} (2014) 172
  doi:10.1007/JHEP10(2014)172
  [\arXiv{arXiv:1407.5975} [hep-th]].

  \bibitem{Das:2014lda}
  S.~R.~Das and T.~Morita,
   ``Kibble-Zurek Scaling in Holographic Quantum Quench: Backreaction,''
  JHEP {\bf 1501} (2015) 084
  doi:10.1007/JHEP01(2015)084
  [\arXiv{arXiv:1409.7361} [hep-th]].

  \bibitem{Camilo:2015wea}
  G.~Camilo, B.~Cuadros-Melgar and E.~Abdalla,
  ``Holographic quenches towards a Lifshitz point,''
  JHEP {\bf 1602} (2016) 014
  doi:10.1007/JHEP02(2016)014
  [\arXiv{arXiv:1511.08843} [hep-th]].

  \bibitem{Zeng:2016api}
  H.~B.~Zeng, Y.~Tian, Z.~Y.~Fan and C.~M.~Chen,
   ``Nonlinear Transport in a Two Dimensional Holographic Superconductor,''
  Phys.\ Rev.\ D {\bf 93} (2016) no.12,  121901
  doi:10.1103/PhysRevD.93.121901
  [\arXiv{arXiv:1604.08422} [hep-th]].
 
\bibitem{Zeng:2016gqj}
  H.~B.~Zeng, Y.~Tian, Z.~Fan and C.~M.~Chen,
   ``Nonlinear Conductivity of a Holographic Superconductor Under Constant Electric Field,''
  Phys.\ Rev.\ D {\bf 95} (2017) no.4,  046014
  doi:10.1103/PhysRevD.95.046014
  [\arXiv{arXiv:1611.06798} [hep-th]].
  
  \bibitem{Withers:2016lft}
  B.~Withers,
 ``Nonlinear conductivity and the ringdown of currents in metallic holography,''
  JHEP {\bf 1610} (2016) 008
    doi:10.1007/JHEP10(2016)008
  [\arXiv{arXiv:1606.03457} [hep-th]].
  
\bibitem{Orenstein}
J.~Orenstein,``Ultrafast spectroscopy of quantum materials,''
Physics Today {\bf 65} 9 (2012) 44
doi: 10.1063/PT.3.1717

\bibitem{DalConte}
S.~Dal Conte et al., 
``Snapshots of the retarded interaction of charge carriers with ultrafast fluctuations in cuprates,"
Nature Physics {\bf 11}, 421-426 (2015)
doi: 10.1038/nphys3265
[\arXiv{arXiv:1501.03833} [cond-mat.supr-con]].

\bibitem{Giannettietal}
C.~Giannetti, M.~Capone, D.~Fausti, M.~Fabrizio, F.~Parmigiani, D.~Mihailovic,
 ``Ultrafast optical spectroscopy of strongly correlated materials and high-temperature superconductors: a non-equilibrium approach,''
Advances in Physics, 65:2, 58-238 (2016) 
doi: 10.1080/00018732.2016.1194044
[\arXiv{arXiv:1601.07204} [cond-mat.supr-con]].

\bibitem{Freericks}
J.~K.~Freericks, O.~P.~Matveev, W.~Shen, T.~P.~Devereaux,
``Theoretical description of pump/probe experiments in  electron mediated charge-density-wave insulators,''
Physica Scripta, 92 3 034007 (2017),
doi: 10.1088/1402-4896/aa5b6c
[\arXiv{arXiv:1610.02613} [cond-mat.supr-con]].
  
\bibitem{Andrade:2013gsa}
  T.~Andrade and B.~Withers,
 ``A simple holographic model of momentum relaxation,''
 doi:10.1007/JHEP05(2014)101
  JHEP {\bf 1405} (2014) 101
  [\arXiv{arXiv:1311.5157} [hep-th]].
  
\bibitem{Horowitz:2012ky}
  G.~T.~Horowitz, J.~E.~Santos and D.~Tong,
  ``Optical Conductivity with Holographic Lattices,''
  JHEP {\bf 1207} (2012) 168
  doi:10.1007/JHEP07(2012)168
  [\arXiv{arXiv:1204.0519} [hep-th]].

  \bibitem{Vegh:2013sk}
  D.~Vegh,
  ``Holography without translational symmetry,''
  \arXiv{arXiv:1301.0537} [hep-th].
  
\bibitem{Blake:2013owa}
  M.~Blake, D.~Tong and D.~Vegh,
  ``Holographic Lattices Give the Graviton an Effective Mass,''
  Phys.\ Rev.\ Lett.\  {\bf 112} (2014) no.7,  071602
  doi:10.1103/PhysRevLett.112.071602
  [\arXiv{arXiv:1310.3832} [hep-th]].
  
\bibitem{Donos:2013eha}
  A.~Donos and J.~P.~Gauntlett,
  ``Holographic Q-lattices,''
  JHEP {\bf 1404} (2014) 040
  doi:10.1007/JHEP04(2014)040
  [\arXiv{arXiv:1311.3292} [hep-th]].  
  
 \bibitem{Horowitz:2013mia}
  G.~T.~Horowitz, N.~Iqbal and J.~E.~Santos,
  ``Simple holographic model of nonlinear conductivity,''
  Phys.\ Rev.\ D {\bf 88} (2013) no.12,  126002
  doi:10.1103/PhysRevD.88.126002
  [\arXiv{arXiv:1309.5088} [hep-th]].
  
  \bibitem{Bardoux:2012aw}
  Y.~Bardoux, M.~M.~Caldarelli and C.~Charmousis,
  ``Shaping black holes with free fields,''
  JHEP {\bf 1205} (2012) 054
  doi:10.1007/JHEP05(2012)054
  [\arXiv{arXiv:1202.4458} [hep-th]].

\bibitem{Bagrov:2017tqn}
  A.~Bagrov, B.~Craps, F.~Galli, V.~Ker\"anen, E.~Keski-Vakkuri and J.~Zaanen,
  ``Holography and thermalization in optical pump-probe spectroscopy,''
  Phys.\ Rev.\ D {\bf 97} (2018) 086005
    doi:10.1103/PhysRevD.97.086005
    [\arXiv{arXiv:1708.08279} [hep-th]].
  
\bibitem{Davison:2015bea}
  R.~A.~Davison and B.~Gout\'eraux,
  ``Dissecting holographic conductivities,''
  JHEP {\bf 1509} (2015) 090
   doi:10.1007/JHEP09(2015)090
  [\arXiv{arXiv:1505.05092} [hep-th]].
  %

  \bibitem{Davison:2013jba} 
  R.~A.~Davison,
 ``Momentum relaxation in holographic massive gravity,''
  Phys.\ Rev.\ D {\bf 88} (2013) 086003
  doi:10.1103/PhysRevD.88.086003
  [\arXiv{arXiv:1306.5792} [hep-th]].
 
\bibitem{Chesler:2008hg}
  P.~M.~Chesler and L.~G.~Yaffe,
 ``Horizon formation and far-from-equilibrium isotropization in supersymmetric Yang-Mills plasma,''
  Phys.\ Rev.\ Lett.\  {\bf 102} (2009) 211601
    doi:10.1103/PhysRevLett.102.211601
  [\arXiv{arXiv:0812.2053} [hep-th]].
 
\bibitem{Heller:2013oxa}
  M.~P.~Heller, D.~Mateos, W.~van der Schee and M.~Triana,
  ``Holographic isotropization linearized,''
  JHEP {\bf 1309} (2013) 026
    doi:10.1007/JHEP09(2013)026
  [\arXiv{arXiv:1304.5172} [hep-th]].
 
\bibitem{Chesler:2013lia}
  P.~M.~Chesler and L.~G.~Yaffe,
  ``Numerical solution of gravitational dynamics in asymptotically anti-de Sitter spacetimes,''
  JHEP {\bf 1407} (2014) 086
    doi:10.1007/JHEP07(2014)086
  [\arXiv{arXiv:1309.1439} [hep-th]].
 
 \bibitem{Ecker:2015kna}
  C.~Ecker, D.~Grumiller and S.~A.~Stricker,
  ``Evolution of holographic entanglement entropy in an anisotropic system,''
  JHEP {\bf 1507} (2015) 146
    doi:10.1007/JHEP07(2015)146
  [\arXiv{arXiv:1506.02658} [hep-th]].
  
\bibitem{Buchel:2013gba}
  A.~Buchel, R.~C.~Myers and A.~van Niekerk,
  ``Universality of Abrupt Holographic Quenches,''
  Phys.\ Rev.\ Lett.\  {\bf 111} (2013) 201602
  doi:10.1103/PhysRevLett.111.201602
  [\arXiv{arXiv:1307.4740} [hep-th]].

 \bibitem{Lenarcic2014}
 Z.~Lenarcic, D.~Golez, J.~Bonca, P.~Prelovsek,
 ``Optical response of highly excited particles in a strongly correlated system,"
Phys. Rev. B \textbf{89}, 125123 (2014)
[\arXiv{arXiv:1312.1962} [cond-mat.str-el]]
  
\bibitem{Cremonini:2016avj} 
  S.~Cremonini, H.~S.~Liu, H.~Lu and C.~N.~Pope,
  ``DC Conductivities from Non-Relativistic Scaling Geometries with Momentum Dissipation,''
  JHEP {\bf 1704}, 009 (2017)
  doi:10.1007/JHEP04(2017)009
  [\arXiv{arXiv:1608.04394} [hep-th]].
 
\bibitem{Davison:2013txa}
  R.~A.~Davison, K.~Schalm and J.~Zaanen,
   ``Holographic duality and the resistivity of strange metals,''
  Phys.\ Rev.\ B {\bf 89} (2014) no.24,  245116
  doi:10.1103/PhysRevB.89.245116
  [\arXiv{arXiv:1311.2451} [hep-th]].
 
\bibitem{Gouteraux:2014hca}
    B.~Gout\'eraux,
  ``Charge transport in holography with momentum dissipation,''
  JHEP {\bf 1404} (2014) 181
  doi:10.1007/JHEP04(2014)181
  [\arXiv{arXiv:1401.5436} [hep-th]].
  

  \bibitem{Andrade:2016tbr}
  T.~Andrade,
  ``A simple model of momentum relaxation in Lifshitz holography,''
 \arXiv{arXiv:1602.00556} [hep-th].
  
\bibitem{Anous:2016kss} 
  T.~Anous, T.~Hartman, A.~Rovai and J.~Sonner,
  ``Black Hole Collapse in the 1/c Expansion,''
  JHEP {\bf 1607}, 123 (2016)
    doi:10.1007/JHEP07(2016)123
  [\arXiv{arXiv:1603.04856} [hep-th]].

\bibitem{He:2017vyf} 
  S.~He, F.~L.~Lin and J.~j.~Zhang,
   ``Subsystem eigenstate thermalization hypothesis for entanglement entropy in CFT,''
  JHEP {\bf 1708}, 126 (2017)
  doi:10.1007/JHEP08(2017)126
  [\arXiv{arXiv:1703.08724} [hep-th]].

\bibitem{Srednicki}
M.~Srednicki,
``Chaos and quantum thermalization,''
Phys.\ Rev.\ E {\bf 50} (1994) 888
doi: 10.1103/PhysRevE.50.888
[\arXiv{arXiv:9403051} [cond-mat]].

\bibitem{Kovtun:2004de}
  P.~Kovtun, D.~T.~Son and A.~O.~Starinets,
  ``Viscosity in strongly interacting quantum field theories from black hole physics,''
  Phys.\ Rev.\ Lett.\  {\bf 94} (2005) 111601
    doi:10.1103/PhysRevLett.94.111601
   [\arXiv{hep-th/0405231}].
 
\bibitem{Schafer:2009dj}
  T.~Schaefer and D.~Teaney,
  ``Nearly Perfect Fluidity: From Cold Atomic Gases to Hot Quark Gluon Plasmas,''
  Rept.\ Prog.\ Phys.\  {\bf 72} (2009) 126001
  doi: 10.1088/0034-4885/72/12/126001
  [\arXiv{arXiv:0904.3107} [hep-ph]].
   


\end{thebibliography}
\end{document}